\documentclass[fleqn,usenatbib]{mnras}

\usepackage{newtxtext,newtxmath}

\usepackage[T1]{fontenc}

\DeclareRobustCommand{\VAN}[3]{#2}
\let\VANthebibliography\thebibliography
\def\thebibliography{\DeclareRobustCommand{\VAN}[3]{##3}\VANthebibliography}

\usepackage{graphicx}	
\usepackage{amsmath}	
\usepackage{gensymb}
\usepackage{verbatim}

\title[Structural Analysis of 170 SMC SCs]{STEP Survey II: Structural Analysis of 170 star clusters in the SMC\thanks{This work is based on INAF-VST guaranteed observing time under ESO programmes: 099.D-0673(A); 098.D-0579(A); 097.D-0209(A); 096.D-0535(A); 095.D-0132(A); 094.D-0492(A); 093.D-0174(A); 092.D-0214(A); 091.D-0574(A); 090.D-0172(A); 089.D-0258(A); 088.D-4014(A).}}

\author[M. Gatto et al.]{
M. Gatto,$^{1,2}$\thanks{E-mail: massimiliano.gatto@inaf.it}
V. Ripepi,$^{1}$
M. Bellazzini,$^{3}$
M. Tosi,$^{3}$
M. Cignoni,$^{3,4,5}$
C. Tortora,$^{1}$
\newauthor{S. Leccia$^{1}$, G. Clementini$^{3}$, E.K. Grebel$^{6}$,
G. Longo$^{2}$, M. Marconi$^{1}$, I. Musella$^{1}$}
\\
$^{1}$ INAF-Osservatorio Astronomico di Capodimonte, Via Moiariello
 16, 80131, Naples, Italy \\
$^{2}$ Dept. of Physics, University of Naples Federico II, C.U. Monte Sant'Angelo, Via Cinthia, 80126, Naples, Italy\\
$^{3}$ INAF-Osservatorio di Astrofisica e Scienza dello Spazio, Via Gobetti 93/3, I-40129 Bologna, Italy \\
$^{4}$ Physics Departement, University of Pisa, Largo Bruno Pontecorvo, 3, I-56127 Pisa, Italy\\
$^{5}${INFN, Largo B. Pontecorvo 3, 56127, Pisa, Italy}\\
$^{6}${Astronomisches Rechen-Institut, Zentrum f$\ddot{u}$r Astronomie der Universit$\ddot{a}$t Heidelberg, M$\ddot{o}$nchhofstr 12–14, D-69120 Heidelberg, Germany}\\
}

\date{Accepted XXX. Received YYY; in original form ZZZ}

\pubyear{2020}

\begin{document}
\label{firstpage}
\pagerange{\pageref{firstpage}--\pageref{lastpage}}
\maketitle

\begin{abstract}
We derived surface brightness profiles in the \emph{g} band for 170 Small Magellanic Cloud (SMC) star clusters (SCs) mainly located in the central region of the galaxy. We provide a set of homogeneous structural parameters obtained by fitting Elson, Fall \& Freeman (EFF) and King models. Through a careful analysis of their colour-magnitude diagrams (CMDs) we also supply the ages for a subsample of 134 SCs.
For the first time, such a large sample of SCs in the SMC is homogeneously characterized in terms of their sizes, luminosities and masses, widening the probed region of the parameter space, down to hundreds of solar masses.
We used these data to explore the evolution of the SC's structural parameters with time. 
In particular, we confirm the existence of a physical mechanism that induces an increase of the core radius after 0.3-1.0 Gyr. We suggest that cluster mass could be the main parameter driving the inner expansion, as none of the SCs having $\log (M/M_{\odot}) \leq 3.5$~dex analysed in this work undergoes to such an expansion.
We also detected a mass-size relationship almost over the entire range of SCs masses investigated here. 
Finally, our data suggest that globally the SMC SC system is dynamically evolved.
\end{abstract}

\begin{keywords}
(galaxies:) Magellanic Clouds - galaxies: star clusters: general - (stars:) Hertzsprung-Russell and colour-magnitude diagrams - surveys
\end{keywords}



\section{Introduction}

The Magellanic Clouds (MCs) are an excellent benchmark for the study of the evolution of star clusters (SCs). Their cluster system includes also very massive ($\log(M/M_{\odot}) \sim 5$~dex) SCs of young ($t \leq$ 1 Gyr) or intermediate age (1 Gyr $\leq t \leq$ 7 Gyr), which are absent in our Galaxy, enabling us to probe a broader region of the parameter space, and in turn, making it possible to investigate the evolutionary processes working in a cluster over cosmic time.
Thanks to several studies aimed at the analysis of clusters in the Large (LMC) and Small Magellanic Cloud (SMC) key progress in this field has been achieved in the last decades. 
\citet{Elson-1989} found a trend between the core radius $r_{\rm c}$ (the distance from the cluster centre where the surface brightness drops down to half its central value) and the age of clusters, namely, the older the SC, the larger the spread in core radius. 
This relationship was confirmed by subsequent works in the LMC \citep[e.g.][]{Elson-1991,Elson-1992,Mackey&Gilmore2003,Carvalho-2008}, SMC \citep[e.g.][]{Mackey&Gilmore2003b,Carvalho-2008} and in other systems like M33, M51, M83, NGC628 and NGC1313 \citep[e.g.][]{SanRoman-2012,Bastian2012,Ryon-2015,Chandar2016,Ryon-2017}. 
\citet{Mackey&Gilmore2003} explored the possibility that some bias due to selection effects could explain this trend, but they ruled it out, concluding that the apparent behaviour reflects a real evolutionary process.
In this scenario, all SCs formed with small core radii ($\sim$ 2-3 pc), then some of them remained compact, while others, after hundreds of Myr, expanded their inner regions.
However, the physical mechanism inducing the core expansion is still debated. 
Energy transfer from the inner to outer regions due to stellar mass black holes \citep{Merritt-2004,Mackey-2007,Mackey-2008}, mass loss due to stellar evolution \citep{Mackey-2007, Mackey-2008}, and/or residual gas expulsion \citep{Kroupa-2001a, Goodwin&Bastian2006, Baumgardt&Kroupa2007, Banerjee&Kroupa2017} are the main processes invoked to explain the phenomenon. Also the presence of a central intermediate-mass black hole (IMBH) can lead to a core expansion \citep{Baumgardt2004a, Baumgardt2004b}.
More likely, a combination of two or three of the above processes could provide the observed evolution in the core radius \citep[e.g.][]{Bastian-2008}.\par
A key ingredient for the study of the dynamical evolution of star clusters is the availability of accurate and homogeneously derived surface brightness profiles (SBPs), extending as far as possible from the cluster centre.
Here we focus on the SMC, which contains a large sample of SCs. The most complete and recent catalog by \citet{Bica-2020} reports about 850 SMC objects classified as SCs.
Nowadays, thanks to many surveys covering both the MCs \citep[e.g. MCPS, SMASH, STEP, YMCA, VISCACHA;][]{Zaritsky1997,nidever-2017,Ripepi2014,Maia2019}, it is feasible to collect detailed SBPs for a statistically significant sample of SCs to better understand the mechanisms ruling SC evolution.
King profiles are employed to approximate SBP shapes of most of the Galactic globular clusters (GGC) \citep[e.g.][]{King1962}.
However, \citet{Elson1987} pointed out that a different kind of analytical function, nowadays called Elson, Fall and Freeman  (EFF) profile, better describes the SBP of the LMC SCs. The main difference between the two quoted profiles is that the latter does not present a ``truncation'' at large radii that separates the SC from the field.
Which of the two profiles better assesses the SBP shape of the MCs SCs is still debated.
\citet{Hill&Zaritsky2006} (HZ06 hereafter), and later \citet{werchan-2011}, analyzed structural parameters of 204 SMC and 1066 LMC SCs, respectively, fitting the SBP with both the above mentioned profiles.
Despite the significant statistical sample of analysed SCs these authors did not find any favored profile, both of them were satisfactory for the majority of the SCs, with King profiles performing slightly better.
Very recently, \citet{Santos-2020} carried out a detailed analysis of 83 SCs located at the periphery of both the MCs in the context of the VIsible Soar photometry of star Clusters in tApii and Coxi HuguA \citep[VISCACHA -][]{Maia2019} survey, with the SOuthern Astrophysical Research (SOAR) telescope. The analysis of the SBPs of these 83 SCs, fitted with both King and EFF profiles, allowed these authors to confirm that the SPBs can be matched almost equally well by both EFF and King models. At the same time, they assessed the presence of an evolution of $r_{\rm c}$ for older SCs in both the LMC and SMC, as already suggested in the literature  \citep[][]{Santos-2020}.\par
In this paper we infer the dynamical evolution of a sample of SCs through an extensive work based on an accurate assessment of the SC main properties by means of the study of their SBPs in the \emph{g} band and colour-magnitude diagrams (CMDs) in the \emph{g} and \emph{i} bands.
Our aim is to provide the scientific community with new accurate structural parameters like sizes, luminosities, masses derived via a model fitting of both King and EFF profiles. We also supply age estimates for a subsample of SCs.
Our sample consists of 170 SMC SCs, whose photometry was obtained with the {\it ``The SMC in Time: Evolution of a Prototype interacting late-type dwarf galaxy''} \citep[STEP: PI V. Ripepi,][]{Ripepi2014} survey, which are already listed in the catalogue by \citet{Bica-2020}, located throughout the inner parts (most of them reside within $\sim$ 2-3 deg) of the SMC, and spanning a wide range of ages and masses. 
Of these 170 SCs, 62 have never been examined before. Our sample extends the range of masses towards values lower than those already present in the literature.
As far as we are aware, it is the first time that a wide-range of structural parameters, from sizes to masses, have been derived homogeneously in the SMC. Moreover, these features are used to study the dynamical phase of SCs, probing the age of the SMC SC system, from the dynamical point of view.
Finally, SC properties are also used as a tool to gain hints on the SMC environment by inspecting how they depend on the distance from the SMC centre.
We exploit the data from the photometric STEP survey.
This survey reaches \emph{g}$\sim$25 mag (in non-heavily crowded regions), well below the main sequence turn-off (TO) of the oldest stellar population, allowing us to investigate accurately even SCs with very low surface brightness.
The high resolution of STEP even at faint magnitudes makes it feasible to achieve a reliable accuracy even for SCs of $\sim$ 100 solar masses, probing in detail a locus of the parameter space barely explored with a conspicuous number of SCs.\par
The paper is structured as follows: we present the observations and data reduction strategy along with the construction of the sample in Sec.~\ref{sec:data}. In Sec.~\ref{sec:method} we describe the procedure adopted to derive the SBPs, while Sec.~\ref{sec:fitting_procedure} presents the derivation of SC structural parameters. We report the age assessment in Sec.~\ref{sec:cmd}. In Sec.~\ref{sec:results} we present the main results of this work, while in Sec.~\ref{sec:discussion} we discuss them. Finally, a brief summary is reported in Sec.~\ref{sec:conclusion}.

\section{Data}
\label{sec:data}

\subsection{The STEP survey}

As mentioned above, the data used in this work were obtained in the context of the STEP survey, carried out with the VLT Survey Telescope as part of the Guaranteed Time Observations (GTO) awarded by the European Southern Observatory (ESO) to the Istituto Nazionale di Astrofisica (INAF). The telescope is equipped with OmegaCAM, a mosaic camera of 32-CCD, $16k\times16k$ detectors with a pixel scale of 0.214 arcsec/pixel and with a field-of-view (FOV) of 1 deg$^2$. 
STEP extends over 53 deg$^2$ covering the SMC and the Bridge connecting the two Clouds. Full details of the observing strategy can be found in \citet{Ripepi2014}. That paper also presents a full description of the data reduction, including procedures of pre-reduction, astrometry and stacking of the different dithered frames into a single mosaic image, which were performed with the VST--Tube imaging pipeline \citep{Grado2012}. The point spread function (PSF) photometry was carried out using DAOPHOT IV/ALLSTAR \citep[][]{Stetson1987,Stetson1992}. As for the photometric calibration, it was improved with respect to \citet{Ripepi2014}, with the adoption of the local standard stars made available by the AAVSO Photometric All-Sky Survey (APASS). The different steps of the photometric calibration are sketched in \citet{Gatto2020}, whereas full details will be provided elsewhere (Ripepi et al. in preparation). Here we only recall that the overall accuracy of the photometry is $\sim$0.02 and $\sim$0.025 mag in $g$ and $i$, respectively. 
Finally, the photometrically calibrated catalogues were cleaned from extended/spurious objects by keeping only objects with typically -0.6$\leq SHARPNESS \leq$0.7, where $SHARPNESS$ is an output parameter of the DAOPHOT package adopted to identify extended objects.   

\subsection{Building the star cluster sample}

To achieve our goals, we built a statistically significant sample of SCs, covering as much as possible different regions of the SMC.
We focused our research on all objects classified as star clusters (i.e., those indicated with the letter \emph{C} in the column {\sc type} of the catalogue) by \citet{Bica2008}.
Then, we restricted the list to those SCs whose spatial positions overlapped with tiles 3\_3, 3\_4, 3\_5, 3\_6, 3\_7, 3\_8, 4\_4, 4\_5, 4\_6, 5\_5, 5\_6 \citep[see Fig.~2 and Table~2 in][for the definition of the tiles]{Ripepi2014}.\par
Despite our efforts, we could not analyse all the 404 SCs present in \citeauthor{Bica2008}'s \citeyearpar{Bica2008} catalogue and falling on the above quoted tiles. In fact, we had to discard almost all the SCs with a minor axis smaller than $\sim$ 0.5\arcmin~, corresponding to about 40\% of the starting list, because they showed too noisy SBPs.
Moreover, we had to reject also those SCs whose centres were too close to the edges of the tile ($\sim$10\%) or which on the images did not appear as real SCs ($\sim$10\%). At the end, we were left with 170 usable SCs.
Figure~\ref{fig:cluster_positions} shows the SC relative positions with respect to the SMC centre defined by classical Cepheids variables: ($\alpha_0, \delta_0$) = (12.54, -73.11), \citet{Ripepi2017}. This centre is very similar to the one adopted by \citet{Gonidakis2009} based on K and M giants, i.e. ($\alpha_0, \delta_0$) = (12.75, -73.10).

\section{Method}
\label{sec:method}

For each SC we derived the SBP (i.e. mag/arcsec$^2$ as a function of the cluster-centric distance) by means of integrated aperture photometry.
After a careful analysis of SPBs in both the $g$ and $i$ bands, we choose to work with the $g-$band, since the data in this filter provide significantly less noisy SPBs for the large majority of the SCs in our sample.
Detection of individual stars in such distant and crowded clusters is biased by significant incompleteness, with strong variation with distance from the cluster center. Integrated photometry overcomes this problem, allowing a safe tracing of the light density distribution.

\subsection{Star cluster centre estimation}
\label{sec:centre_estimation}

The key to compute accurately the SBP of a SC is to carefully determine its centre.
Our sample contains many SCs that are very patchy and irregular, making the centre estimation procedure rather tricky.
Since the majority is also asymmetric, it is unsuitable to make use of algorithms, such as the mirror-autocorrelation method described by \citet{Djorgovski1988}, which takes advantage of the symmetry to properly determine the centre of an object \citep{Mackey&Gilmore2003}.
Therefore, we preferred to use individual star positions to locate the ``centroid" of the cluster and define it as SC centre.
To this aim, we employed an iterated two dimensional Kernel Density Estimation (KDE)\footnote{A KDE is a non-parametric technique that works by smoothing data through a kernel function in order to estimate the probability density function of a random variable. We used the version available in the scikit-learn package \citep{scikit-learn}} by using the star coordinates R.A. and Dec as inputs to find the SC centre.
As a starting point, we used the SC centres and major axes by \citet{Bica-2020} to build surface density maps via KDE, obtained by calculating its value in a circular region centred on the SC with a radius twice its major axis.
For each SC, we updated the SC centre by assigning it the coordinates where the surface density map has its maximum, and we repeated the same procedure until two successive estimated centres differ by less than 1\arcsec.
To better assess the SC centre, we ran the KDE many times by varying the bandwidth\footnote{The bandwidth of the kernel is the only parameter that must be set in a KDE.} of the kernel function in the range 0.01\arcmin- 0.4\arcmin.
Finally, we took the mean of all centroids as our best estimate for each SC centre.\par
To check the validity of our procedure, we compared our SC centres with those derived by \citet{Carvalho-2008}, who have 19 SCs in common with our sample. The mean difference between the two centre evaluations is below 2\arcsec.
In the second and third column of Tab.~\ref{tab:results_fit_params} we listed all SC centres determined with the described procedure.

\begin{figure}
    \centering
    \includegraphics[scale = 0.35]{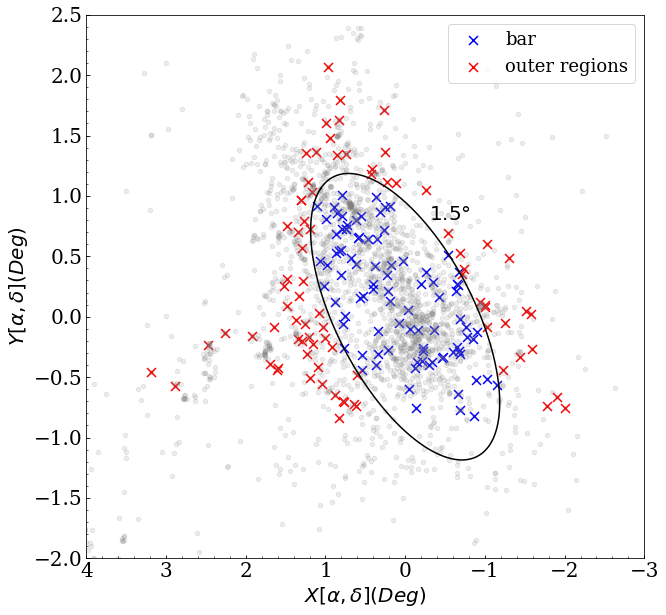}
    \caption{Positions of all 170 SCs studied in this paper, in a zenithal equidistant projection, with respect to the SMC centre \citep{Ripepi2017}. Small grey dots show the whole SC catalogue by \citet{Bica-2020}.
    Blue crosses represent SCs within an ellipse with its semimajor axis $a = 1.5 \degree$ and ellipticity $e = 0.5$, whereas red crosses are the SCs outside the ellipse (see Appendix \ref{App:SMC_distance}).}
    \label{fig:cluster_positions}
\end{figure}

\subsection{Surface brightness determination.}
\label{sec:annulus_construction}

We followed the procedure developed by \citet{Djorgovski1988}, which consists of dividing the entire annulus in eight sectors of equal area and evaluating the flux in each of them. The flux in each annulus was obtained through the open source {\sc photutils} package \citep{bradley2019}, written in {\sc Python}.
The SB of an annulus is then the median of the eight fluxes, and we adopted $\Delta_{\mu} = 1.4826 \cdot MAD$, where MAD is the median absolute deviation, as their estimated uncertainty.
This technique mitigates the impact of very bright or foreground stars on the SBP.
Indeed, bright stars will increase the SB, and might generate artificial bumps in its shape, leading to wrong physical interpretations.
This effect could be dramatic in our selected sample of SCs, since many of them have a low SB.
The median ensures that sectors with such stars do not dominate the budget of the estimated SB.\par
In order to evenly sample inner and outer regions, many authors \citep[e.g.][]{Mackey&Gilmore2003,Mackey&Gilmore2003b,Carvalho-2008} utilized four sets of annuli of different thickness.
Since many SCs in our sample have a SB only marginally above the background level, in order to better assess the SB value at a given radius we preferred to deal with 16 sets of annuli, having widths between 1.0\arcsec~and 4.0\arcsec~with a step size of 0.2\arcsec.
Since smaller annuli aim to sample the inner regions, we used those with sizes between 1.0\arcsec-1.5\arcsec~and 1.5\arcsec-2.0\arcsec~up to 20\arcsec~and to 30\arcsec~ from the centre, respectively.
The remaining sets of annuli, i.e. those with sizes between 2.2\arcsec up to 4.0\arcsec, need to probe the outer regions, hence we constructed them up to 5\arcmin~from the centre of the cluster.
Once all annuli were produced, and SB was derived in each of them, we averaged all the values in regular intervals of $\log_{10} (r/arcsec) = 0.1$.
We performed a weighted mean to obtain our best estimate of the SB and its error. Similarly, we used as cluster-centric radius the mean of all radii within each interval of the binning procedure.\par
\begin{figure}
    \centering
    \includegraphics[width = 0.45\textwidth]{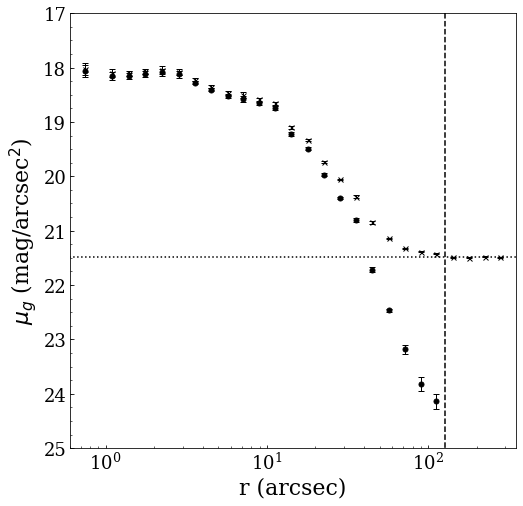}
    \caption{Example of the background subtraction procedure for NGC419. The SBP calculated with the background contribution is displayed with crosses, while the decontaminated one is in black dots. Errorbars mark the uncertainties on the SBP as defined in the text.
    The horizontal line represents the estimated background as described in the text. The vertical line shows the cluster-centric distance where the contaminated SBP intersects the background level.}
    \label{fig:background_subtraction}
\end{figure}

The SBP derived as described above is still affected by background contamination from field stars and sky background that needs to be corrected before proceeding with the analysis.
The background level needs to be determined in a region well away from the cluster, and the VST data are suitable for such requirements due to the large field of view.
To this purpose, we extended the annuli construction up to 300\arcsec~for every SC and 
adopted as local background the average flux measured between 150\arcsec~ and 300\arcsec.
This range of values 
is large enough to be both statistically significant and robust against fluctuations (i.e. a rich SC located within the background estimation area would have increased the sky level if the sampling regions were smaller).
Finally, the background level was subtracted from each annulus in order to get a decontaminated SBP.
Figure~\ref{fig:background_subtraction} shows an example of the background subtraction in the case of NGC419. In detail, the figure shows the SBPs before (crosses) and after (filled circles) the background subtraction, while the horizontal line shows the estimated background level.\par
\begin{figure}
    \centering
    \includegraphics[width=0.45\textwidth]{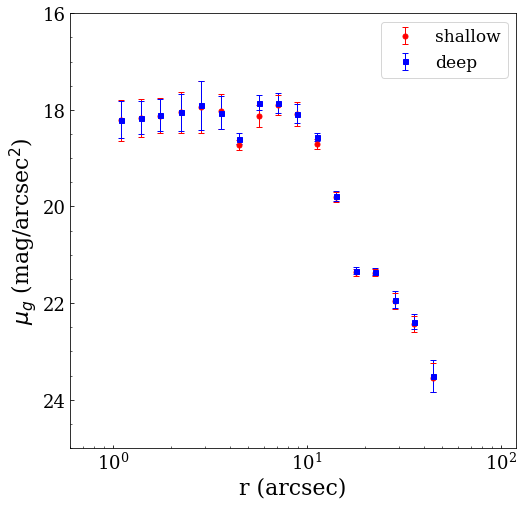}
    \caption{SBPs derived with shallow (red points) and deep (blue squares) images for NGC376.}
    \label{fig:shallow_vs_deep}
\end{figure}

Even though the procedure applied to obtain the SBP prevents saturated stars from dominating the luminosity budget (see \S\ref{sec:annulus_construction}), their effect must be carefully investigated. 
Shallow images provided us the opportunity to evaluate the impact of their presence in 11 SCs containing more than 5 saturated stars in the deep images, but not in the shallow ones.
Figure~\ref{fig:shallow_vs_deep} shows the SBP derived with either shallow (red points) or deep (blue squares) images for the SC NGC376, and, apart from some barely visible differences, the overall SBP shape is unchanged.
We inspected shallow and deep images for all the 11 SCs with more than 5 saturated stars, not finding any remarkable difference, which led us to conclude that, with our methodology, saturated stars do not alter our SBP shapes.

\section{Fitting procedure}
\label{sec:fitting_procedure}

We adopted both EFF and King profiles to fit SBPs and to derive the SC structural parameters. As the results obtained with the two different profiles are very close to each other, in the following we focus our discussion on the EFF profile in order to be consistent with previous works, since most of them adopted only the EFF profile. The interested reader can find a detailed report about the King profile in Appendix~\ref{app:king_model}.

\subsection{EFF model}
\label{sec:fitting_eff}

The EFF formulation is the following:
\begin{equation}
\label{eq:eff_profile}
    \mu(r) = \mu_0 + 1.25\gamma\log\left(1 + \frac{r^2}{\alpha^2}\right)
\end{equation}
\noindent
where $\mu(r)$ is the SB expressed in mag/arcsec$^2$ at a given distance $r$ from the cluster centre.
$\mu_0$, $\gamma$ and $\alpha$ are, respectively, the central SB, the dimensionless slope of the power-law and a parameter (expressed in arcsec) related to the core radius $r_{\rm c}$, which is the cluster-centric distance where the surface brightness has half of its central value, through the relation
\begin{equation}
\label{eq:eff_rcore}
    r_{\rm c} = \alpha\sqrt{2^{2/\gamma} - 1}
\end{equation}
\noindent
These three parameters have been derived via a non-linear least squares method through a $\chi^2$ minimization.
As it would not be appropriate to perform the fit all along the SBP, since it extends well beyond the cluster limit, we set a limiting radius, hereafter called fitting radius ($r_{\rm f}$), within which we executed the fit. 
To estimate its value we adopted the local background estimated for each cluster, and set as $r_{\rm f}$ the distance from the cluster centre where the original SBP (i.e. not subtracted) reaches it or when $\mu(r) - \Delta_{\mu}(r)$ is smaller than the background level in two consecutive bins.
These requirements may fail 
when two SCs are too close in projection: 
the SBP increases again at the distance where the annuli begin to incorporate stars belonging to the nearby cluster. 
Overall, 
the net effect is an unrealistically large $r_{\rm f}$, thus forcing us to set the fitting radius by hand for these SCs.
The SC luminosity comes from the integration of eq.~\ref{eq:eff_profile}, at the limit of $r_{\rm f}$ to infinity and yields:
\begin{equation}
\label{eq:eff_luminosity}
    L_{\infty} = \frac{2\pi10^{-0.4\mu}\alpha^2}{\gamma - 2}
\end{equation}
\noindent
provided that $\gamma > 2$, otherwise the luminosity becomes infinite.\footnote{We set $\gamma=2$ as lowest bound limit in the curve fitting to avoid negative luminosities.}
To calculate the extinction-corrected total luminosity, we adopted the extinction values obtained through the procedure described in Sect.~\ref{sec:mass_estimation} when available, or the extinction maps provided by \citet{Schlegel-1998}, re-calibrated by \citet{Schlafly2011}, otherwise.\par
To convert $r_{\rm c}$ from arcsec to a real physical size expressed in parsec (pc), we need the distance modulus (DM) of the cluster.
As a first approximation, we adopted the same distance modulus (DM) for all SCs, i.e. $DM = 18.98$ mag, corresponding to the SMC centre distance, recently estimated by \citet{Graczyk2020} (see also discussion in Sec.~\ref{sec:cmd}). To be conservative, we consider an error on the DM of $\Delta DM = 0.4$ mag, corresponding to about 25 kpc at the SMC distance, in order to take into account also the depth of the SMC along the line of sight.
The derived values of $\mu_0,~\alpha,~\gamma$ for the EFF models are listed in Table~\ref{tab:results_fit_params}, while $r_{\rm c}$ and $L_{\infty}$ are reported in the first columns of Table~\ref{tab:results_lum_mass}.\par
We point out that for H86-6 and H86-11 we were not able to perform the fit of the $g$ SBP because it was too noisy. For these SCs, we adopted the $i$-band SBP to derive their structural parameters.

\subsection{Comparison with literature}
\label{sec:comparison_sbp}

\begin{figure*}
    \centering
    \includegraphics[scale=0.32]{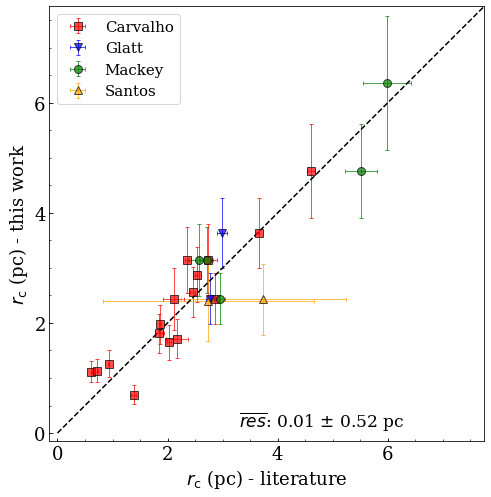}
    \includegraphics[scale=0.35]{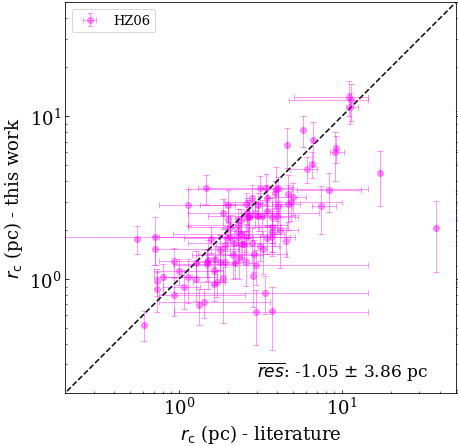}
    \caption{\emph{Left:} comparison of our core radii (y-axis) with values in the literature (x-axis). Different works are marked with different colours. \emph{Right: } comparison of our core radii with those taken from HZ06. Note that this panel is in logarithmic scale.}
    \label{fig:rc_our_lit}
\end{figure*}

We compared our results on the main properties of SCs with those existing in the literature for SCs in common with our sample.
Since 62 SCs were analyzed for the first time in this work, the comparison was carried out with the remaining $\sim$ 100.
The literature studies considered here include \citet{Mackey&Gilmore2003b, Hill&Zaritsky2006, Carvalho-2008, Glatt-2009, Santos-2020}, who derived the main structural parameters in a homogeneous way for several SMC SCs using EFF profiles. 
The left panel of Fig.~\ref{fig:rc_our_lit} displays the comparison between our $r_{\rm c}$ obtained via the EFF profile and the literature ones, with the exception of \citet{Hill&Zaritsky2006}, which will be discussed separately.
To make the comparison with previous works meaningful, we homogenized the core radii, recalculating the literature values for the SMC distance modulus we have adopted. \footnote{\citet{Glatt-2009}, \citet{Mackey&Gilmore2003b} and \citet{Santos-2020} set the SMC distance modulus to $DM = 18.88 - 18.9 - 18.96$ mag, respectively. \citet{Carvalho-2008} did not provide information on their adopted SC distances.}
An inspection of Fig.~\ref{fig:rc_our_lit} reveals that our core radii are nicely consistent within the errors with those estimated in the literature. Indeed, the average residual between our and the literature values is $0.01 \pm 0.52$ pc, corresponding to less than 2\% in terms of fractional residuals.\par
The right panel of Fig.~\ref{fig:rc_our_lit} shows the comparison between our and \citeauthor{Hill&Zaritsky2006}'s \citeyearpar{Hill&Zaritsky2006} core radii values.
In this case, we were not able to adjust their core radii to our SMC distance, since they did not give information on their adopted distance.
The figure shows that our $r_{\rm c}$ are systematically lower ($-1.05 \pm 3.86$ pc). Although the large uncertainties prevent us from determining whether the offset is real or not, we speculate that this systematic difference might originate from the different assumption of the SMC DM by \citet{Hill&Zaritsky2006}. Furthermore, their work is based on the Magellanic Clouds Photometric Survey (MCPS), which is significantly shallower than the STEP survey\footnote{The MCPS has a completeness at $\sim$50\% for V$\sim$21 \citep[][]{Zaritsky1997}.}. This occurrence might be responsible for the large scatter observed between the results based on the two surveys.
Indeed, also \citet{Santos-2020}, using data from the VISCACHA survey, found significant residuals ($1.1 \pm 2.7$ pc, see their Fig.6), with respect to \citet{Hill&Zaritsky2006}.

\section{Analysis of the CMD}
\label{sec:cmd}

In the previous section we estimated the SC structural parameters from the study of their SBPs. In this section we exploit the SC CMDs to extract additional fundamental physical parameters, such as the age and the mass. To this aim, the common procedure is to use the CMD to estimate the SC age from isochrone fitting and then to convert the integrated SC luminosity into total mass by adopting a mass-to-light ratio (MLR)-age-metallicity relation \citep[e.g.][]{Santos-2020} or alternatively to use integrated colors \citep[e.g.][]{Roediger2015}. However, these procedures are not usable for SCs with masses lower than $\sim$5000 $M_{\odot}$ as the stochasticity in the stellar luminosity function can produce substantial systematics in the inferred SC mass \citep{Fouesneau-2010,Silva-Villa2011,Fouesneau-2014,Krumholz-2015}.
Since our sample includes many low mass SCs, we decided to follow a different procedure, adopting the Automated Stellar Cluster Analysis (ASteCA) open source python package \citep{Perren-2015}. This tool generates synthetic SCs from a set of theoretical isochrones and performs an automatic isochrone fitting procedure. A genetic algorithm is used to find the best solution \citep[see][for full details]{Perren-2015}.
The great advantage of using ASteCA is that in practice it provides simultaneously an estimate of the age, reddening, mass and their uncertainties by taking into account the inherently stochastic process of the isochrone fitting procedure, as a synthetic cluster is generated from a stochastically sampled initial mass function (IMF)\footnote{We adopted a \citet{Kroupa-2001b} IMF.} \citep[see Sect. 2.9.1 in ][]{Perren-2015}.
In the following two sections we describe the procedure employed to derive SC ages and masses.

\subsection{Visual isochrone fitting}
\label{sec:visual_isochrone_fitting}

We carefully examined the CMD of each SC in our sample before running ASteCA, with the aim of performing in advance a visual isochrone fitting procedure. Such results have been used as priors for ASteCA in order to make it quicker, and to avoid that the software could catch in a local minimum, far away from the actual solution. This section describes the steps we performed to get SC ages from a visual ischrone fitting.\par
The CMD of a SC is contaminated by background/foreground stars belonging to the field of the host galaxy (the SMC in our case). In order to obtain reliable ages, non-cluster members should be statistically removed, especially for less populous SCs where the expected fraction of field stars is relatively high.
To this aim we adopted the procedure developed by \citet{Piatti2012}, which allows us to estimate the membership probability $P$ of each SC star, permitting us to carry out the isochrone comparison procedure using only stars with the highest membership probability.
This method has been extensively used in literature \citep[see e.g.][and references therein]{Piatti2014,Piatti2015b,Piatti2016,Ivanov2017,Gatto2020} and proven to reduce the uncertainties in the SC parameter estimation. 
We made use of the PARSEC models  \citep{Bressan2012}\footnote{http://stev.oapd.inaf.it/cgi-bin/cmd} to visually identify the isochrone of a given age and metallicity\footnote{We corrected the isochrones for the adopted distance and the colour excess through the relations $g = g_{\rm iso} + DM + R_{\rm g} \times E(B - V)$ and $E(g - i) = (R_{\rm g} - R_{\rm i}) \times E(B - V)$ with $R_{\rm g} = 3.303$ and $R_{\rm i} =  1.698$ \citep{Schlafly2011}.}
that best fits the distribution of SC stars with $P$ > 60\% 
, in the CMD.
To reduce the wide space of parameters affecting the shape of an isochrone, we fixed the distance modulus (DM) for all SCs to $DM = 18.98$ mag \citep{Graczyk2020}, and we made use of an age-metallicity relation (AMR) derived for the SMC SCs \citep[][]{parisi2015} to fix the metal-content of the SCs, as well.
Although the SMC has a considerable extension along the line of sight\citep[see e.g.][and references therein]{Haschke2012,Subramanian2012,Jacy2016,Ripepi2017}, we estimated the total age uncertainty of the visual fitting procedure to be $\Delta\log(t) = 0.2$ dex, which comprises both a statistical error of 0.1 dex originating from the method, and a further 0.1 dex to take into account the fixed DM for the galaxy.\footnote{An error of 0.1 dex in the age estimate is the result of a $\Delta$DM$\sim$0.4 mag at the SMC distance.}
We varied the age and reddening of the isochrones looking for the one best matching the following key evolutionary sequences: main sequence (MS), turn-off (TO) point, subgiant branch, blue loop (BL) and red clump (RC). These sequences have been used to aid the visual fitting procedure, which was performed by the authors independently, in order to check the resulting reliability/uncertainty.
For 36 SCs we were not able to derive even a rough estimate of the age, as their CMDs present none or too few stars on the above quoted key evolutionary stellar phases. 
Therefore, we conservatively decided to exclude them from the analysis with ASteCA. For these 36 clusters we only provide the structural parameters derived through the study of their SBPs.


\subsection{Analysis of the CMD with ASteCA}
\label{sec:mass_estimation}

The 134 SCs with an age estimate from visual isochrone fitting of their CMD have been further analyzed with ASteCA with the main scope of deriving their ages and masses as well as the proper uncertainties by means of an objective methodology.\footnote{We adopted the same set of PARSEC isochrones utilized in Sec.~\ref{sec:visual_isochrone_fitting}.}
To use the ASteCA package we have first to define a list of priors for the relevant SCs quantities to be estimated.
In more detail, we fixed the metallicity as already done in Sect.~\ref{sec:visual_isochrone_fitting} but we let the DM vary in the $18.6 \leq DM \leq 19.2$~mag interval\footnote{All priors indicated in the text are flat.}, in order to correct any bias we introduced in the visual fitting procedure by fixing it. The reddening values were allowed to vary in the range $0 \leq E(B-V) \leq 0.3$~mag. Priors on the SC ages were of $\log(t) \pm 0.6$~dex, (i.e. $3\sigma$), around the age estimated in the previous section through the visual fitting procedure. Finally, we allowed the SC total mass to vary over a large interval, i.e. $10 \leq (M/M_{\odot}) \leq 10^6$.\footnote{We checked that estimated masses did not change within uncertainties with a different prior selection.} 
Results from the CMD fitting by means of the ASteCA package are listed in Table~\ref{tab:results_lum_mass}.

\subsection{Comparison with previous studies}

\begin{figure*}
    \centering
    \includegraphics[scale=0.5]{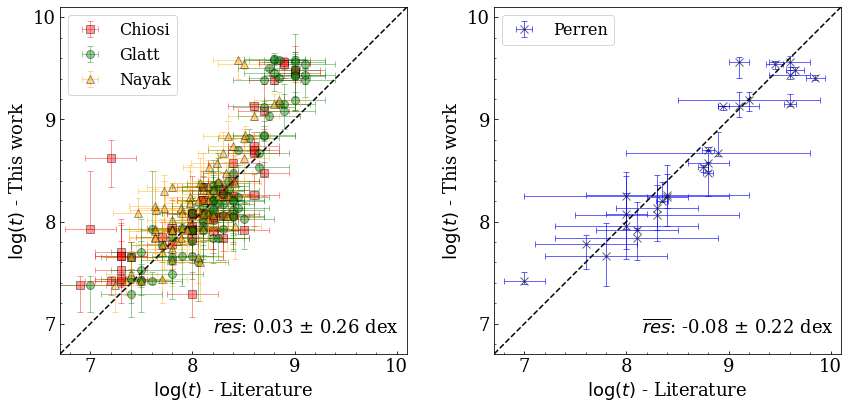}
    \caption{Comparison between our age estimates (y-axis) with those taken from the literature (x-axis). Different works are marked with different colours. The black dashed line indicates the one-to-one relationship. Residuals are to be read as our work minus literature. Note that residuals in the left panel have been calculated using only SCs younger than 1 Gyr.}
   \label{fig:params_our_lit}
\end{figure*}
\begin{figure*}
    \centering
    \includegraphics[scale=0.5]{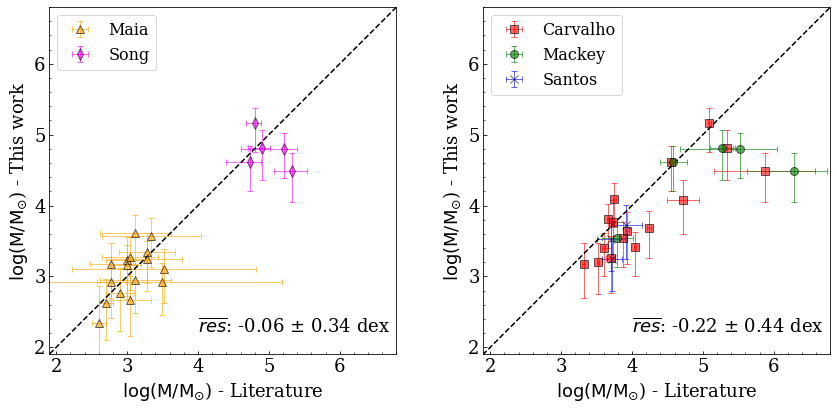}
    \caption{Comparison between our mass estimates (y-axis) with those taken from the literature (x-axis). Different works are marked with different colours. The black dashed line indicates the one-to-one relationship. Residuals are to be read as our work minus literature.}
   \label{fig:params_our_lit_mass}
\end{figure*}

We compared the ages derived from our CMDs with those present in the literature. To this aim we selected the works by \citet{Chiosi-2006,Glatt2010, Perren2017} and \citet{Nayak2018} who presented ages for a significant number of SCs using the isochrone fitting procedure (visual or automatic). 
We note that \citeauthor{Nayak2018}'s \citeyearpar{Nayak2018} errors are between 0.24 and 0.26 dex in $\log(t)$, hence we adopted a mean of 0.25 dex for all their SCs, while for \citeauthor{Glatt2010}'s \citeyearpar{Glatt2010} SCs we adopted a mean error of 0.3 dex, i.e. their lowest error.
The left panel of Fig.~\ref{fig:params_our_lit} shows the comparison between our results and those by \citet{Chiosi-2006,Glatt2010} and \citet{Nayak2018}. It can be clearly seen that, except for a few cases, our age estimates are in very good agreement with those from the literature up to 1 Gyr, whereas for older ages, the difference becomes noticeable. This discrepancy is likely due to the too shallow photometry analysed in the quoted works, which makes it unfeasible to correctly detect the MS turn-off for clusters older than 1 Gyr. Indeed, \citet{Glatt2010} pointed out the difficulty to derive ages of intermediate-age clusters with their MCPS data (see their Sec. 2). 
In contrast, the comparison with \citet{Perren2017},\footnote{Their data set consists of $CT_1$ Washington photometry, compiled on the basis of 19 previous works.} displayed in the right panel of Fig.~\ref{fig:params_our_lit}, shows very good agreement for the whole range of ages derived in this work.
Concerning the estimated total masses, Fig.~\ref{fig:params_our_lit_mass} (left panel) shows the comparison  between our results and those by \citet{Maia2014} and \citet{song-2021} for the samples of SCs in common. \citet{Maia2014} provided SC masses derived through star counts integrated down to 0.1 $M_{\odot}$, while \citet{song-2021} obtained dynamical masses through spectroscopic measurements of radial velocities of the individual member stars. The figure shows a very good agreement between this work and the two quoted investigations, with an average of the residuals (our work minus literature) equal to -0.06$\pm$0.34 dex, i.e. without any systematic difference over a very broad range of masses, down to few hundreds of solar masses. In addition, we also considered the works by \citet{Mackey&Gilmore2003b,Carvalho-2008} and \citet{Santos-2020}, which provided mass estimates trough a MLR by using the total cluster luminosity estimated from the EFF model fitting. The result of this comparison is presented in the right panel of Fig.~\ref{fig:params_our_lit_mass}, which shows an overall overestimate of the masses in the aforementioned investigations with respect to this work, even if a relatively high scatter. The comparison of the masses with 4 SCs in common between \citet{song-2021}, \citet{Mackey&Gilmore2003b} and \citet{Carvalho-2008}, revealed that the latter two works derived larger masses also with respect to the dynamical estimates, i.e. $\Delta \log M=0.37$~dex for \citet{Mackey&Gilmore2003b} and $\Delta \log M=0.27$~dex for \citet{Carvalho-2008}. 
As dynamical masses are more reliable than those derived through integrated properties, the good agreement displayed in the left panel of the figure makes us confident about our mass assessment. 


\section{Results}
\label{sec:results}

The SC parameters derived from the SBP fitting are shown in Table~\ref{tab:results_fit_params}, along with the radius fitting, the coordinates and the tile hosting the SC. The physical SC parameters, like their sizes, luminosities and masses are, instead, listed in Table~\ref{tab:results_lum_mass}. 
The best fit profiles for all SCs are depicted in Fig.~\ref{fig:best_fit_both_profiles} for both EFF and King models. 
Again, throughout this section we provide results obtained with the EFF profiles. For completeness, a discussion about quantities obtained with the King profile is presented in Appendix~\ref{app:king_model}.

\begin{table*}
    \centering
    \vspace*{-0.5cm}
    \tiny
    \begin{tabular}{l|c|c|c|c|c|c|c|c|c|c|c|c|}
    \hline
     & \multicolumn{2}{c}{EFF} & \multicolumn{4}{c}{KING} & \multicolumn{4}{c}{CMD}\\
     \hline
    ID & $r_{\rm c}$ & $\log(L/L_{\odot})$ & $r_{\rm c}$ & $c$ & $r_{\rm h}$ & $\log(L/L_{\odot})$ & $\log$ (t) & E(B-V) & Z & $\log(M/M_{\odot})$ & $\log(M/L)$ & $\log(t_{\rm rh})$\\ 
     & (pc) &  & (pc) &  & (pc) & & & (mag) & &  & & \\
    \hline
B10 & $0.2 \pm 0.2$ & $3.95 \pm 0.59$ & $0.2 \pm 0.2$ & $1.6 \pm 0.8$ & $0.7 \pm 0.3$ & $3.87 \pm 0.30$ & $9.48_{-0.39}^{+0.17}$ & 0.11 & 0.002 & $3.03_{-0.58}^{+0.39}$ & -0.92 & 6.86\\
B100 & $0.7 \pm 0.1$ & $4.18 \pm 0.34$ & $0.6 \pm 0.1$ & $1.0 \pm 0.1$ & $1.1 \pm 0.2$ & $4.16 \pm 0.18$ & $8.00_{-0.20}^{+0.15}$ & 0.09 & 0.004 & $3.12_{-0.42}^{+0.23}$ & -1.06 & 7.05\\
B103 & $0.9 \pm 0.2$ & - & $1.4 \pm 0.5$ & $0.8 \pm 0.3$ & $1.8 \pm 0.5$ & $3.87 \pm 0.28$ & $8.24_{-0.10}^{+0.32}$ & 0.20 & 0.004 & $2.95_{-0.46}^{+0.28}$ & -0.92 & 7.33\\
B105 & $0.8 \pm 0.2$ & $4.11 \pm 0.88$ & $0.9 \pm 0.2$ & $1.1 \pm 0.2$ & $1.6 \pm 0.4$ & $3.83 \pm 0.20$ & $7.67_{-0.23}^{+0.31}$ & 0.05 & 0.004 & $3.03_{-0.44}^{+0.26}$ & -1.08 & 7.26\\
B108 & $1.3 \pm 0.4$ & - & $2.0 \pm 1.0$ & $0.7 \pm 0.6$ & $2.1 \pm 0.7$ & $3.42 \pm 0.34$ & $9.39_{-0.04}^{+0.01}$ & 0.06 & 0.002 & $3.62_{-0.40}^{+0.21}$ & 0.20 & 7.79\\
B111 & $0.8 \pm 0.2$ & - & $1.2 \pm 0.4$ & $0.9 \pm 0.4$ & $1.5 \pm 0.4$ & $3.00 \pm 0.25$ & $9.13_{-0.11}^{+0.13}$ & 0.12 & 0.002 & $2.62_{-0.52}^{+0.34}$ & -0.38 & 7.19\\
B113 & $1.0 \pm 0.3$ & $3.52 \pm 0.39$ & $1.2 \pm 0.4$ & $0.7 \pm 0.2$ & $1.3 \pm 0.3$ & $3.47 \pm 0.27$ & $9.04_{-0.05}^{+0.04}$ & 0.05 & 0.002 & $3.29_{-0.43}^{+0.25}$ & -0.22 & 7.35\\
B115 & $1.0 \pm 0.4$ & - & $1.6 \pm 0.8$ & $0.8 \pm 0.6$ & $2.0 \pm 0.7$ & $4.11 \pm 0.35$ & - & - & - & - & - & - \\
B117 & $1.0 \pm 0.2$ & $4.19 \pm 0.75$ & $1.3 \pm 0.3$ & $0.9 \pm 0.2$ & $1.8 \pm 0.4$ & $3.94 \pm 0.20$ & $8.17_{-0.24}^{+0.15}$ & 0.09 & 0.004 & $3.08_{-0.59}^{+0.40}$ & -1.11 & 7.37\\
B119 & - & - & $2.9 \pm 2.4$ & $0.1 \pm 0.4$ & $1.4 \pm 0.4$ & $2.95 \pm 5.09$ & - & - & - & - & - & - \\
B12 & - & - & $2.4 \pm 2.0$ & $0.3 \pm 0.6$ & $1.7 \pm 0.6$ & $3.20 \pm 0.88$ & $9.46_{-0.09}^{+0.26}$ & 0.14 & 0.002 & $3.13_{-0.64}^{+0.46}$ & -0.07 & 7.49\\
B121 & $1.3 \pm 0.4$ & $4.33 \pm 0.66$ & $2.7 \pm 0.8$ & $0.3 \pm 0.1$ & $1.9 \pm 0.4$ & $4.16 \pm 0.60$ & $7.92_{-0.13}^{+0.18}$ & 0.11 & 0.004 & $2.93_{-0.47}^{+0.28}$ & -1.40 & 7.36\\
B122 & $2.1 \pm 1.0$ & - & $6.2 \pm 4.7$ & $0.4 \pm 0.7$ & $4.8 \pm 1.7$ & $3.83 \pm 0.50$ & $8.21_{-0.34}^{+0.32}$ & 0.07 & 0.004 & $3.30_{-0.43}^{+0.24}$ & -0.52 & 8.11\\
B124 & $1.2 \pm 0.3$ & $3.63 \pm 0.37$ & $2.0 \pm 0.6$ & $0.4 \pm 0.2$ & $1.6 \pm 0.3$ & $3.57 \pm 0.46$ & $8.08_{-0.35}^{+0.20}$ & 0.10 & 0.004 & $2.35_{-0.70}^{+0.51}$ & -1.29 & 7.06\\
B128 & - & - & $2.9 \pm 1.9$ & $0.3 \pm 0.4$ & $2.0 \pm 0.6$ & $3.52 \pm 0.98$ & $8.02_{-0.29}^{+0.15}$ & 0.12 & 0.004 & $2.71_{-0.52}^{+0.33}$ & -0.81 & 7.33\\
B137 & $0.8 \pm 0.2$ & - & $1.2 \pm 0.3$ & $1.0 \pm 0.3$ & $1.8 \pm 0.4$ & $3.65 \pm 0.21$ & - & - & - & - & - & - \\
B165 & $1.1 \pm 0.3$ & $3.09 \pm 0.32$ & $1.5 \pm 0.5$ & $0.6 \pm 0.2$ & $1.5 \pm 0.3$ & $3.08 \pm 0.32$ & - & - & - & - & - & - \\
B18 & $0.7 \pm 0.2$ & $3.29 \pm 0.43$ & $0.8 \pm 0.2$ & $0.7 \pm 0.3$ & $0.8 \pm 0.2$ & $3.20 \pm 0.23$ & $9.72_{-0.65}^{+0.27}$ & 0.10 & 0.002 & $2.93_{-0.86}^{+0.67}$ & -0.36 & 6.97\\
B21 & $0.8 \pm 0.4$ & $3.39 \pm 2.32$ & $2.5 \pm 1.2$ & $0.0 \pm 0.2$ & $1.1 \pm 0.2$ & $2.85 \pm 6.51$ & $9.49_{-0.49}^{+0.11}$ & 0.05 & 0.002 & $2.81_{-0.63}^{+0.45}$ & -0.57 & 7.05\\
B22 & $0.2 \pm 0.1$ & - & $0.3 \pm 0.1$ & $1.2 \pm 0.3$ & $0.6 \pm 0.1$ & $3.27 \pm 0.21$ & $9.97_{-0.24}^{+0.09}$ & 0.12 & 0.002 & $3.42_{-0.71}^{+0.53}$ & 0.16 & 6.95\\
B26 & $0.6 \pm 0.2$ & $2.98 \pm 0.87$ & $1.5 \pm 0.5$ & $0.2 \pm 0.2$ & $0.9 \pm 0.2$ & $2.67 \pm 1.09$ & - & - & - & - & - & - \\
B34A & $0.8 \pm 0.3$ & $3.55 \pm 0.30$ & $1.7 \pm 0.9$ & $0.3 \pm 0.3$ & $1.1 \pm 0.3$ & $3.55 \pm 1.29$ & - & - & - & - & - & - \\
B39 & $1.5 \pm 0.3$ & - & $2.2 \pm 0.6$ & $0.9 \pm 0.3$ & $2.9 \pm 0.6$ & $4.09 \pm 0.22$ & $8.67_{-0.04}^{+0.20}$ & 0.13 & 0.004 & $3.39_{-0.45}^{+0.26}$ & -0.70 & 7.85\\
B52 & $4.1 \pm 1.3$ & - & $10.6 \pm 8.1$ & $0.3 \pm 0.6$ & $6.3 \pm 2.1$ & $4.40 \pm 1.15$ & - & - & - & - & - & - \\
B65 & $1.3 \pm 0.2$ & $4.49 \pm 0.26$ & $1.4 \pm 0.3$ & $0.6 \pm 0.1$ & $1.5 \pm 0.3$ & $4.48 \pm 0.22$ & - & - & - & - & - & - \\
B70 & $0.5 \pm 0.1$ & $3.74 \pm 1.01$ & $0.5 \pm 0.1$ & $1.7 \pm 0.1$ & $1.4 \pm 0.2$ & $3.45 \pm 0.15$ & $9.64_{-0.19}^{+0.14}$ & 0.07 & 0.002 & $2.97_{-0.61}^{+0.42}$ & -0.77 & 7.30\\
B71 & $1.8 \pm 0.4$ & $4.65 \pm 0.24$ & $2.8 \pm 0.7$ & $0.4 \pm 0.1$ & $2.2 \pm 0.4$ & $4.66 \pm 0.51$ & $7.53_{-0.18}^{+0.28}$ & 0.13 & 0.004 & $3.01_{-0.44}^{+0.25}$ & -1.63 & 7.45\\
B79 & $1.3 \pm 0.3$ & $4.56 \pm 0.60$ & $1.3 \pm 0.4$ & $1.1 \pm 0.3$ & $2.3 \pm 0.5$ & $4.45 \pm 0.22$ & $7.70_{-0.25}^{+0.32}$ & 0.05 & 0.004 & $2.98_{-0.43}^{+0.25}$ & -1.57 & 7.49\\
B80 & $0.6 \pm 0.3$ & - & $0.9 \pm 0.4$ & $1.0 \pm 0.6$ & $1.4 \pm 0.5$ & $3.52 \pm 0.32$ & $9.08_{-0.18}^{+0.27}$ & 0.14 & 0.002 & $2.84_{-0.62}^{+0.44}$ & -0.68 & 7.20\\
B9 & $1.0 \pm 0.5$ & $3.73 \pm 1.56$ & $2.7 \pm 1.1$ & $0.3 \pm 0.2$ & $1.8 \pm 0.4$ & $3.36 \pm 0.92$ & - & - & - & - & - & - \\
B96 & $2.6 \pm 0.5$ & $4.24 \pm 0.28$ & $4.2 \pm 1.0$ & $0.4 \pm 0.1$ & $3.1 \pm 0.6$ & $4.21 \pm 0.51$ & $8.27_{-0.16}^{+0.00}$ & 0.18 & 0.004 & $3.26_{-0.44}^{+0.26}$ & -0.98 & 7.82\\
B97 & $1.4 \pm 0.5$ & - & $3.1 \pm 2.1$ & $0.3 \pm 0.5$ & $1.9 \pm 0.6$ & $3.14 \pm 1.05$ & - & - & - & - & - & - \\
B99 & $1.4 \pm 0.4$ & $4.43 \pm 0.99$ & $1.8 \pm 0.6$ & $0.8 \pm 0.3$ & $2.3 \pm 0.6$ & $4.11 \pm 0.26$ & $8.25_{-0.32}^{+0.24}$ & 0.19 & 0.004 & $3.23_{-0.51}^{+0.32}$ & -1.20 & 7.62\\
BS102 & $1.2 \pm 0.4$ & $3.73 \pm 2.07$ & $1.7 \pm 0.7$ & $0.7 \pm 0.5$ & $1.7 \pm 0.4$ & $3.17 \pm 0.29$ & $8.37_{-0.32}^{+0.28}$ & 0.06 & 0.004 & $2.56_{-0.53}^{+0.35}$ & -1.17 & 7.18\\
BS128 & - & - & $1.1 \pm 0.7$ & $0.8 \pm 1.3$ & $1.0 \pm 0.2$ & $3.11 \pm 0.37$ & $8.35_{-0.36}^{+0.32}$ & 0.10 & 0.004 & $2.23_{-0.62}^{+0.44}$ & -0.89 & 6.73\\
BS131 & $1.3 \pm 0.3$ & $4.04 \pm 0.62$ & $2.3 \pm 1.1$ & $0.3 \pm 0.3$ & $1.5 \pm 0.3$ & $3.83 \pm 0.84$ & - & - & - & - & - & - \\
BS138 & $3.5 \pm 0.9$ & - & $6.0 \pm 4.8$ & $0.4 \pm 0.8$ & $4.1 \pm 1.2$ & $3.85 \pm 0.55$ & $9.53_{-0.04}^{+0.03}$ & 0.12 & 0.002 & $3.68_{-0.41}^{+0.22}$ & -0.17 & 8.26\\
BS2 & $1.1 \pm 0.3$ & $2.94 \pm 0.43$ & $1.1 \pm 0.4$ & $0.9 \pm 0.4$ & $1.6 \pm 0.4$ & $2.89 \pm 0.27$ & - & - & - & - & - & - \\
BS276 & $0.5 \pm 0.1$ & $4.19 \pm 0.50$ & $0.5 \pm 0.1$ & $1.2 \pm 0.1$ & $1.0 \pm 0.2$ & $4.15 \pm 0.17$ & $8.38_{-0.32}^{+0.24}$ & 0.10 & 0.004 & $3.01_{-0.47}^{+0.28}$ & -1.18 & 6.99\\
BS76 & $1.7 \pm 0.4$ & $3.78 \pm 0.48$ & $1.9 \pm 0.5$ & $0.8 \pm 0.3$ & $2.4 \pm 0.6$ & $3.69 \pm 0.22$ & $8.15_{-0.27}^{+0.17}$ & 0.05 & 0.004 & $2.95_{-0.49}^{+0.30}$ & -0.83 & 7.53\\
BS88 & $1.2 \pm 0.4$ & - & $2.3 \pm 1.7$ & $0.3 \pm 0.6$ & $1.3 \pm 0.4$ & $3.13 \pm 0.85$ & $9.56_{-0.16}^{+0.05}$ & 0.12 & 0.002 & $3.20_{-0.55}^{+0.37}$ & 0.07 & 7.35\\
H86-11 & $0.5 \pm 0.3$ & - & $0.9 \pm 0.5$ & $0.9 \pm 0.5$ & $1.3 \pm 0.5$ & $2.80 \pm 0.37$ & - & - & - & - & - & - \\
H86-114 & $1.4 \pm 0.3$ & - & $2.2 \pm 0.6$ & $0.7 \pm 0.3$ & $2.5 \pm 0.6$ & $4.05 \pm 0.25$ & $8.34_{-0.37}^{+0.26}$ & 0.06 & 0.004 & $2.79_{-0.42}^{+0.24}$ & -1.26 & 7.50\\
H86-137 & $0.8 \pm 0.2$ & $4.28 \pm 1.05$ & $0.9 \pm 0.2$ & $1.3 \pm 0.2$ & $1.5 \pm 0.3$ & $3.87 \pm 0.16$ & $9.14_{-0.01}^{+0.02}$ & 0.13 & 0.002 & $3.52_{-0.47}^{+0.28}$ & -0.76 & 7.50\\
H86-146 & $0.4 \pm 0.2$ & - & $0.6 \pm 0.3$ & $1.3 \pm 0.4$ & $1.4 \pm 0.4$ & $4.11 \pm 0.26$ & - & - & - & - & - & - \\
H86-150 & $0.9 \pm 0.3$ & - & $1.5 \pm 0.6$ & $0.8 \pm 0.4$ & $1.8 \pm 0.5$ & $3.91 \pm 0.31$ & $7.62_{-0.09}^{+0.16}$ & 0.12 & 0.004 & $2.92_{-0.44}^{+0.26}$ & -0.99 & 7.31\\
H86-159 & $1.6 \pm 0.4$ & $3.60 \pm 0.43$ & $2.0 \pm 0.7$ & $0.6 \pm 0.3$ & $2.0 \pm 0.6$ & $3.51 \pm 0.30$ & - & - & - & - & - & - \\
H86-165 & $1.7 \pm 0.7$ & - & $2.5 \pm 1.4$ & $0.6 \pm 0.6$ & $2.6 \pm 0.9$ & $3.79 \pm 0.39$ & $8.08_{-0.26}^{+0.23}$ & 0.10 & 0.004 & $2.94_{-0.46}^{+0.27}$ & -0.85 & 7.57\\
H86-174 & $1.3 \pm 0.3$ & $4.64 \pm 2.49$ & $1.8 \pm 0.6$ & $0.7 \pm 0.3$ & $2.0 \pm 0.5$ & $3.85 \pm 0.25$ & $8.58_{-0.08}^{+0.13}$ & 0.19 & 0.002 & $2.92_{-0.51}^{+0.32}$ & -1.73 & 7.44\\
H86-175 & $0.9 \pm 0.2$ & $3.56 \pm 0.19$ & $1.3 \pm 0.3$ & $0.4 \pm 0.1$ & $0.9 \pm 0.2$ & $3.55 \pm 0.51$ & - & - & - & - & - & - \\
H86-179 & $1.2 \pm 0.5$ & $3.72 \pm 0.81$ & $2.6 \pm 1.9$ & $0.2 \pm 0.4$ & $1.6 \pm 0.4$ & $3.53 \pm 1.73$ & - & - & - & - & - & - \\
H86-181 & $0.5 \pm 0.2$ & $4.23 \pm 0.81$ & $0.6 \pm 0.2$ & $1.2 \pm 0.3$ & $1.1 \pm 0.3$ & $4.01 \pm 0.22$ & - & - & - & - & - & - \\
H86-182 & $2.3 \pm 0.6$ & $4.44 \pm 0.71$ & $3.1 \pm 1.1$ & $0.6 \pm 0.3$ & $3.2 \pm 0.7$ & $4.22 \pm 0.29$ & - & - & - & - & - & - \\
H86-186 & $1.1 \pm 0.2$ & $4.11 \pm 0.33$ & $1.2 \pm 0.2$ & $0.8 \pm 0.1$ & $1.5 \pm 0.3$ & $4.08 \pm 0.19$ & $8.21_{-0.18}^{+0.28}$ & 0.11 & 0.004 & $3.15_{-0.46}^{+0.28}$ & -0.96 & 7.30\\
H86-190 & $0.6 \pm 0.2$ & - & $1.0 \pm 0.5$ & $0.7 \pm 0.5$ & $1.1 \pm 0.3$ & $3.54 \pm 0.32$ & $7.66_{-0.29}^{+0.32}$ & 0.05 & 0.004 & $2.30_{-0.56}^{+0.38}$ & -1.24 & 6.75\\
H86-191 & - & - & $2.0 \pm 1.5$ & $0.3 \pm 0.5$ & $1.3 \pm 0.4$ & $3.28 \pm 1.16$ & $8.04_{-0.31}^{+0.44}$ & 0.08 & 0.004 & $2.53_{-0.43}^{+0.24}$ & -0.75 & 7.00\\
H86-193 & $1.8 \pm 0.4$ & $3.75 \pm 0.49$ & $2.3 \pm 0.7$ & $0.6 \pm 0.3$ & $1.9 \pm 0.4$ & $3.62 \pm 0.25$ & $8.06_{-0.34}^{+0.31}$ & 0.14 & 0.004 & $2.78_{-0.45}^{+0.27}$ & -0.98 & 7.31\\
H86-194 & $0.7 \pm 0.4$ & - & $2.3 \pm 1.1$ & $0.1 \pm 0.2$ & $1.1 \pm 0.2$ & $3.27 \pm 2.82$ & $8.07_{-0.27}^{+0.32}$ & 0.07 & 0.004 & $2.36_{-0.60}^{+0.42}$ & -0.92 & 6.85\\
H86-213 & $0.3 \pm 0.1$ & - & $0.5 \pm 0.2$ & $1.3 \pm 0.3$ & $1.0 \pm 0.2$ & $2.99 \pm 0.20$ & $8.40_{-0.28}^{+0.20}$ & 0.12 & 0.004 & $2.74_{-0.45}^{+0.26}$ & -0.25 & 6.92\\
H86-6 & $1.2 \pm 0.3$ & $2.57 \pm 0.37$ & $2.4 \pm 0.9$ & $0.2 \pm 0.2$ & $1.4 \pm 0.3$ & $2.48 \pm 1.43$ & $9.44_{-0.22}^{+0.07}$ & 0.08 & 0.002 & $2.86_{-0.52}^{+0.34}$ & 0.29 & 7.24\\
H86-60 & - & - & $1.5 \pm 1.1$ & $0.3 \pm 0.5$ & $0.9 \pm 0.3$ & $2.84 \pm 1.24$ & - & - & - & - & - & - \\
H86-74 & $0.9 \pm 0.2$ & $4.77 \pm 0.82$ & $1.0 \pm 0.2$ & $1.1 \pm 0.2$ & $1.7 \pm 0.3$ & $4.51 \pm 0.18$ & $7.92_{-0.14}^{+0.12}$ & 0.14 & 0.004 & $2.66_{-0.56}^{+0.37}$ & -2.10 & 7.20\\
H86-86 & $0.3 \pm 0.1$ & $2.83 \pm 0.62$ & $0.4 \pm 0.2$ & $0.9 \pm 0.4$ & $0.6 \pm 0.1$ & $2.68 \pm 0.26$ & - & - & - & - & - & - \\
H86-87 & $1.7 \pm 0.6$ & - & $4.1 \pm 3.5$ & $0.3 \pm 0.6$ & $2.5 \pm 0.8$ & $3.81 \pm 1.07$ & $8.26_{-0.30}^{+0.21}$ & 0.15 & 0.004 & $2.92_{-0.46}^{+0.27}$ & -0.89 & 7.54\\
H86-97 & $1.5 \pm 0.4$ & - & $2.7 \pm 0.9$ & $0.5 \pm 0.2$ & $2.2 \pm 0.5$ & $4.39 \pm 0.35$ & $7.92_{-0.00}^{+0.27}$ & 0.22 & 0.004 & $3.10_{-0.47}^{+0.29}$ & -1.29 & 7.52\\
HW10 & $3.0 \pm 0.7$ & - & $4.4 \pm 1.4$ & $0.8 \pm 0.3$ & $5.7 \pm 1.4$ & $3.92 \pm 0.26$ & $9.59_{-0.09}^{+0.25}$ & 0.09 & 0.002 & $4.29_{-0.56}^{+0.37}$ & 0.38 & 8.72\\
HW11 & $3.6 \pm 1.3$ & - & $5.4 \pm 2.5$ & $0.9 \pm 0.6$ & $7.1 \pm 2.3$ & $4.22 \pm 0.36$ & - & - & - & - & - & - \\
HW14 & $7.1 \pm 2.2$ & - & $13.6 \pm 10.0$ & $0.4 \pm 0.7$ & $9.6 \pm 2.7$ & $4.27 \pm 0.52$ & $9.50_{-0.02}^{+0.01}$ & 0.12 & 0.002 & $4.23_{-0.43}^{+0.24}$ & -0.04 & 9.03\\
HW18 & $0.5 \pm 0.1$ & - & $0.8 \pm 0.3$ & $0.8 \pm 0.3$ & $1.0 \pm 0.3$ & $2.72 \pm 0.27$ & - & - & - & - & - & - \\
HW22 & $1.0 \pm 0.2$ & $3.86 \pm 0.58$ & $0.9 \pm 0.2$ & $1.3 \pm 0.1$ & $1.9 \pm 0.4$ & $3.78 \pm 0.17$ & $9.41_{-0.03}^{+0.03}$ & 0.02 & 0.002 & $3.59_{-0.45}^{+0.26}$ & -0.27 & 7.73\\
HW26 & $0.2 \pm 0.0$ & $4.11 \pm 0.49$ & - & - & - & - & - & - & - & - & - & - \\
HW34 & - & - & $1.9 \pm 1.4$ & $0.2 \pm 0.5$ & $1.0 \pm 0.3$ & $2.90 \pm 2.17$ & $9.58_{-0.09}^{+0.07}$ & 0.06 & 0.002 & $3.35_{-0.56}^{+0.38}$ & 0.45 & 7.20\\
HW35 & $0.9 \pm 0.2$ & $3.61 \pm 0.62$ & $0.9 \pm 0.2$ & $1.1 \pm 0.2$ & $1.4 \pm 0.3$ & $3.47 \pm 0.17$ & $8.18_{-0.33}^{+0.31}$ & 0.12 & 0.004 & $2.81_{-0.48}^{+0.29}$ & -0.81 & 7.14\\
HW36 & $0.9 \pm 0.2$ & $3.12 \pm 0.29$ & $1.2 \pm 0.5$ & $0.5 \pm 0.2$ & $1.1 \pm 0.3$ & $3.09 \pm 0.40$ & $9.12_{-0.11}^{+0.19}$ & 0.09 & 0.002 & $2.74_{-0.67}^{+0.48}$ & -0.37 & 7.04\\
HW37 & - & - & $3.4 \pm 2.7$ & $0.2 \pm 0.4$ & $2.1 \pm 0.6$ & $4.15 \pm 2.30$ & $7.60_{-0.38}^{+0.28}$ & 0.11 & 0.004 & $2.30_{-0.56}^{+0.38}$ & -1.84 & 7.21\\
HW38 & $2.4 \pm 0.7$ & $3.64 \pm 0.99$ & $3.3 \pm 1.8$ & $0.6 \pm 0.6$ & $2.9 \pm 0.9$ & $3.38 \pm 0.33$ & $9.59_{-0.31}^{+0.03}$ & 0.06 & 0.002 & $3.72_{-0.48}^{+0.29}$ & 0.07 & 8.06\\
HW40 & $1.6 \pm 0.3$ & $3.59 \pm 0.53$ & $1.7 \pm 0.3$ & $1.1 \pm 0.1$ & $2.9 \pm 0.5$ & $3.51 \pm 0.17$ & $9.43_{-0.04}^{+0.08}$ & 0.04 & 0.002 & $3.53_{-0.43}^{+0.24}$ & -0.06 & 7.96\\
\hline
    \end{tabular}
    \caption{In the first column we report the SC ID as taken from \citet{Bica-2020}. SC properties derived through EFF models: core radius and total luminosity are in the second and third column, respectively. SC properties obtained through King's profile are listed in the fourth to the seventh column, i.e., the core radius, concentration parameter, the half-light radius and luminosity, respectively. Then we list parameters derived through ASteCA, namely, age, reddening, metallicity and mass. Mass-to-light ratio was derived by using the total luminosity derived from the EFF model when available, or through the King model, otherwise. 
    Finally, in the last columns, we indicate the relaxation time.}
    \label{tab:results_lum_mass}
\end{table*}

\begin{table*}
    \centering
    \tiny
    \begin{tabular}{l|c|c|c|c|c|c|c|c|c|c|c|c|}
    \hline
     & \multicolumn{2}{c}{EFF} & \multicolumn{4}{c}{KING} & \multicolumn{4}{c}{CMD}\\
     \hline
    ID & $r_{\rm c}$ & $\log(L/L_{\odot})$ & $r_{\rm c}$ & $c$ & $r_{\rm h}$ & $\log(L/L_{\odot})$ & $\log$ (t) & E(B-V) & Z & $\log(M/M_{\odot})$ & $\log$ (M/L) & $\log(t_{\rm rh})$\\ 
     & (pc) &  & (pc) &  & (pc) & & & (mag) & &  & & \\
    \hline
HW41 & $2.8 \pm 0.9$ & - & $5.6 \pm 2.3$ & $0.8 \pm 0.5$ & $6.4 \pm 1.7$ & $3.89 \pm 0.31$ & $9.56_{-0.08}^{+0.06}$ & 0.08 & 0.002 & $3.85_{-0.43}^{+0.24}$ & -0.04 & 8.63\\
HW43 & $1.6 \pm 0.4$ & - & $2.6 \pm 0.9$ & $0.6 \pm 0.3$ & $2.6 \pm 0.7$ & $3.71 \pm 0.29$ & $8.08_{-0.09}^{+0.19}$ & 0.06 & 0.004 & $3.11_{-0.45}^{+0.26}$ & -0.60 & 7.63\\
HW44 & $1.4 \pm 0.6$ & - & $4.3 \pm 3.5$ & $0.2 \pm 0.6$ & $2.5 \pm 0.8$ & $3.57 \pm 1.22$ & $8.63_{-0.36}^{+0.14}$ & 0.11 & 0.004 & $2.71_{-0.53}^{+0.35}$ & -0.87 & 7.51\\
HW48 & $1.0 \pm 0.2$ & $4.06 \pm 0.74$ & $0.9 \pm 0.2$ & $1.4 \pm 0.2$ & $1.9 \pm 0.4$ & $3.90 \pm 0.16$ & $8.11_{-0.23}^{+0.07}$ & 0.10 & 0.004 & $3.01_{-0.48}^{+0.30}$ & -1.05 & 7.40\\
HW50 & $1.1 \pm 0.7$ & - & $3.6 \pm 3.4$ & $0.5 \pm 1.0$ & $3.0 \pm 1.0$ & $3.46 \pm 0.49$ & $8.14_{-0.26}^{+0.07}$ & 0.11 & 0.004 & $2.65_{-0.44}^{+0.25}$ & -0.81 & 7.57\\
HW52 & $1.9 \pm 0.5$ & $3.98 \pm 0.64$ & $4.3 \pm 1.6$ & $0.2 \pm 0.2$ & $2.6 \pm 0.5$ & $3.80 \pm 1.19$ & $8.14_{-0.22}^{+0.32}$ & 0.04 & 0.004 & $2.77_{-0.54}^{+0.35}$ & -1.21 & 7.52\\
HW53 & $1.3 \pm 0.3$ & $3.19 \pm 0.74$ & $2.2 \pm 0.9$ & $0.3 \pm 0.3$ & $1.3 \pm 0.3$ & $2.88 \pm 0.70$ & $9.58_{-0.19}^{+0.03}$ & 0.03 & 0.002 & $3.27_{-0.57}^{+0.38}$ & 0.08 & 7.37\\
HW54 & $1.9 \pm 0.8$ & - & - & - & - & - & $9.43_{-0.01}^{+0.24}$ & 0.11 & 0.002 & $3.35_{-0.43}^{+0.25}$ & -2.03 & 8.13\\
HW55 & $2.1 \pm 0.6$ & - & $5.3 \pm 2.8$ & $0.2 \pm 0.4$ & $3.1 \pm 0.8$ & $3.50 \pm 1.07$ & $9.19_{-0.11}^{+0.07}$ & 0.08 & 0.002 & $2.86_{-0.47}^{+0.28}$ & -0.64 & 7.75\\
HW59 & - & - & $0.6 \pm 0.2$ & $0.2 \pm 0.1$ & $0.4 \pm 0.1$ & $2.57 \pm 1.60$ & - & - & - & - & - & - \\
HW61 & $1.8 \pm 0.6$ & $4.36 \pm 2.27$ & $3.7 \pm 1.6$ & $0.2 \pm 0.3$ & $2.3 \pm 0.5$ & $3.76 \pm 1.10$ & $7.93_{-0.00}^{+0.24}$ & 0.11 & 0.004 & $3.02_{-0.49}^{+0.31}$ & -1.35 & 7.50\\
HW68 & $0.7 \pm 0.3$ & - & $1.2 \pm 0.7$ & $0.8 \pm 0.6$ & $1.5 \pm 0.5$ & $3.06 \pm 0.42$ & - & - & - & - & - & - \\
HW74 & $2.3 \pm 0.8$ & $4.08 \pm 1.81$ & $3.7 \pm 1.8$ & $0.5 \pm 0.4$ & $3.3 \pm 0.9$ & $3.61 \pm 0.42$ & $7.43_{-0.32}^{+0.16}$ & 0.09 & 0.004 & $2.67_{-0.51}^{+0.32}$ & -1.41 & 7.59\\
HW78 & $0.8 \pm 0.2$ & $3.66 \pm 0.32$ & $0.8 \pm 0.3$ & $0.9 \pm 0.2$ & $1.2 \pm 0.2$ & $3.64 \pm 0.24$ & $7.13_{-0.11}^{+0.45}$ & 0.09 & 0.004 & $2.31_{-0.57}^{+0.38}$ & -1.35 & 6.79\\
HW8 & $1.0 \pm 0.3$ & - & $1.4 \pm 0.4$ & $1.4 \pm 0.3$ & $3.3 \pm 0.8$ & $4.24 \pm 0.22$ & $7.81_{-0.03}^{+0.18}$ & 0.07 & 0.004 & $3.49_{-0.43}^{+0.24}$ & -0.75 & 7.90\\
HW82 & $0.6 \pm 0.2$ & - & $0.8 \pm 0.2$ & $1.3 \pm 0.3$ & $1.9 \pm 0.4$ & $3.83 \pm 0.20$ & $7.73_{-0.36}^{+0.30}$ & 0.07 & 0.004 & $2.10_{-0.55}^{+0.36}$ & -1.72 & 7.06\\
HW9 & $1.2 \pm 0.3$ & - & $2.1 \pm 0.8$ & $0.6 \pm 0.4$ & $2.1 \pm 0.6$ & $3.18 \pm 0.30$ & $9.52_{-0.04}^{+0.08}$ & 0.12 & 0.002 & $3.48_{-0.49}^{+0.30}$ & 0.29 & 7.74\\
IC1611 & $1.8 \pm 0.4$ & $5.28 \pm 0.98$ & $1.9 \pm 0.4$ & $1.5 \pm 0.3$ & $4.4 \pm 0.9$ & $4.97 \pm 0.17$ & $8.21_{-0.28}^{+0.22}$ & 0.10 & 0.004 & $4.07_{-0.47}^{+0.29}$ & -1.21 & 8.34\\
IC1612 & $3.3 \pm 1.1$ & - & $8.5 \pm 6.1$ & $0.4 \pm 0.7$ & $6.2 \pm 1.9$ & $4.65 \pm 0.50$ & $8.08_{-0.17}^{+0.15}$ & 0.08 & 0.004 & $3.25_{-0.49}^{+0.31}$ & -1.39 & 8.25\\
IC1624 & $2.6 \pm 0.5$ & $4.86 \pm 0.43$ & $2.7 \pm 0.5$ & $0.9 \pm 0.1$ & $3.9 \pm 0.7$ & $4.81 \pm 0.18$ & $8.31_{-0.26}^{+0.05}$ & 0.10 & 0.004 & $3.80_{-0.40}^{+0.22}$ & -1.06 & 8.17\\
IC1662 & $2.6 \pm 0.6$ & $4.22 \pm 0.46$ & $2.7 \pm 0.7$ & $0.9 \pm 0.2$ & $4.0 \pm 1.0$ & $4.16 \pm 0.23$ & $8.01_{-0.24}^{+0.22}$ & 0.10 & 0.004 & $3.22_{-0.59}^{+0.41}$ & -1.00 & 7.94\\
K1 & $7.8 \pm 1.6$ & $4.44 \pm 0.42$ & $9.1 \pm 2.4$ & $0.7 \pm 0.3$ & $10.2 \pm 2.0$ & $4.37 \pm 0.23$ & $9.84_{-0.20}^{+0.03}$ & 0.10 & 0.001 & $4.44_{-0.42}^{+0.23}$ & -0.00 & 9.19\\
K11 & $3.2 \pm 0.6$ & $3.87 \pm 0.31$ & $3.5 \pm 0.8$ & $0.7 \pm 0.1$ & $4.1 \pm 0.8$ & $3.85 \pm 0.23$ & $9.32_{-0.04}^{+0.09}$ & 0.16 & 0.002 & $3.54_{-0.44}^{+0.26}$ & -0.33 & 8.19\\
K13 & $6.0 \pm 2.0$ & - & $19.1 \pm 15.5$ & $0.2 \pm 0.5$ & $9.8 \pm 3.1$ & $4.28 \pm 2.74$ & $9.54_{-0.04}^{+0.05}$ & 0.05 & 0.002 & $4.63_{-0.43}^{+0.24}$ & 0.35 & 9.21\\
K15 & $1.8 \pm 0.3$ & $4.76 \pm 0.98$ & $1.8 \pm 0.3$ & $1.6 \pm 0.2$ & $4.0 \pm 0.7$ & $4.46 \pm 0.16$ & $8.84_{-0.39}^{+0.15}$ & 0.14 & 0.002 & $3.43_{-0.49}^{+0.31}$ & -1.33 & 8.08\\
K16 & $2.1 \pm 0.4$ & $3.46 \pm 0.31$ & $2.9 \pm 0.8$ & $0.5 \pm 0.2$ & $2.3 \pm 0.5$ & $3.40 \pm 0.35$ & $9.46_{-0.02}^{+0.08}$ & 0.05 & 0.002 & $3.55_{-0.47}^{+0.28}$ & 0.08 & 7.83\\
K17 & $2.0 \pm 0.4$ & $4.78 \pm 0.69$ & $1.9 \pm 0.3$ & $1.4 \pm 0.1$ & $4.6 \pm 0.8$ & $4.66 \pm 0.16$ & $8.82_{-0.13}^{+0.08}$ & 0.11 & 0.002 & $3.64_{-0.46}^{+0.27}$ & -1.15 & 8.26\\
K21 & $8.3 \pm 1.7$ & - & $10.7 \pm 2.7$ & $0.9 \pm 0.3$ & $14.8 \pm 3.1$ & $4.91 \pm 0.22$ & $9.48_{-0.31}^{+0.02}$ & 0.05 & 0.002 & $4.96_{-0.39}^{+0.20}$ & 0.05 & 9.61\\
K25 & $1.8 \pm 0.6$ & - & $2.8 \pm 1.2$ & $0.8 \pm 0.5$ & $3.5 \pm 1.2$ & $4.38 \pm 0.33$ & $8.25_{-0.32}^{+0.22}$ & 0.15 & 0.004 & $3.40_{-0.43}^{+0.25}$ & -0.97 & 7.96\\
K27 & $3.6 \pm 0.7$ & - & $5.2 \pm 1.4$ & $0.9 \pm 0.3$ & $7.1 \pm 1.6$ & $4.56 \pm 0.23$ & $9.13_{-0.07}^{+0.06}$ & 0.11 & 0.002 & $4.03_{-0.45}^{+0.27}$ & -0.53 & 8.72\\
K28 & $5.1 \pm 0.9$ & $4.99 \pm 0.68$ & $5.2 \pm 1.0$ & $1.2 \pm 0.2$ & $8.9 \pm 1.6$ & $4.83 \pm 0.17$ & $9.57_{-0.04}^{+0.03}$ & 0.08 & 0.002 & $4.70_{-0.41}^{+0.23}$ & -0.28 & 9.18\\
K30 & $6.6 \pm 1.8$ & - & $10.0 \pm 3.6$ & $0.7 \pm 0.4$ & $11.7 \pm 3.2$ & $4.89 \pm 0.30$ & $8.30_{-0.30}^{+0.09}$ & 0.09 & 0.004 & $3.87_{-0.45}^{+0.27}$ & -1.02 & 8.91\\
K31 & $13.2 \pm 3.2$ & - & $20.8 \pm 8.8$ & $0.5 \pm 0.4$ & $18.1 \pm 5.7$ & $5.28 \pm 0.32$ & $8.63_{-0.29}^{+0.17}$ & 0.06 & 0.004 & $4.24_{-0.44}^{+0.26}$ & -1.05 & 9.37\\
K34 & $2.9 \pm 0.5$ & $4.80 \pm 0.31$ & $3.4 \pm 0.6$ & $0.6 \pm 0.1$ & $3.6 \pm 0.6$ & $4.77 \pm 0.22$ & $8.74_{-0.10}^{+0.06}$ & 0.11 & 0.002 & $3.69_{-0.42}^{+0.24}$ & -1.12 & 8.11\\
K38 & $12.6 \pm 3.2$ & - & $19.7 \pm 8.1$ & $0.5 \pm 0.4$ & $17.7 \pm 4.8$ & $4.92 \pm 0.32$ & $9.48_{-0.05}^{+0.03}$ & 0.05 & 0.002 & $4.69_{-0.43}^{+0.24}$ & -0.23 & 9.61\\
K4 & $4.3 \pm 0.9$ & - & $5.3 \pm 1.2$ & $1.0 \pm 0.3$ & $7.7 \pm 1.5$ & $4.17 \pm 0.19$ & $9.70_{-0.01}^{+0.01}$ & 0.14 & 0.002 & $4.11_{-0.45}^{+0.26}$ & -0.06 & 8.86\\
K42 & $0.7 \pm 0.1$ & $4.85 \pm 0.91$ & $0.7 \pm 0.1$ & $1.5 \pm 0.2$ & $1.7 \pm 0.3$ & $4.61 \pm 0.16$ & $7.85_{-0.12}^{+0.14}$ & 0.13 & 0.002 & $2.83_{-0.48}^{+0.30}$ & -2.02 & 7.25\\
K43 & $2.3 \pm 0.5$ & - & $3.5 \pm 1.1$ & $0.8 \pm 0.3$ & $4.5 \pm 1.1$ & $4.40 \pm 0.26$ & $8.21_{-0.02}^{+0.18}$ & 0.16 & 0.004 & $3.61_{-0.43}^{+0.25}$ & -0.79 & 8.19\\
K44 & $11.4 \pm 2.5$ & $4.77 \pm 0.48$ & $13.3 \pm 3.9$ & $0.8 \pm 0.3$ & $15.7 \pm 3.7$ & $4.68 \pm 0.24$ & $9.57_{-0.06}^{+0.08}$ & 0.09 & 0.002 & $5.09_{-0.47}^{+0.29}$ & 0.32 & 9.71\\
K45w & $1.6 \pm 0.4$ & $3.82 \pm 0.37$ & $2.2 \pm 0.6$ & $0.6 \pm 0.2$ & $2.2 \pm 0.5$ & $3.78 \pm 0.31$ & $8.20_{-0.41}^{+0.17}$ & 0.08 & 0.004 & $2.72_{-0.47}^{+0.29}$ & -1.10 & 7.39\\
K47 & $1.7 \pm 0.4$ & $4.47 \pm 0.33$ & $1.8 \pm 0.5$ & $0.8 \pm 0.2$ & $2.4 \pm 0.5$ & $4.44 \pm 0.23$ & $7.42_{-0.03}^{+0.09}$ & 0.03 & 0.004 & $3.17_{-0.48}^{+0.29}$ & -1.29 & 7.55\\
K5 & $3.6 \pm 0.7$ & $4.83 \pm 0.69$ & $3.7 \pm 0.7$ & $1.2 \pm 0.2$ & $6.6 \pm 1.2$ & $4.67 \pm 0.17$ & $9.14_{-0.01}^{+0.03}$ & 0.15 & 0.002 & $4.18_{-0.42}^{+0.23}$ & -0.65 & 8.73\\
K50 & $2.6 \pm 0.5$ & $4.74 \pm 0.60$ & $3.4 \pm 1.1$ & $0.6 \pm 0.3$ & $3.0 \pm 0.7$ & $4.54 \pm 0.27$ & $7.38_{-0.26}^{+0.09}$ & 0.09 & 0.004 & $3.36_{-0.41}^{+0.23}$ & -1.38 & 7.77\\
K53 & $2.3 \pm 0.8$ & - & $8.3 \pm 2.5$ & $0.1 \pm 0.1$ & $4.0 \pm 0.7$ & $4.28 \pm 2.83$ & $8.07_{-0.09}^{+0.09}$ & 0.07 & 0.004 & $2.93_{-0.43}^{+0.25}$ & -1.35 & 7.85\\
K54 & $2.9 \pm 0.7$ & $4.58 \pm 0.31$ & $3.1 \pm 0.9$ & $0.7 \pm 0.2$ & $3.8 \pm 1.0$ & $4.56 \pm 0.26$ & - & - & - & - & - & - \\
K55 & $2.4 \pm 0.6$ & $4.67 \pm 1.03$ & $2.9 \pm 0.8$ & $1.1 \pm 0.3$ & $4.9 \pm 1.1$ & $4.32 \pm 0.22$ & $8.48_{-0.23}^{+0.11}$ & 0.13 & 0.004 & $3.35_{-0.45}^{+0.26}$ & -1.32 & 8.16\\
K56 & $2.2 \pm 0.5$ & - & $3.1 \pm 0.9$ & $0.8 \pm 0.3$ & $3.8 \pm 0.8$ & $4.26 \pm 0.24$ & $7.90_{-0.21}^{+0.23}$ & 0.11 & 0.004 & $3.33_{-0.43}^{+0.24}$ & -0.93 & 7.95\\
K57 & $2.4 \pm 0.6$ & - & $4.2 \pm 2.2$ & $0.5 \pm 0.5$ & $3.1 \pm 0.8$ & $3.89 \pm 0.38$ & $8.53_{-0.05}^{+0.02}$ & 0.09 & 0.004 & $3.24_{-0.45}^{+0.26}$ & -0.65 & 7.84\\
K61 & $2.0 \pm 0.6$ & - & $5.0 \pm 2.3$ & $0.3 \pm 0.3$ & $3.4 \pm 0.8$ & $4.10 \pm 0.81$ & $7.96_{-0.33}^{+0.49}$ & 0.13 & 0.004 & $2.67_{-0.52}^{+0.33}$ & -1.43 & 7.64\\
K63 & $1.5 \pm 0.5$ & - & $2.4 \pm 0.9$ & $0.8 \pm 0.3$ & $2.9 \pm 0.8$ & $4.06 \pm 0.29$ & $8.07_{-0.26}^{+0.29}$ & 0.08 & 0.004 & $3.16_{-0.47}^{+0.29}$ & -0.90 & 7.72\\
K8 & $2.4 \pm 0.5$ & - & $3.4 \pm 1.0$ & $0.9 \pm 0.3$ & $4.6 \pm 1.1$ & $3.89 \pm 0.24$ & $9.45_{-0.00}^{+0.02}$ & 0.12 & 0.002 & $3.65_{-0.43}^{+0.25}$ & -0.24 & 8.32\\
K9 & $2.9 \pm 0.9$ & - & $6.3 \pm 2.3$ & $0.6 \pm 0.4$ & $6.5 \pm 1.7$ & $3.91 \pm 0.29$ & $8.85_{-0.06}^{+0.09}$ & 0.09 & 0.002 & $3.52_{-0.45}^{+0.26}$ & -0.39 & 8.44\\
L14 & $2.8 \pm 0.7$ & $4.20 \pm 1.18$ & $3.7 \pm 1.1$ & $0.8 \pm 0.3$ & $4.7 \pm 1.1$ & $3.76 \pm 0.25$ & $9.38_{-0.03}^{+0.11}$ & 0.13 & 0.002 & $3.65_{-0.42}^{+0.24}$ & -0.55 & 8.32\\
L19 & $4.5 \pm 1.7$ & - & - & - & - & - & $9.55_{-0.05}^{+0.03}$ & 0.19 & 0.002 & $3.85_{-0.50}^{+0.32}$ & -2.08 & 8.67\\
L28 & $1.3 \pm 0.3$ & $4.38 \pm 0.62$ & $1.3 \pm 0.3$ & $1.2 \pm 0.1$ & $2.8 \pm 0.5$ & $4.28 \pm 0.17$ & - & - & - & - & - & - \\
L31 & $1.6 \pm 0.3$ & $4.49 \pm 1.03$ & $1.7 \pm 0.3$ & $1.3 \pm 0.2$ & $3.3 \pm 0.6$ & $4.12 \pm 0.17$ & - & - & - & - & - & - \\
L33 & $1.7 \pm 0.4$ & $4.28 \pm 0.71$ & $1.9 \pm 0.5$ & $1.0 \pm 0.3$ & $3.1 \pm 0.7$ & $4.10 \pm 0.23$ & - & - & - & - & - & - \\
L48 & $1.6 \pm 0.3$ & $4.48 \pm 0.31$ & $1.6 \pm 0.4$ & $0.9 \pm 0.1$ & $2.3 \pm 0.4$ & $4.46 \pm 0.20$ & $7.63_{-0.05}^{+0.20}$ & 0.09 & 0.004 & $3.40_{-0.40}^{+0.21}$ & -1.08 & 7.63\\
L51 & $1.3 \pm 0.3$ & $4.93 \pm 1.29$ & $1.7 \pm 0.5$ & $1.0 \pm 0.3$ & $2.6 \pm 0.7$ & $4.45 \pm 0.23$ & $7.48_{-0.24}^{+0.23}$ & 0.02 & 0.004 & $3.21_{-0.44}^{+0.25}$ & -1.72 & 7.64\\
L52 & $1.1 \pm 0.2$ & $4.15 \pm 0.61$ & $1.1 \pm 0.2$ & $1.2 \pm 0.2$ & $2.0 \pm 0.4$ & $4.04 \pm 0.17$ & $8.31_{-0.10}^{+0.08}$ & 0.02 & 0.004 & $3.23_{-0.48}^{+0.30}$ & -0.92 & 7.52\\
L56 & $1.1 \pm 0.2$ & $5.10 \pm 0.38$ & $1.0 \pm 0.2$ & $1.1 \pm 0.1$ & $1.9 \pm 0.4$ & $5.07 \pm 0.18$ & $7.65_{-0.25}^{+0.15}$ & 0.05 & 0.004 & $3.41_{-0.41}^{+0.22}$ & -1.69 & 7.50\\
L65 & $2.0 \pm 0.8$ & - & $4.6 \pm 2.9$ & $0.7 \pm 0.7$ & $5.0 \pm 2.2$ & $4.37 \pm 0.44$ & $7.84_{-0.24}^{+0.18}$ & 0.11 & 0.004 & $3.24_{-0.42}^{+0.23}$ & -1.13 & 8.09\\
L66 & $1.3 \pm 0.2$ & $4.85 \pm 0.33$ & $1.3 \pm 0.3$ & $0.8 \pm 0.1$ & $1.7 \pm 0.3$ & $4.82 \pm 0.19$ & $7.44_{-0.23}^{+0.25}$ & 0.05 & 0.004 & $3.21_{-0.45}^{+0.27}$ & -1.64 & 7.36\\
L80 & $3.2 \pm 0.7$ & - & $4.7 \pm 1.4$ & $0.8 \pm 0.3$ & $5.8 \pm 1.3$ & $4.45 \pm 0.24$ & $8.34_{-0.01}^{+0.04}$ & 0.05 & 0.004 & $3.57_{-0.41}^{+0.23}$ & -0.88 & 8.34\\
L91 & $3.6 \pm 0.8$ & $4.37 \pm 0.69$ & $5.2 \pm 1.8$ & $0.5 \pm 0.3$ & $4.5 \pm 1.0$ & $4.12 \pm 0.30$ & $9.15_{-0.02}^{+0.10}$ & 0.19 & 0.002 & $3.60_{-0.44}^{+0.26}$ & -0.77 & 8.26\\
L93 & $3.4 \pm 0.6$ & $3.93 \pm 0.47$ & $3.8 \pm 0.9$ & $0.8 \pm 0.2$ & $4.3 \pm 0.8$ & $3.83 \pm 0.20$ & $9.43_{-0.21}^{+0.22}$ & 0.06 & 0.002 & $3.32_{-0.45}^{+0.26}$ & -0.60 & 8.16\\
NGC152 & $6.4 \pm 1.2$ & $5.20 \pm 0.74$ & $6.7 \pm 1.4$ & $1.2 \pm 0.3$ & $11.2 \pm 2.1$ & $4.99 \pm 0.18$ & $9.15_{-0.02}^{+0.03}$ & 0.13 & 0.002 & $4.80_{-0.41}^{+0.23}$ & -0.40 & 9.32\\
NGC176 & $3.1 \pm 0.7$ & $4.56 \pm 0.37$ & $3.5 \pm 0.9$ & $0.8 \pm 0.2$ & $4.3 \pm 0.9$ & $4.52 \pm 0.23$ & $8.11_{-0.04}^{+0.01}$ & 0.07 & 0.004 & $3.54_{-0.41}^{+0.22}$ & -1.02 & 8.12\\
NGC220 & $2.9 \pm 0.6$ & $5.11 \pm 0.72$ & $2.9 \pm 0.6$ & $1.3 \pm 0.3$ & $6.2 \pm 1.3$ & $4.95 \pm 0.18$ & $8.08_{-0.26}^{+0.13}$ & 0.08 & 0.004 & $3.90_{-0.43}^{+0.24}$ & -1.21 & 8.49\\
NGC222 & $2.4 \pm 0.6$ & - & $3.2 \pm 0.9$ & $1.1 \pm 0.3$ & $5.8 \pm 1.5$ & $4.90 \pm 0.22$ & $7.92_{-0.24}^{+0.20}$ & 0.09 & 0.004 & $3.79_{-0.42}^{+0.24}$ & -1.12 & 8.40\\
NGC231 & $2.9 \pm 0.7$ & - & $4.3 \pm 1.3$ & $1.0 \pm 0.3$ & $7.2 \pm 2.2$ & $4.82 \pm 0.25$ & $7.94_{-0.15}^{+0.27}$ & 0.06 & 0.004 & $3.84_{-0.41}^{+0.23}$ & -0.99 & 8.55\\
NGC241 & $1.8 \pm 0.4$ & $4.71 \pm 0.62$ & $1.8 \pm 0.4$ & $1.2 \pm 0.3$ & $3.5 \pm 0.8$ & $4.60 \pm 0.20$ & $7.84_{-0.21}^{+0.09}$ & 0.07 & 0.004 & $3.57_{-0.43}^{+0.25}$ & -1.14 & 7.98\\
 \hline
    \end{tabular}
    
    \contcaption{}

\end{table*}

\begin{table*}
    \centering
    \tiny
    \begin{tabular}{l|c|c|c|c|c|c|c|c|c|c|c|c|}
    \hline
     & \multicolumn{2}{c}{EFF} & \multicolumn{4}{c}{KING} & \multicolumn{4}{c}{CMD}\\
     \hline
    ID & $r_{\rm c}$ & $\log(L/L_{\odot})$ & $r_{\rm c}$ & $c$ & $r_{\rm h}$ & $\log(L/L_{\odot})$ & $\log$ (t) & E(B-V) & Z & $\log(M/M_{\odot})$ & $\log$ (M/L) & $\log(t_{\rm rh})$\\ 
     & (pc) &  & (pc) &  & (pc) & & & (mag) & &  & & \\
    \hline
NGC242 & $1.4 \pm 0.4$ & - & $1.9 \pm 0.5$ & $1.1 \pm 0.3$ & $3.5 \pm 0.9$ & $4.59 \pm 0.22$ & $7.78_{-0.24}^{+0.10}$ & 0.07 & 0.004 & $3.27_{-0.46}^{+0.28}$ & -1.31 & 7.87\\
NGC256 & $1.8 \pm 0.4$ & $4.72 \pm 0.23$ & $3.0 \pm 0.7$ & $0.3 \pm 0.1$ & $2.2 \pm 0.4$ & $4.74 \pm 0.56$ & $7.92_{-0.01}^{+0.00}$ & 0.13 & 0.004 & $3.13_{-0.47}^{+0.29}$ & -1.59 & 7.52\\
NGC265 & $2.9 \pm 0.5$ & $5.11 \pm 0.49$ & $3.0 \pm 0.6$ & $1.0 \pm 0.1$ & $4.7 \pm 0.9$ & $5.04 \pm 0.18$ & $8.40_{-0.13}^{+0.05}$ & 0.16 & 0.004 & $3.85_{-0.43}^{+0.25}$ & -1.26 & 8.32\\
NGC269 & $1.8 \pm 0.3$ & $4.80 \pm 0.70$ & $1.8 \pm 0.3$ & $1.3 \pm 0.1$ & $3.7 \pm 0.6$ & $4.66 \pm 0.16$ & - & - & - & - & - & - \\
NGC290 & $0.7 \pm 0.2$ & - & $0.9 \pm 0.3$ & $1.4 \pm 0.3$ & $2.3 \pm 0.5$ & $4.96 \pm 0.20$ & $7.93_{-0.00}^{+0.56}$ & 0.19 & 0.004 & $3.76_{-0.50}^{+0.32}$ & -1.20 & 7.78\\
NGC294 & $2.3 \pm 0.4$ & $4.92 \pm 0.40$ & $2.4 \pm 0.4$ & $0.9 \pm 0.1$ & $3.5 \pm 0.6$ & $4.88 \pm 0.17$ & $8.70_{-0.01}^{+0.03}$ & 0.08 & 0.004 & $3.81_{-0.44}^{+0.25}$ & -1.10 & 8.14\\
NGC299 & $1.3 \pm 0.3$ & $4.98 \pm 0.38$ & $1.3 \pm 0.3$ & $0.8 \pm 0.1$ & $1.8 \pm 0.4$ & $4.94 \pm 0.19$ & $7.65_{-0.12}^{+0.15}$ & 0.04 & 0.004 & $3.05_{-0.43}^{+0.24}$ & -1.93 & 7.36\\
NGC306 & $1.7 \pm 0.4$ & $4.71 \pm 0.41$ & $2.9 \pm 0.9$ & $0.4 \pm 0.2$ & $2.3 \pm 0.5$ & $4.66 \pm 0.46$ & $7.77_{-0.17}^{+0.13}$ & 0.10 & 0.004 & $3.27_{-0.43}^{+0.24}$ & -1.44 & 7.60\\
NGC330 & $3.1 \pm 0.6$ & $6.08 \pm 0.55$ & $3.0 \pm 0.6$ & $1.3 \pm 0.1$ & $6.6 \pm 1.2$ & $6.01 \pm 0.17$ & $7.29_{-0.23}^{+0.22}$ & 0.14 & 0.004 & $4.61_{-0.41}^{+0.23}$ & -1.46 & 8.77\\
NGC361 & $4.8 \pm 0.9$ & $5.09 \pm 0.62$ & $5.0 \pm 0.9$ & $1.1 \pm 0.1$ & $8.3 \pm 1.5$ & $4.95 \pm 0.17$ & $9.49_{-0.01}^{+0.02}$ & 0.07 & 0.002 & $4.49_{-0.44}^{+0.25}$ & -0.60 & 9.03\\
NGC376 & $2.4 \pm 0.6$ & $5.32 \pm 0.37$ & $2.4 \pm 0.7$ & $0.9 \pm 0.2$ & $3.8 \pm 0.8$ & $5.29 \pm 0.22$ & $7.42_{-0.13}^{+0.05}$ & 0.11 & 0.004 & $4.10_{-0.40}^{+0.21}$ & -1.22 & 8.21\\
NGC416 & $2.4 \pm 0.5$ & $5.45 \pm 0.78$ & $2.3 \pm 0.5$ & $1.6 \pm 0.3$ & $6.1 \pm 1.2$ & $5.30 \pm 0.17$ & $9.79_{-0.04}^{+0.03}$ & 0.12 & 0.002 & $4.81_{-0.44}^{+0.26}$ & -0.64 & 9.00\\
NGC419 & $3.6 \pm 0.6$ & $5.72 \pm 0.57$ & $3.3 \pm 0.6$ & $1.3 \pm 0.1$ & $7.9 \pm 1.4$ & $5.65 \pm 0.16$ & $9.13_{-0.04}^{+0.00}$ & 0.09 & 0.002 & $5.16_{-0.40}^{+0.22}$ & -0.56 & 9.25\\
OGLE132 & - & - & $2.5 \pm 1.9$ & $0.4 \pm 0.6$ & $1.8 \pm 0.6$ & $3.55 \pm 0.59$ & $9.09_{-0.09}^{+0.21}$ & 0.14 & 0.002 & $2.95_{-0.58}^{+0.40}$ & -0.60 & 7.42\\
OGLE172 & $0.5 \pm 0.2$ & $2.66 \pm 0.31$ & $1.7 \pm 0.5$ & $0.0 \pm 0.1$ & $0.7 \pm 0.1$ & $2.60 \pm 5.67$ & - & - & - & - & - & - \\
OGLE28 & $0.6 \pm 0.2$ & - & $1.1 \pm 0.6$ & $0.6 \pm 0.5$ & $1.1 \pm 0.3$ & $2.94 \pm 0.39$ & - & - & - & - & - & - \\
OGLE5 & $0.6 \pm 0.2$ & $2.31 \pm 0.39$ & $1.7 \pm 0.7$ & $0.0 \pm 0.2$ & $0.7 \pm 0.1$ & $2.20 \pm 4.91$ & - & - & - & - & - & - \\
OGLE53 & $0.9 \pm 0.3$ & - & $2.4 \pm 1.9$ & $0.1 \pm 0.5$ & $1.0 \pm 0.3$ & $3.39 \pm 3.54$ & - & - & - & - & - & - \\
OGLE6 & $0.2 \pm 0.1$ & - & $0.3 \pm 0.1$ & $1.1 \pm 0.3$ & $0.6 \pm 0.2$ & $2.67 \pm 0.22$ & $9.59_{-0.40}^{+0.14}$ & 0.08 & 0.002 & $2.86_{-0.75}^{+0.57}$ & 0.19 & 6.71\\
RZ140 & $1.3 \pm 0.4$ & - & $2.0 \pm 0.8$ & $0.7 \pm 0.4$ & $2.3 \pm 0.7$ & $3.21 \pm 0.32$ & $9.13_{-0.04}^{+0.05}$ & 0.09 & 0.002 & $3.15_{-0.44}^{+0.25}$ & -0.06 & 7.66\\
RZ82 & $1.1 \pm 0.2$ & $3.24 \pm 0.69$ & $1.3 \pm 0.4$ & $0.8 \pm 0.3$ & $1.6 \pm 0.3$ & $3.02 \pm 0.22$ & - & - & - & - & - & - \\

 \hline
    \end{tabular}
    
    \contcaption{}

\end{table*}


\subsection{Core radius evolution with age: is it mass dependent?}
\label{sec:age_radius}

\begin{figure}
    \centering
    \includegraphics[width = 0.5\textwidth]{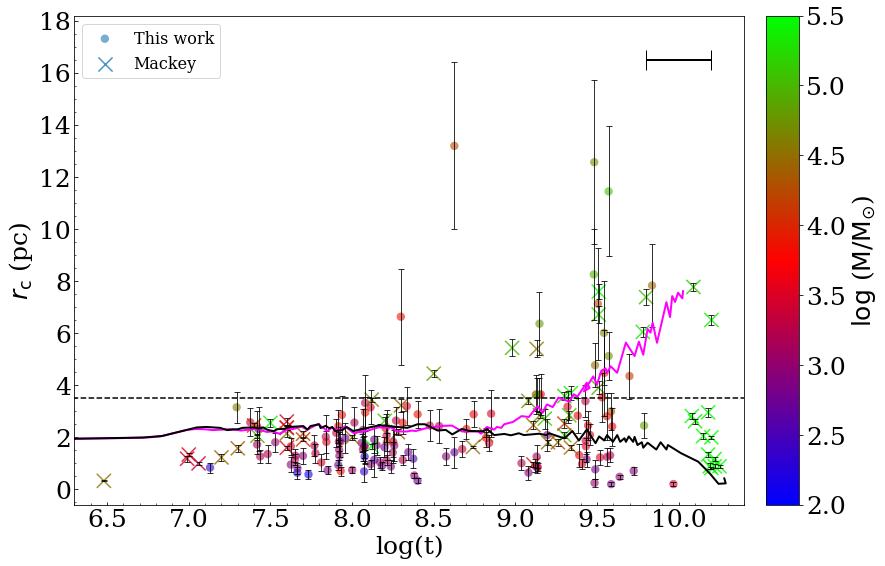}
    \caption{Core radius, estimated via an EFF profile, versus the SC age. Filled circles indicate SCs analyzed in this work while crosses represent SCs studied by \citet{Mackey&Gilmore2003,Mackey&Gilmore2003b}. All SCs are colour coded according to their mass. In order to keep the figure readable, we do not draw the errors on the x-axis. A typical age errorbar of $\Delta \log (t)$ = 0.2 dex is shown in the upper-right corner.} The horizontal dashed line is at 3.5 pc. The two solid lines represent the results of N-body simulations extrapolated from \citet{Mackey-2008}, for a case with no primordial mass segregation and a fraction of retained BH $f_{\rm BH} = 1$ (magenta line) and $f_{\rm BH} = 0$ (black line), respectively.
    \label{fig:rc_logt_eff}
\end{figure}

An expansion of the core radius is expected to take place after a few hundreds of Myr. Besides the MCs, \citet{Ryon-2015} analyzed a large sample of young ($\log(t) \leq 8.5$) and massive ($\log (M/M_{\odot}) \geq 10^4$) SCs in M83, and found evidence of an increasing trend between core radius and age. \citet{Ryon-2017} noticed a similar trend in NGC~628 and NGC~1313 again with a sample of young massive SCs (i.e. $\log (M/M_{\odot}) \geq 5*10^3$ and $t \leq 200$Myr). Both works indicated the mass-loss by stellar evolution as the candidate physical mechanism responsible for the presence of such core radius-age relationship. \citet{Chandar2016} studied 3816 SCs in M51, finding again an expansion during their early life. \citet{SanRoman-2012}, on the basis of the study of 161 SCs in M33 depicted a trend quite similar to the one observed in the LMC by \citet{Elson-1989,Elson-1991,Elson-1992,Mackey&Gilmore2003}, namely an increasing core radius spread with age \citep[see e.g. Fig. 14 in][]{Mackey&Gilmore2003}. A hint for a similar behaviour was found for the first time in the SMC by \citet{Mackey&Gilmore2003b} based on a sample of 10 SCs. More recently, \citet{Carvalho-2008} and \citet{Santos-2020}, investigated a larger but still rather small sample of 23 and 25 SMC SCs, respectively, observing a larger spread of the core radius for SCs older than $\log(t) \sim$8.5 dex with respect to young SCs \citep[e.g. Fig. 14 in][]{Santos-2020}).
The current scenario assumes that, besides an expansion in the early SC life likely due to mass-loss from stellar evolution, another inner physical process triggers some SCs to increase their core radius as they get older than $\log(t) = 8.0-8.5$ dex, but what the actual mechanism is and why it affects only a fraction of the SCs is still an open question.
In Fig.~\ref{fig:rc_logt_eff} we show the $r_{\rm c}$ as a function of the SC ages, colour coded according to their estimated mass for the sample of 134 SCs investigated in this work. To widen the sample and to augment the probed age interval, we also plotted in the same figure LMC and SMC SCs analyzed by \citet{Mackey&Gilmore2003,Mackey&Gilmore2003b}, so that the total sample of SCs displayed in the figure is 185 objects. 
An inspection of Fig.~\ref{fig:rc_logt_eff} confirms that SCs younger than $\sim$ 200-300 Myr ($\log(t) \sim 8.3-8.5$~dex) are all compact with no exception.\footnote{Actually, three SCs reach 3.5 pc at 1$\sigma$.} More precisely, an early expansion in the first 10-20 Myr seems to be present. This early expansion has been detected also  in stellar systems beyond the MCs \citep[e.g.][]{Ryon-2015,Chandar2016,Ryon-2017}, and is thought to be caused by mass-loss from stellar evolution.\par 
After this rapid phase of expansion, for the next 200-300 Myr the core radius does not seem to experience any further alteration. Beyond $t\sim$200 Myr a few SCs with $r_{\rm c} > 3.5$ pc appear, while at $t \geq$~1 Gyr the fraction of SCs with $r_{\rm c} > 3.5$ pc becomes significant. Hence, a different process with respect to that responsible for the early expansion must be invoked to explain the core radius expansion at ages later than $t\sim$200 Myr.
In the following we refer to this feature as core radius-age relationship, as done in previous works.
This behaviour has been explained by \citet{Elson1987} and \citet{Mackey&Gilmore2003} as a real evolution during the SCs lifetime.
Taking advantage of the larger number and wider range in mass of our SC sample we may put constraints on the physics behind the core radius expansion.


First of all, the fraction of SCs having $\log(t) \geq 8.25$~and $r_{\rm c} \geq 3.5$ pc is $\sim$~0.28. If this fraction were constant all along the SC lifetime we would expect about 22 young SCs with a large core radius, hence the complete lack of them is significant at more than the $4 \sigma$ level.\par
\begin{figure}
    \centering
    \includegraphics[width = 0.45\textwidth]{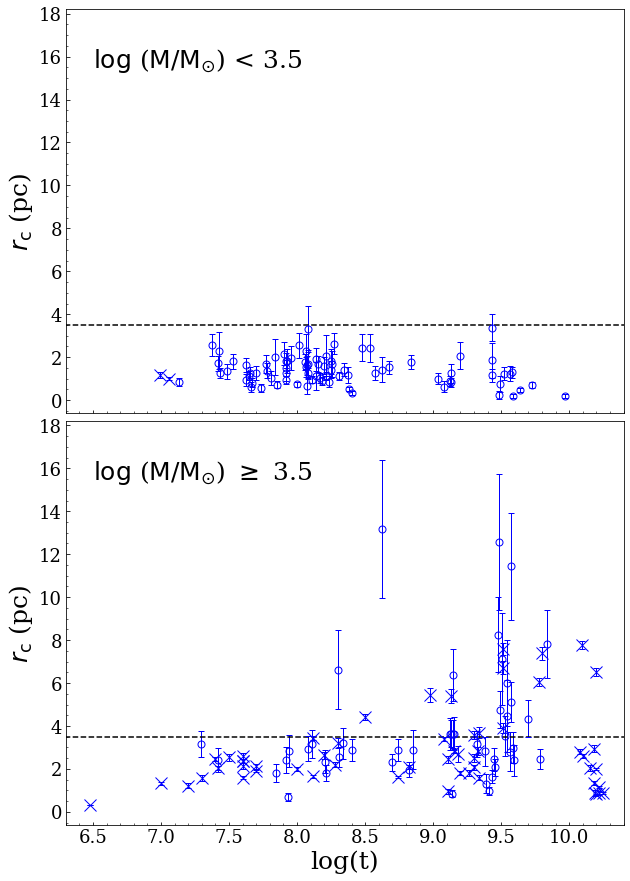}
    \caption{Same as figure \ref{fig:rc_logt_eff} but with SCs splitted into different mass ranges. \emph{Top}: SCs with $M < 10^{3.5} M_{\odot}$. 
    \emph{Bottom}: SCs with $M \geq 10^{3.5} M_{\odot}$.}
    \label{fig:rc_logt_massth}
\end{figure}

A closer inspection of Fig.~\ref{fig:rc_logt_eff} also suggests that the majority of the SCs with large core radii are also very massive. We wonder whether the mass could be, in first approximation, the physical parameter that drives intermediate-age SCs to expand their inner regions. To better investigate this hypothesis, in Fig.~\ref{fig:rc_logt_massth} we split up the $r_{\rm c}$-age plot into two different mass intervals.
The difference between the two panels is remarkable: in the high mass range (i.e. $M \geq 10^{3.5} M_{\odot}$, bottom panel) 40\% of the SCs older than $\log(t) \geq 8.25$ have experienced an inner expansion. Keeping the same percentage also in the low mass interval (i.e. $M < 10^{3.5} M_{\odot}$, top panel) we would expect to find that 13 intermediate-age SCs have $r_{\rm c} \geq 3.5$ pc, while, in practice, none are observed, a result that is significant at more than the $3\sigma$ level. Another interesting feature visible in the same figure, is that SCs older than 10-20 Myr in the high-mass regime seem to have, on average, a larger core radius with respect to those in the low-mass regime. This might indicate that the early expansion could be more severe for massive SCs, or perhaps that low-mass SCs that experience a great expansion do not survive. To summarize, after a rapid expansion during the first 10-20 Myr, all SCs younger than about 200-300 Myr are compact regardless of their mass, while the subsequent evolutionary path appears to be very different for low/high mass SCs as they get older.\par
In principle, we cannot exclude the possibility that only massive clusters are able to survive to old ages even if they are not very compact. However, in our sample a non-negligible number of old and low mass SCs is present. In particular, in the low-mass regime, 18 SCs are older than 1 Gyr.
This confirms our hypothesis that the SC mass is crucial to drive the core expansion.\footnote{One potential selection effect is that low mass SCs, especially with large radii (and/or old ages) are unlikely to end up in cluster catalogues in the first place.}\par
The fact that the core radius-age relationship likely originates from a physical process that begins when the SC is already a few hundred Myr old, and that affects only massive SCs, suggests that gas expulsion or mass loss due to stellar evolution are not the favourite mechanisms responsible for the expansion, as they are only effective in the early phases of a SC's life.
\citet{Mackey-2008} discussed the possibility that a consistent population of retained stellar mass black holes could lead to an expansion of the cluster core. Such a process starts after a few hundreds of Myr and causes a sustained increase of the cluster core, lasting for many Gyrs.
They tested this process through N-body simulations of massive clusters (i.e. initial total cluster masses of $\log M_{\rm tot} \sim 10^{4.75}$). 
By varying the fraction of retained stellar BHs at a fixed SC mass, they found that the expansion rate is strongly dependent on this fraction, and that if it is not large enough, the expansion would not begin at all.
In Fig.~\ref{fig:rc_logt_eff} we report two runs of the N-body simulations performed by \citet{Mackey-2008}, extrapolated from their Fig. 5. The two lines represent two extreme cases: i) with no primordial mass segregation and a fraction of retained BH $f_{\rm BH} = 1$ and ii) with a fraction $f_{\rm BH} = 0$. The former predicts an expansion of the core radius, which never ends, after about 1 Gyr. \citet{Mackey-2008} tested also other possibilities (not displayed in the figure to preserve readability), that foresee different evolutionary paths. For example, by imposing also a primordial mass segregation, the inner expansion starts when a SC has about 10 Myr (see their Fig. 15), and this occurrence might explain the few SCs younger than 1 Gyr observed to have a larger core radius.
As the potential well of low mass SCs is shallower, it would be possible that a large fraction of BHs is able to escape from their parent SC. This hypothesis might explain why we do not observe a core radius growth for low mass SCs.
Nevertheless, N-body simulations with a finer mass resolution are crucial to shed light on this matter. For example, it could clarify whether or not a small fraction of retained BHs in low mass SCs could trigger the inner expansion.\\
Even the presence of a IMBH could drive a prolonged expansion of the SC radius, although \cite{Baumgardt2004b} showed that this effect is anti-correlated with the number of stars belonging to the SC, hence this outcome is at odds with our result.
Nevertheless, we cannot exclude the possibility that also less massive SCs undergo an increase of their $r_{\rm c}$, but due to their low mass (i.e. low gravitational potential) this process ends up with the evaporation of the SC. Therefore we do not observe such kind of SCs, since they dissolve faster with respect to the more massive ones.\\
Unfortunately, all previous investigations based on N-body simulations studied only the evolution of SCs with $M\sim10^4 - 10^5 M_{\odot}$ or larger.
N-body studies probing lower mass SCs (down to hundreds of solar masses as in the present study) are urgently needed to understand how the different physical mechanisms affect differentially low and high mass SCs.

\subsection{Mass-radius relationship}
\label{sec:mass-radius}

\begin{figure}
    \centering
    \includegraphics[scale = 0.4]{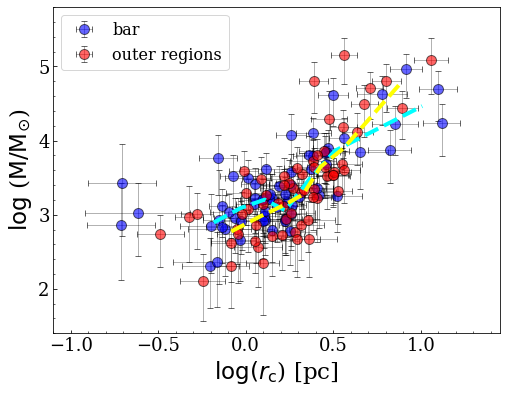}
    \caption{Mass-size relationship. The core radius was derived with an EFF profile. Blue and red points define the SCs belonging to the SMC bar and to the outer regions, respectively. Cyan and yellow dashed lines indicate the median of the SC mass calculated in bin intervals built in order to have a minimum number of 10 objects for the bar and outer SCs, respectively.}
    \label{fig:mass-size}
\end{figure}

We delved into the mass-radius relationship. Figure~\ref{fig:mass-size} depicts the estimated SC mass as a function of the core radius derived through the EFF profile.
We notice a tight linear trend in the $\log - \log$ space along almost the whole range of core radii, even if we can recognise a few very compact but massive SCs.
A weighted linear fit (i.e. $\log M = a + b \cdot \log r_{\rm c}$) \footnote{We adopted the RANSAC (RANdom SAmple Consensus) algorithm implemented by the {\sc scikit-learn} python package \citep{Pedregosa2011}, in conjunction with a bootstrap procedure} 
yielded a = $2.91 \pm 0.08$ and b = $1.54 \pm 0.30$.
We derived a Spearman rank of 0.43, with a significance larger than  99\%.
This outcome is at odds with \citet{Mackey&Gilmore2003b} and their sample of 10 SMC SCs, while a linear trend has been found by \citet{Carvalho-2008}, although their range of masses was considerably smaller than ours. In fact, it is the first time that such a trend is observed in the SMC for a conspicuous number of SCs spanning an interval of masses of $\sim$ 4 order of magnitudes, down to hundreds of solar masses.\par
\noindent
In Figure~\ref{fig:mass-size} we also analyzed separately the behaviour of the SCs located within the SMC bar and in the outer regions (see also Fig.~\ref{fig:cluster_positions}). 
An inspection of this figure suggests that the most compact objects are in the bar. Looking at the median masses that are outlined with dashed lines in Fig.~\ref{fig:mass-size} for both inner and outer SCs, we observed, on average, a similar mass at any given radius between the two sub-populations, indicating that the bar, despite being a denser environment, might not have a primary role in the evolutionary path of the SCs in the SMC. 

\subsection{SC parameters as function of their spatial positions}

\begin{figure}
    \centering
    \includegraphics[width=0.45\textwidth]{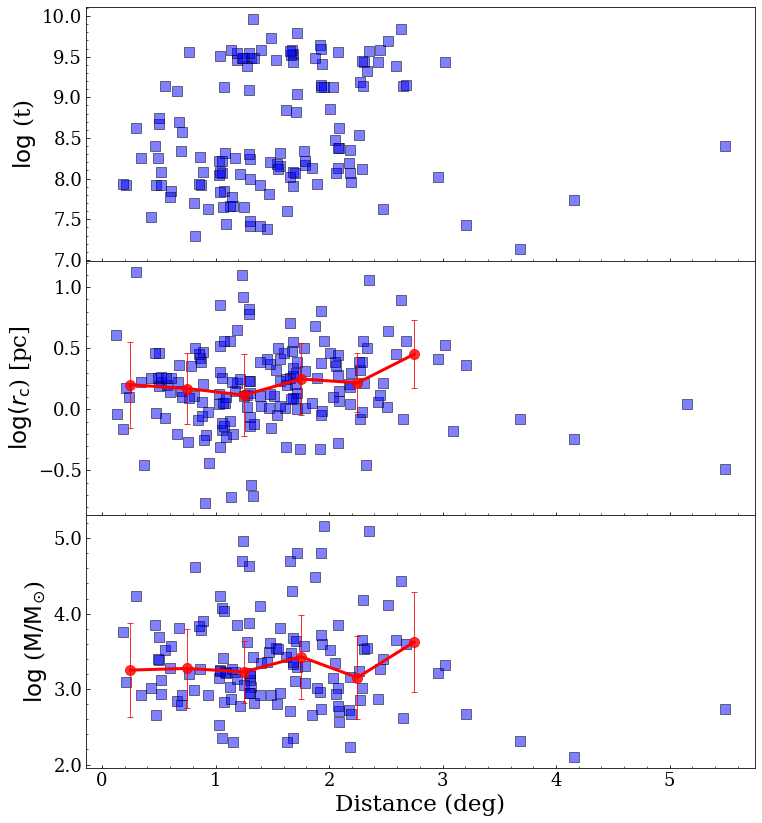}
    \caption{Age (top), core radius (middle) and mass (bottom) as a function of galactocentric distance, expressed in degrees. Red circles mark the medians estimated in bins of 0.5\degree.}
    \label{fig:params_vs_distance}
\end{figure}

We investigated the distribution of the SC main parameters as a function of their galactocentric distance. 
Figure~\ref{fig:params_vs_distance} shows the distribution of the age (top panel), core radius (middle panel) and mass (bottom panel) with the distance.
Regarding the age, the majority of SCs younger than $\sim \log(t) = 8.5$ dex is placed in the bar. Vice versa, beyond $\sim 1.5\degree$ the SC population is preferentially composed by intermediate-age objects. This result is consistent with the literature \citep[e.g.][]{Maia2014,Dias2016,Bica-2020,Piatti2021}.
No apparent trend stands out clearly in the last two panels, as is evident also by the medians of these physical quantities.
The SCs located beyond 3\degree~belong to the SMC Bridge and  are younger than 1 Gyr. Moreover, their size and mass are smaller than the averages estimated within 3\degree~(i.e. $\sim$ 1 pc and $\leq$ 1000 solar masses, respectively). We can speculate that this reflects their birth in a low-density region.

\section{Discussion}
\label{sec:discussion}

A mass-radius relation is expected to be present since the early stages of SC formation. Indeed, a relation of the form $R \propto M^\alpha$ has been found in the MW's SC forming regions, where they are present as dense clumps within giant molecular clouds (GMCs). The same kind of relation was found very recently also in the LMC \citep[see][and references therein]{Mok2021}. The exponent of this power-law estimated from the  proto-clusters is in the range $0.3 \leq \alpha \leq 0.6$.
A mass-radius direct proportionality holds also in the already formed SCs, and it has been observed in other environments beside the MW, such as nearby galaxies \citep[e.g. M31, M51, M82, etc., see the detailed review by][and references therein]{Krumholz2019}.
In figure 9 in \citet{Krumholz2019}, the authors drew on the same plot the half-light radius as a function of the mass for SCs belonging to many different galaxies in the Local Group, concluding that the relation seems to have the general form $r_{\rm h} \propto M^{1/3}$\footnote{In this work (see also figure \ref{fig:mass-size}) we obtained a relation of $r_{\rm c} \propto M^{0.34}$.}.
Nevertheless, the authors also pointed out that the scatter is too broad to devise a linear fit, either by considering only SCs hosted by the same galaxy or all SCs belonging to different systems.
As a cluster grows older, new physical mechanisms become dominant and the mass-radius trend could change or even reverse its sign.\par
\begin{figure*}
    \centering
    \includegraphics[scale = 0.4]{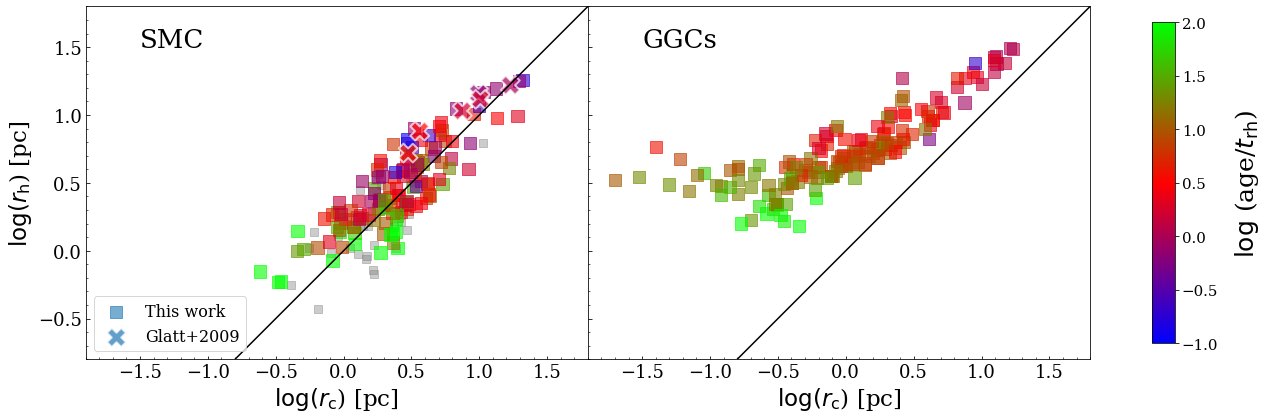}
    \caption{\emph{Left:} half-light radius as a function of core radius, estimated in this work using King profiles. \emph{Right:} same plot for GGCs taken from Baumgardt's GGC database. 
    Crosses indicate six SCs analysed by \citet{Glatt-2009}.
    The black solid line indicates the one-to-one relationship.
    SCs are colour-coded according their dynamical age (see definition in the text). SCs with no estimated age are plotted in gray}
    \label{fig:rh_vs_rc}
\end{figure*}

We can take advantage of our large sample of SCs to investigate the dynamical status of our SCs. Mass segregation and core collapse may be signs of an advanced evolutionary phase.
A good proxy for the dynamical evolutionary status of a SC is represented by the ratio between the age and the relaxation time of the SC, i.e. $age/t_{\rm rh}$. According to \citet{Spitzer1987}, $t_{\rm rh}$ is expressed as:  
\begin{equation}
    t_{\rm rh} = 8.933 \times 10^5 yr \times \frac{1}{\ln (0.4 N_{*})} <m_*>^{-1} M^{1/2} r_{\rm h}^{3/2}
    \end{equation}
\noindent
where $N_{*} = M/<m_*>$ is the estimated number of stars belonging to the SC and $<m_*>$ is the average stellar mass in solar units. To get an appropriate $<m_*>$, we integrated IMF models of different ages, in order to derive a very tight $<m_*>$-age relationship.
\citet{Glatt2011} determined the present day mass function of six intermediate-age SMC SCs (in the age interval between 6 and 10.5 Gyr) using deep data from the HST, finding that all six SCs show signs of mass segregation to different degrees, suggesting that they are dynamically evolved and/or had primordial mass segregation.
They also estimated the present day relaxation time for the six SCs to be a few Gyrs, i.e comparable to their ages, with the exception of one object.
In Fig.~\ref{fig:rh_vs_rc} we plotted the estimated half-light radius as a function of the core radius, with SCs painted with different colours according to their evolutionary agefor the SMC SCs investigated in this work (left panel).
The six SCs studied by \citet{Glatt2011} are also overplotted in Fig~\ref{fig:rh_vs_rc}\footnote{Their sizes have been taken from \citet{Glatt-2009}.} and their dynamical ages are very similar to most SCs investigated in this work, hence we expect that also other SMC SCs might show signs of mass segregation.\par
In the right panel of the same figure, we also plotted the GGCs.\footnote{Here we exploited data from Baumgardt's GGC database (https://people.smp.uq.edu.au/HolgerBaumgardt/globular/) which contains an up-to-date compilation of 158 GGCs fundamental parameters.} The GGC population is quite different with respect to the SMC SC system, as it contains much more massive and much older objects. 
In particular, the SMC contains only one GC (i.e. NGC~121), while the MW does not have populous intermediate-age and young SCs as the SMC. Also, GGCs are objects typically older than 10 Gyr.
Nevertheless, it would be instructive comparing the two systems. In addition, the MW is also representative of a very different environment with respect to the SMC (i.e. stronger tidal fields, the presence of a bulge and a thick and thin discs where the GGCs pass through, etc.).
A remarkable feature visible in Figure~\ref{fig:rh_vs_rc} for the GGCs is a noteworthy deviation from the identity relationship for $r_{\rm c} \leq 5$ pc ($\log r_{\rm c} \sim 0.7$). While going towards lower $r_{\rm c}$ values the trend flattens out, settling around $r_{\rm h} \sim 1-3$ pc ($\log r_{\rm h} \sim 0.0-0.5$) at the lowest end of the relation. As suggested by \citet{Djorgovski1996}, $r_{\rm c}$ and $r_{\rm h}$ are fairly similar as long as $t_{\rm rc} \gg$ SC age, after which the core collapses with little changes in the half-light radius. A completely different behaviour is visible for the SMC clusters studied here. Indeed, a linear trend emerges for the entire range of observed radii.
This reveals that, unlike SMC SCs, the GGCs have entered a core collapse phase despite a similar $age/t_{\rm rh}$ ratio.
suggesting that the SC evolutionary path in the two systems is noticeable different.\par
\begin{figure}
    \centering
    \includegraphics[width = 0.5\textwidth]{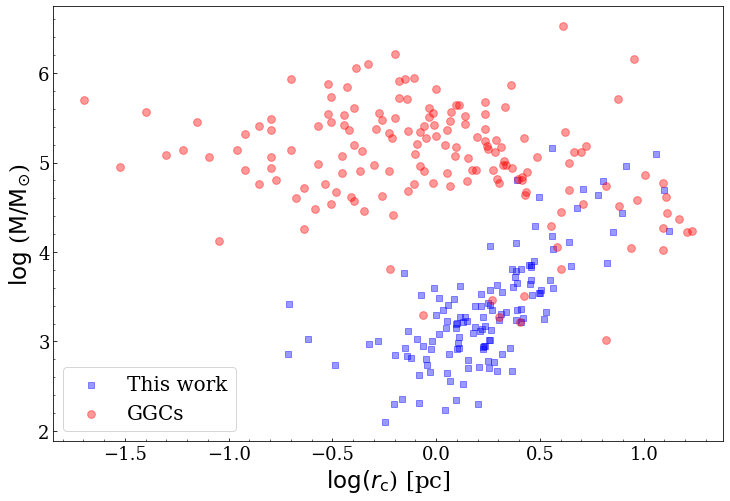}
    \caption{Core radius - mass relationship with GGCs (red points) and our SMC sample (blue squares).}
    \label{fig:Minf_vs_rc_ggc}
\end{figure}

In order to better investigate how a different evolutionary status leads to different mass-size trends, in Fig.~\ref{fig:Minf_vs_rc_ggc} we compare mass vs core radius for our sample of SCs with those of GGCs. This plot shows that, in contrast to the SMC case, the average mass of GGCs is quite constant (i.e. $\log(M/M_{\odot} \sim 10^5-10^{5.5}$~dex) up to $r_{\rm c} \leq 1$ pc, then it begins to decrease at larger core radii. As anticipated, the break of the mass-size relationship, or even the reverse sign at the upper end, reveals that the dynamical state of GGCs entered the so-called core collapse phase \citep[][]{Harris1996,McLaughlin-2005}.
Additionally, the mass spectrum of the two galaxies is remarkably different. The SCs analyzed in this work cover a broad range of masses, from hundreds up to a million of solar masses, while the majority of the GGCs have $M \geq 10^4 M_{\odot}$ \citep[e.g.][]{Harris1996,Baumgardt&Hilker2018}. It is evident that tidal effects of the MW had a great impact on the survival of the low mass GGCs, as expected, while this effect is not visible in the SMC, at least in our sample. In their landmark paper, \citet{Gnedin&Ostriker1997} investigated the destruction of the GGCs and showed that in the inner regions, only fairly massive and compact clusters can survive. This is largely due to external effects such as tides, disk shocks, etc., which should be largely absent in the SMC. Thus, environment can play a
considerable role, but is probably much less important in a dwarf
irregular like the SMC. Another key difference is the age: with the exception of NGC~121, the SMC SCs are much younger than the GGCs and hence had less time to evolve dynamically.
To summarize, although the SMC SCs seem dynamically evolved, showing in a few cases signs of mass segregation \citep[][]{Glatt2011}, they are not in a core-collapse phase yet.

\section{Conclusions}
\label{sec:conclusion}

We provide SBPs from homogeneous g band integrated photometry for 170 SMC star clusters. For 62 of them this is the first SBP appearing in the literature.
The derived profiles were fitted with EFF and King models. We list for all the clusters the structural parameters obtained from the SBP fits (core and half-light radii, total luminosity, etc.). We estimated ages and masses for 134 SCs by means of the open source python package ASteCA. 
Finally, we derived the projected distances of all SCs with respect to the SMC centre to investigate if/how their properties vary with galactocentric distance. The main results of this study are the following:
\begin{itemize}
    \item 
    We find that young SCs are compact, while clusters older than $\log(t) \sim 8.3$ show a wide range of $r_{\rm c}$ values. This occurrence has been already noted and discussed in the literature, here we confirm and strengthen it on the basis of a much larger SC sample. 
    The quoted feature has been observed also in other galaxies (see discussion in the Introduction), suggesting that the evolutionary path of SCs is the same regardless of the hosting galaxy. However, the physical mechanism responsible for this phenomenon is still unclear.
    We speculate that the SC mass is the main parameter driving the inner expansion, as none of the SCs with $\log(M/M_{\odot}) < 3.5$~dex have been observed to go through such an expansion.
    \item A mass-size relation clearly stands out over almost the entire range of radii covered in this work 
    Analysing this relationship separately for SCs placed in the bar (i.e. with distance $d \leq 1.5 \degree$ from the SMC centre) and those outside, we did not find any remarkable difference.
    \item We do not find any trend of the relevant SC parameters with distance from the SMC centre, except that young SCs are mainly concentrated in the bar and vice-versa.
    We caution that projection effects may play a role given the large depth extent of the SMC.
    \item We derived the dynamical age of our sample and we excluded that any of the studied SCs is in a core collapse phase. However, we cannot rule out that some of them present mass segregation to different degrees, as observed by \citet{Glatt2011}, in a sub-sample of six  old ($t > 6$ Gyr) SMC SCs.
    
\end{itemize}




\section*{Acknowledgements}
We thank the referee for their helpful comments and suggestions which helped us to improve the manuscript.
M. G. and V. R. acknowledge support from the INAF grant ‘Funzionamento
VST’ (1.05.03.02.04).\\
This work has been partially supported by INAF through the “Main Stream SSH program" (1.05.01.86.28).\\
M. C. acknowledges the support of INFN “Iniziativa specifica TAsP”.
We thank L. Limatola and A. Grado for their help with the data reduction.\\
We sincerely thank M.-R. L. Cioni for a critical reading of the paper.\\
This research was made possible through the use of the AAVSO Photometric All-Sky Survey (APASS), funded by the Robert Martin Ayers Sciences Fund and NSF AST-1412587. \\ 

\section*{Data availability statement}

The data underlying this article are available in the article and in its online supplementary material.



\bibliographystyle{mnras}
\bibliography{mybibliography_sbp} 




\appendix

\section{King model}
\label{app:king_model}


The empirical King's profile \citep{King1962}, expressed in mag/arcsec$^2$, is given by
\begin{equation}
\label{eq:king_profile}
    \mu(r) = \mu' - 5\log_{10}\left(\frac{1}{\sqrt{1 + \left(\frac{r}{r_{\rm c}}\right)^2}} - \frac{1}{\sqrt{1 + \left(\frac{r_{\rm t}}{r_{\rm c}}\right)^2}}\right)
\end{equation}
\begin{equation}
    \mu' = \mu_{0} + 5\log_{10}\left(1 - \frac{1}{\sqrt{1 + \left(\frac{r_{\rm t}}{r_{\rm c}}\right)^2}}\right)
\end{equation}

\noindent
where $\mu_{0}$ is the central SB, $r_{\rm c}$ is the core radius and $r_{\rm t}$ is the tidal radius. With these values, it is possible to calculate the SC concentration, defined as $c = \log(r_{\rm t}/r_{\rm c})$.
It is worth to point out that $r_{\rm c}$ defined through eq.~\ref{eq:king_profile} is similar (it indicates the distance at which the SB is half the central value) to $r_{\rm c}$ derived by means of eq.~\ref{eq:eff_profile} only if $r_{\rm t} \gg r_{\rm c}$ (high concentration). For low concentrations, the $r_{\rm c}$ value obtained from the King profile is larger than its counterpart determined from the EFF function \citep[see also Fig. 10 in][]{Glatt-2009}.\\
Another useful parameter is the half-light radius ($r_{\rm h}$), which is the radius enclosing half of the cluster total light. 
We derived it from the best model achieved through eq.~\ref{eq:eff_profile}. We calculated the projected luminosity as a function of radius (substantially the growth curve), normalized by the total projected luminosity.The total projected luminosity is obtained by integrating up to the fitting radius.
The errors are calculated with a Monte Carlo-like approach, perturbing each parameter with Gaussians with the best fitted value as mean, and the error on the parameter as sigma, and these are ``propagated” in the calculation.\\
Regarding the fitting procedure, and the choice of $r_{\rm f}$ for the King's profiles, we proceeded as discussed in the previous section for the EFF models.


Again, an integration of the equation \ref{eq:king_profile} by using as extreme values of the integral $0$ and $r_{\rm t}$ provides the total luminosity $L$ of the SC.
\begin{equation}
\label{eq:king_luminosity}
    L = \pi r_{\rm c}^2k \{\ln \left[1 + \left(\frac{r_{\rm t}}{r_{\rm c}}\right)^2\right] + \frac{\left(\frac{r_{\rm t}}{r_{\rm c}}\right)^2 + 4\sqrt{1 + \left(\frac{r_{\rm t}}{r_{\rm c}}\right)^2} - 4\left[1 + \left(\frac{r_{\rm t}}{r_{\rm c}}\right)^2\right]}{1 + \left(\frac{r_{\rm t}}{r_{\rm c}}\right)^2}\}
\end{equation}
\begin{equation}
    k = 10^{-0.4\mu}\left(1 - \frac{1}{\sqrt{1 + \left(\frac{r_{\rm t}}{r_{\rm c}}\right)^2}}\right)
\end{equation}
\noindent
In Table~\ref{tab:results_fit_params} we reported the values of $\mu_0$, $r_{\rm c}$ and $r_{\rm t}$ (in arcsec), from fitting all SCs in our sample with King models.
We listed luminosities, $r_{\rm c} - r_{\rm h}$ (in pc) and the concentration parameter in table~\ref{tab:results_lum_mass}.

\begin{figure}
    \centering
    \includegraphics[width = 0.45\textwidth]{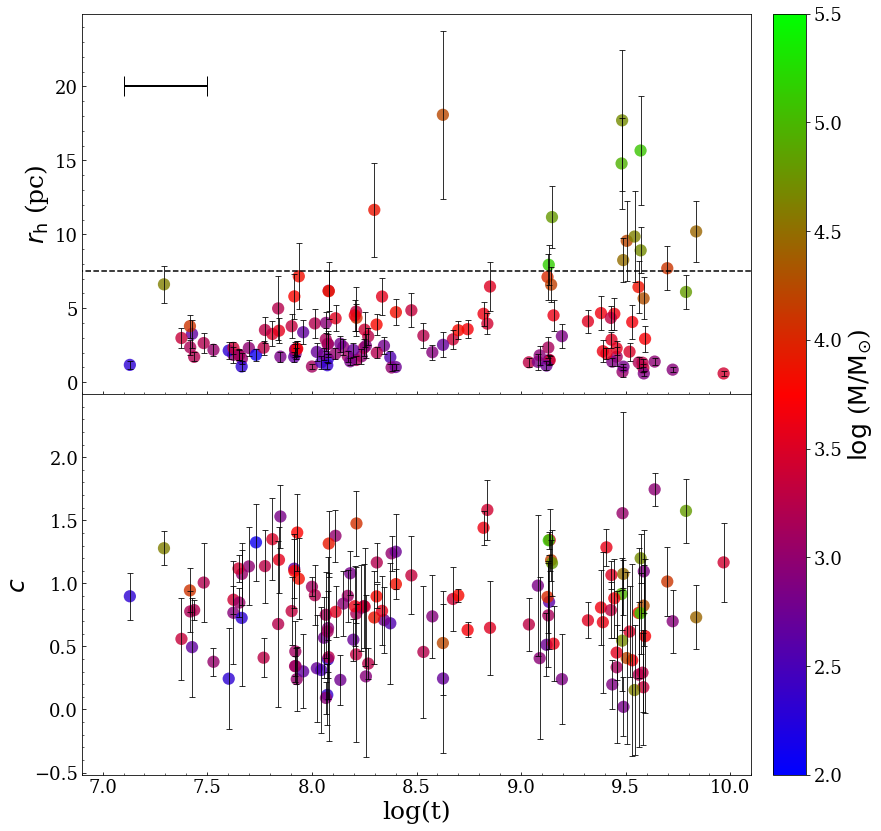}
    \caption{\emph{Top:} half-light radius, and \emph{bottom:} concentration parameter as a function of the SC age. Points are colour coded according their mass. Diamonds are SCs flagged as {\sc old} (see text for details). In the lower right cornel of the bottom panel we sketched the size of the age uncertainties, equal for all SCs.}
    \label{fig:radii_vs_logt_king}
\end{figure}

We also briefly discuss the parameters derived through the King profile in order to show that they provide the same results achieved with the EFF model.
In Fig.~\ref{fig:radii_vs_logt_king} we display the half-light radius as a function of the age (top panel).
Overall, $r_{\rm h}$ seems to show a trend with the SC age akin that observed in Fig.~\ref{fig:rc_logt_eff}.
In particular, all SCs younger than 100-300 Myr have $r_{\rm h} \leq$ 7.5 pc. Then, some SCs undergo an increase of their inner regions while the majority remains compact.
We find also in this case that the bulk of the SCs having large half-light radius are also the most massive.
Figure~\ref{fig:radii_vs_logt_king} (bottom panel) shows the concentration parameter as a function of the age. No particular trend can be detected in this case.

\section{Deprojected distance from the SMC centre}
\label{App:SMC_distance}

To obtain the 2D distance from the SMC centre we followed the recipe developed by \citet{Piatti2005} and successively used in other works \citep[e.g.][]{Dias2014,Dias2016}.
The core idea is to define the distance from the SMC through ellipses, as they better reproduce the SMC 2D geometry. 
Following \citet{Piatti2005}, we draw ellipses with an ellipticity of $e = 0.5$, having the semi-major axis direction at 45\degree, starting from the north in a counterclockwise pattern. For each SC, we looked for the ellipse intersecting its position and we set the major-axis of the ellipse as the SC galactocentric distance.
Moreover, following \citet{Dias2016} we also splitted our SC sample into subpopulations based on their 2D distance: i) the bar, constituted by SCs with a distance lower than 1.5\degree and (ii) the outer regions, made by the remaining objects.
Note that while \citet{Dias2016} called bar those SCs within 2\degree, we utilized a separation of 1.5\degree~in order to have a similar number of SCs in each of the two subpopulations.

\section{Parameters of the fitting procedure}
\label{app:parameters}

In Table~\ref{tab:results_fit_params} we reported all the SC parameters derived by fitting the SBPs with EFF and King profiles.
In the first column the SC's name is displayed, while in the second and third columns SCs' coordinates are reported. From the fourth to the sixth columns we inserted fitting parameters estimated through the EFF profile, namely $\mu_0 - \alpha - \gamma $ which are described in \S\ref{sec:fitting_eff}. From the seventh to the nineth columns we put parameters obtained via the King profile, or $\mu_0 - r_{\rm c} - c$, described in Appendix~\ref{app:king_model}.
The last two columns display the fitting radius (see discussion in \S\ref{sec:fitting_eff}) and the tile where SCs reside.

\begin{table*}
    \centering
    \tiny
    \begin{tabular}{l|c|c|c|c|c|c|c|c|c|c}
    \hline
     & & & \multicolumn{3}{c}{EFF} & \multicolumn{3}{c}{KING} & & \\
     \hline
    ID & R.A. & Dec & $\mu_0$ & $\alpha$ & $\gamma$ & $\mu_0$ & $r_{\rm c}$ & $r_{\rm t}$ & $r_{\rm f}$ & Tile\\ 
     & (J2000) & (J2000) & (mag/arcsec$^2$) & (arcsec) & & (mag/arcsec$^2$) & (arcsec) & (arcsec) & (arcsec) &\\
    \hline
B10 & 9.4345 & -73.2108 &  $16.78 \pm 1.33$ & $1.0 \pm 0.8$ & $2.8 \pm 0.6$ & $17.19 \pm 1.68$ & $0.8 \pm 0.8$ & $28.8 \pm 25.8$ & 12.0 & 3\_4\\
B100 & 15.0999 & -72.0888 &  $17.96 \pm 0.10$ & $4.0 \pm 0.4$ & $4.5 \pm 0.4$ & $17.79 \pm 0.16$ & $2.1 \pm 0.2$ & $19.9 \pm 1.2$ & 16.0 & 4\_6\\
B103 & 15.2375 & -73.1541 &  $20.02 \pm 0.21$ & $2.9 \pm 1.0$ & $2.0 \pm 0.4$ & $20.24 \pm 0.24$ & $4.5 \pm 1.4$ & $28.8 \pm 13.4$ & 16.0 & 3\_5\\
B105 & 15.4037 & -72.4073 &  $19.41 \pm 0.23$ & $3.0 \pm 0.9$ & $2.4 \pm 0.4$ & $19.49 \pm 0.19$ & $3.1 \pm 0.6$ & $37.3 \pm 14.2$ & 16.0 & 4\_6\\
B108 & 15.4637 & -72.1821 &  $21.39 \pm 0.23$ & $4.3 \pm 2.6$ & $2.0 \pm 1.0$ & $21.51 \pm 0.23$ & $6.5 \pm 3.0$ & $31.7 \pm 26.0$ & 16.0 & 4\_6\\
B111 & 15.5094 & -71.0208 &  $21.78 \pm 0.18$ & $2.7 \pm 0.9$ & $2.0 \pm 0.5$ & $21.95 \pm 0.19$ & $3.9 \pm 1.1$ & $28.1 \pm 16.2$ & 12.0 & 5\_6\\
B113 & 15.7273 & -73.3380 &  $20.25 \pm 0.22$ & $5.1 \pm 2.3$ & $3.9 \pm 1.7$ & $20.25 \pm 0.20$ & $4.0 \pm 1.0$ & $18.8 \pm 4.4$ & 12.0 & 3\_6\\
B115 & 15.8500 & -72.6516 &  $19.35 \pm 0.27$ & $3.4 \pm 1.8$ & $2.0 \pm 0.7$ & $19.57 \pm 0.27$ & $5.4 \pm 2.5$ & $33.6 \pm 30.1$ & 16.0 & 4\_6\\
B117 & 16.0545 & -73.2603 &  $19.64 \pm 0.12$ & $4.0 \pm 0.9$ & $2.5 \pm 0.4$ & $19.69 \pm 0.13$ & $4.1 \pm 0.7$ & $33.0 \pm 10.2$ & 16.0 & 3\_6\\
B119 & 16.0813 & -73.1678 &  - & - & - & $21.96 \pm 0.19$ & $9.7 \pm 7.7$ & $12.3 \pm 2.9$ & 10.0 & 3\_6\\
B12 & 9.5871 & -73.2707 &  - & - & - & $21.96 \pm 0.24$ & $8.0 \pm 6.5$ & $17.4 \pm 10.1$ & 13.0 & 3\_4\\
B121 & 16.1257 & -72.6191 &  $19.45 \pm 0.21$ & $5.3 \pm 2.6$ & $2.8 \pm 1.3$ & $19.65 \pm 0.11$ & $8.7 \pm 2.0$ & $19.3 \pm 2.0$ & 16.0 & 4\_6\\
B122 & 16.1007 & -71.7144 &  $21.92 \pm 0.35$ & $6.8 \pm 5.4$ & $2.0 \pm 1.1$ & $22.47 \pm 0.18$ & $20.5 \pm 15.1$ & $56.1 \pm 48.3$ & 32.0 & 5\_6\\
B124 & 16.2608 & -73.0427 &  $20.57 \pm 0.13$ & $6.5 \pm 2.5$ & $4.2 \pm 1.8$ & $20.66 \pm 0.11$ & $6.7 \pm 1.5$ & $16.8 \pm 2.2$ & 13.0 & 3\_6\\
B128 & 16.4628 & -71.9534 &  - & - & - & $21.45 \pm 0.26$ & $9.6 \pm 6.2$ & $20.4 \pm 6.9$ & 16.0 & 4\_6\\
B137 & 17.6174 & -72.9574 &  $20.18 \pm 0.12$ & $2.8 \pm 0.5$ & $2.0 \pm 0.2$ & $20.35 \pm 0.16$ & $3.8 \pm 0.8$ & $38.8 \pm 18.6$ & 16.0 & 3\_6\\
B165 & 22.7115 & -73.4340 &  $21.46 \pm 0.25$ & $6.3 \pm 2.5$ & $4.8 \pm 1.9$ & $21.55 \pm 0.23$ & $5.0 \pm 1.2$ & $18.3 \pm 1.8$ & 16.0 & 3\_8\\
B18 & 10.2765 & -72.7259 &  $20.35 \pm 0.14$ & $3.4 \pm 1.2$ & $3.6 \pm 1.4$ & $20.34 \pm 0.13$ & $2.7 \pm 0.6$ & $13.7 \pm 4.9$ & 6.0 & 4\_4\\
B21 & 10.3142 & -72.8316 &  $21.29 \pm 0.41$ & $2.8 \pm 2.7$ & $2.3 \pm 2.0$ & $21.53 \pm 0.11$ & $8.2 \pm 3.7$ & $8.6 \pm 0.7$ & 8.0 & 4\_4\\
B22 & 10.3361 & -72.8244 &  $18.32 \pm 0.66$ & $0.7 \pm 0.3$ & $2.0 \pm 0.2$ & $18.96 \pm 0.39$ & $1.1 \pm 0.3$ & $15.9 \pm 6.8$ & 6.0 & 4\_4\\
B26 & 10.1890 & -73.7390 &  $21.23 \pm 0.29$ & $2.4 \pm 1.5$ & $2.6 \pm 1.4$ & $21.51 \pm 0.10$ & $4.8 \pm 1.3$ & $8.2 \pm 0.9$ & 6.0 & 3\_4\\
B34A & 11.1054 & -72.9446 &  $19.50 \pm 0.43$ & $5.0 \pm 4.0$ & $5.4 \pm 4.8$ & $19.77 \pm 0.31$ & $5.6 \pm 2.7$ & $10.5 \pm 1.3$ & 10.0 & 3\_4\\
B39 & 11.3626 & -73.4800 &  $20.49 \pm 0.09$ & $5.1 \pm 1.0$ & $2.0 \pm 0.3$ & $20.61 \pm 0.12$ & $7.1 \pm 1.4$ & $53.8 \pm 21.2$ & 25.0 & 3\_4\\
B52 & 12.4189 & -73.0582 &  $21.41 \pm 0.16$ & $13.5 \pm 8.8$ & $2.0 \pm 1.2$ & $21.65 \pm 0.12$ & $35.0 \pm 25.9$ & $62.5 \pm 33.4$ & 40.0 & 3\_5\\
B65 & 13.1845 & -72.9795 &  $17.93 \pm 0.08$ & $8.6 \pm 1.4$ & $6.6 \pm 1.3$ & $17.88 \pm 0.10$ & $4.6 \pm 0.5$ & $20.0 \pm 1.3$ & 16.0 & 3\_5\\
B70 & 13.3609 & -71.7489 &  $19.68 \pm 0.11$ & $1.7 \pm 0.2$ & $2.2 \pm 0.1$ & $19.64 \pm 0.10$ & $1.5 \pm 0.1$ & $84.2 \pm 20.0$ & 20.0 & 5\_5\\
B71 & 13.3237 & -72.7660 &  $18.63 \pm 0.13$ & $13.2 \pm 4.9$ & $7.5 \pm 3.7$ & $18.71 \pm 0.12$ & $9.4 \pm 1.8$ & $22.4 \pm 1.5$ & 20.0 & 4\_5\\
B79 & 13.6992 & -72.4661 &  $18.73 \pm 0.20$ & $5.3 \pm 1.7$ & $2.8 \pm 0.6$ & $18.69 \pm 0.21$ & $4.2 \pm 0.9$ & $57.7 \pm 28.3$ & 25.0 & 4\_5\\
B80 & 13.7084 & -73.2241 &  $20.13 \pm 0.40$ & $2.1 \pm 1.2$ & $2.0 \pm 0.6$ & $20.39 \pm 0.34$ & $3.1 \pm 1.3$ & $29.8 \pm 26.4$ & 12.0 & 3\_5\\
B9 & 9.3155 & -72.9626 &  $20.97 \pm 0.42$ & $3.7 \pm 3.0$ & $2.3 \pm 1.3$ & $21.30 \pm 0.20$ & $8.8 \pm 3.4$ & $17.8 \pm 2.2$ & 16.0 & 4\_4\\
B96 & 14.8078 & -72.6089 &  $20.75 \pm 0.08$ & $16.7 \pm 5.7$ & $5.8 \pm 2.7$ & $20.79 \pm 0.07$ & $13.7 \pm 2.5$ & $31.7 \pm 3.2$ & 25.0 & 4\_6\\
B97 & 14.9158 & -71.7446 &  $21.87 \pm 0.26$ & $4.5 \pm 3.9$ & $2.0 \pm 1.6$ & $22.08 \pm 0.16$ & $10.1 \pm 6.8$ & $18.9 \pm 8.6$ & 12.0 & 5\_6\\
B99 & 15.1180 & -73.0865 &  $20.21 \pm 0.20$ & $5.3 \pm 2.3$ & $2.4 \pm 0.8$ & $20.27 \pm 0.19$ & $6.0 \pm 1.7$ & $39.8 \pm 19.4$ & 20.0 & 3\_5\\
BS102 & 15.3120 & -73.7949 &  $21.64 \pm 0.22$ & $4.2 \pm 2.5$ & $2.2 \pm 1.2$ & $21.74 \pm 0.18$ & $5.5 \pm 2.1$ & $26.4 \pm 18.9$ & 12.0 & 3\_6\\
BS128 & 16.3894 & -73.4904 &  - & - & - & $21.29 \pm 0.22$ & $3.5 \pm 2.2$ & $23.3 \pm 54.7$ & 6.0 & 3\_6\\
BS131 & 16.4978 & -72.3411 &  $19.70 \pm 0.14$ & $5.6 \pm 3.3$ & $3.1 \pm 2.2$ & $19.79 \pm 0.12$ & $7.7 \pm 3.4$ & $14.9 \pm 4.3$ & 10.0 & 4\_6\\
BS138 & 16.8159 & -72.1005 &  $22.30 \pm 0.12$ & $11.7 \pm 8.5$ & $2.0 \pm 1.6$ & $22.38 \pm 0.13$ & $19.8 \pm 15.5$ & $48.7 \pm 47.6$ & 25.0 & 4\_6\\
BS2 & 7.1291 & -73.0137 &  $21.86 \pm 0.26$ & $5.0 \pm 2.3$ & $3.6 \pm 1.4$ & $21.82 \pm 0.29$ & $3.5 \pm 1.1$ & $29.9 \pm 14.7$ & 16.0 & 3\_3\\
BS276 & 17.5233 & -72.7370 &  $17.63 \pm 0.18$ & $2.3 \pm 0.4$ & $3.1 \pm 0.3$ & $17.47 \pm 0.26$ & $1.5 \pm 0.3$ & $26.0 \pm 3.8$ & 16.0 & 4\_6\\
BS76 & 13.9092 & -71.9203 &  $20.97 \pm 0.12$ & $7.7 \pm 2.4$ & $3.3 \pm 1.0$ & $20.96 \pm 0.12$ & $6.2 \pm 1.2$ & $42.7 \pm 16.8$ & 20.0 & 5\_6\\
BS88 & 14.4583 & -72.9444 &  $21.69 \pm 0.17$ & $4.1 \pm 3.4$ & $2.0 \pm 1.8$ & $21.79 \pm 0.11$ & $7.6 \pm 5.4$ & $14.4 \pm 9.0$ & 8.0 & 4\_6\\
H86-11 & 8.2348 & -73.5038 &  $20.48 \pm 0.58$ & $1.6 \pm 1.1$ & $2.0 \pm 0.6$ & $21.00 \pm 0.44$ & $3.0 \pm 1.5$ & $23.0 \pm 16.8$ & 12.0 & 3\_3\\
H86-114 & 12.6414 & -72.6520 &  $20.08 \pm 0.12$ & $4.7 \pm 1.3$ & $2.0 \pm 0.4$ & $20.22 \pm 0.13$ & $7.2 \pm 1.7$ & $37.1 \pm 13.5$ & 20.0 & 4\_5\\
H86-137 & 13.1392 & -72.6804 &  $19.85 \pm 0.06$ & $3.0 \pm 0.3$ & $2.2 \pm 0.2$ & $19.86 \pm 0.06$ & $3.0 \pm 0.2$ & $65.3 \pm 32.2$ & 12.0 & 4\_5\\
H86-146 & 13.4132 & -72.3935 &  $17.61 \pm 0.65$ & $1.2 \pm 0.7$ & $2.0 \pm 0.3$ & $18.16 \pm 0.56$ & $2.0 \pm 0.8$ & $41.3 \pm 25.4$ & 20.0 & 4\_5\\
H86-150 & 13.7475 & -73.4191 &  $19.83 \pm 0.23$ & $3.1 \pm 1.3$ & $2.0 \pm 0.5$ & $20.06 \pm 0.25$ & $5.0 \pm 1.9$ & $29.4 \pm 16.7$ & 16.0 & 3\_5\\
H86-159 & 13.7995 & -72.6830 &  $21.15 \pm 0.17$ & $7.8 \pm 4.1$ & $3.7 \pm 2.2$ & $21.17 \pm 0.17$ & $6.6 \pm 2.2$ & $27.7 \pm 12.1$ & 16.0 & 4\_5\\
H86-165 & 13.9242 & -72.8798 &  $21.07 \pm 0.27$ & $5.7 \pm 4.3$ & $2.1 \pm 1.3$ & $21.16 \pm 0.25$ & $8.3 \pm 4.4$ & $36.6 \pm 28.0$ & 20.0 & 4\_5\\
H86-174 & 14.3251 & -72.9343 &  $20.63 \pm 0.16$ & $4.3 \pm 1.6$ & $2.1 \pm 0.6$ & $20.74 \pm 0.15$ & $5.9 \pm 1.6$ & $32.3 \pm 15.6$ & 16.0 & 4\_6\\
H86-175 & 14.4596 & -72.4401 &  $19.57 \pm 0.09$ & $9.2 \pm 3.6$ & $14.4 \pm 9.0$ & $19.56 \pm 0.10$ & $4.2 \pm 0.6$ & $9.6 \pm 0.5$ & 8.0 & 4\_6\\
H86-179 & 14.4882 & -72.4456 &  $20.72 \pm 0.42$ & $4.8 \pm 4.7$ & $2.9 \pm 2.6$ & $20.95 \pm 0.28$ & $8.7 \pm 6.0$ & $15.3 \pm 4.0$ & 12.0 & 4\_6\\
H86-181 & 14.5811 & -72.2996 &  $17.97 \pm 0.35$ & $1.8 \pm 0.7$ & $2.4 \pm 0.4$ & $18.05 \pm 0.35$ & $1.8 \pm 0.5$ & $27.3 \pm 12.2$ & 12.0 & 4\_6\\
H86-182 & 14.6073 & -72.6656 &  $20.42 \pm 0.16$ & $9.2 \pm 4.4$ & $2.7 \pm 1.3$ & $20.48 \pm 0.15$ & $10.2 \pm 3.3$ & $44.7 \pm 20.1$ & 25.0 & 4\_6\\
H86-186 & 14.9814 & -72.3719 &  $19.11 \pm 0.06$ & $6.2 \pm 0.7$ & $4.7 \pm 0.5$ & $19.06 \pm 0.08$ & $3.9 \pm 0.4$ & $22.5 \pm 1.7$ & 16.0 & 4\_6\\
H86-190 & 15.1377 & -72.2581 &  $19.50 \pm 0.32$ & $2.1 \pm 1.2$ & $2.0 \pm 0.8$ & $19.72 \pm 0.26$ & $3.2 \pm 1.4$ & $17.1 \pm 13.6$ & 8.0 & 4\_6\\
H86-191 & 15.2430 & -72.5398 &  - & - & - & $21.03 \pm 0.30$ & $6.5 \pm 4.9$ & $13.3 \pm 5.3$ & 10.0 & 4\_6\\
H86-193 & 15.3260 & -72.2283 &  $21.35 \pm 0.08$ & $8.4 \pm 3.7$ & $3.5 \pm 2.0$ & $21.36 \pm 0.08$ & $7.6 \pm 2.0$ & $28.0 \pm 13.4$ & 12.0 & 4\_6\\
H86-194 & 15.3105 & -72.5506 &  $20.25 \pm 0.54$ & $2.3 \pm 2.2$ & $2.0 \pm 1.4$ & $20.74 \pm 0.11$ & $7.7 \pm 3.2$ & $10.0 \pm 1.4$ & 8.0 & 4\_6\\
H86-213 & 23.6721 & -73.2746 &  $20.22 \pm 0.36$ & $1.1 \pm 0.3$ & $2.0 \pm 0.2$ & $20.68 \pm 0.32$ & $1.7 \pm 0.4$ & $29.7 \pm 12.9$ & 12.0 & 3\_8\\
H86-6 & 7.3467 & -72.9988 &  $22.25 \pm 0.19$ & $6.2 \pm 3.5$ & $4.3 \pm 2.9$ & $22.42 \pm 0.11$ & $8.0 \pm 2.6$ & $12.7 \pm 1.5$ & 10.0 & 3\_3\\
H86-60 & 10.1858 & -73.1180 &  - & - & - & $21.34 \pm 0.24$ & $4.9 \pm 3.7$ & $9.2 \pm 4.0$ & 6.0 & 3\_4\\
H86-74 & 11.3058 & -73.2199 &  $18.31 \pm 0.12$ & $3.5 \pm 0.7$ & $2.4 \pm 0.3$ & $18.34 \pm 0.12$ & $3.4 \pm 0.5$ & $44.4 \pm 17.9$ & 16.0 & 3\_4\\
H86-86 & 11.7602 & -73.3945 &  $20.27 \pm 0.42$ & $1.4 \pm 0.7$ & $2.8 \pm 0.9$ & $20.34 \pm 0.41$ & $1.4 \pm 0.5$ & $10.8 \pm 5.2$ & 5.0 & 3\_4\\
H86-87 & 11.7713 & -73.3706 &  $21.07 \pm 0.23$ & $5.6 \pm 4.4$ & $2.0 \pm 1.5$ & $21.34 \pm 0.15$ & $13.4 \pm 11.2$ & $24.5 \pm 15.8$ & 16.0 & 3\_4\\
H86-97 & 11.9695 & -73.2217 &  $19.64 \pm 0.13$ & $5.0 \pm 2.0$ & $2.0 \pm 0.7$ & $19.80 \pm 0.09$ & $8.9 \pm 2.4$ & $25.8 \pm 7.5$ & 16.0 & 3\_4\\
HW10 & 9.1296 & -72.9866 &  $22.23 \pm 0.12$ & $9.9 \pm 3.0$ & $2.0 \pm 0.4$ & $22.34 \pm 0.15$ & $14.5 \pm 3.8$ & $96.4 \pm 45.7$ & 50.0 & 3\_4\\
HW11 & 9.3803 & -73.6131 &  $21.78 \pm 0.23$ & $11.9 \pm 6.9$ & $2.0 \pm 0.8$ & $21.91 \pm 0.24$ & $17.7 \pm 7.6$ & $126.4 \pm 107.4$ & 64.0 & 3\_4\\
HW14 & 10.0640 & -73.8710 &  $23.00 \pm 0.14$ & $23.6 \pm 16.8$ & $2.0 \pm 1.5$ & $23.13 \pm 0.13$ & $44.7 \pm 32.0$ & $114.3 \pm 97.1$ & 60.0 & 3\_4\\
HW18 & 10.7491 & -72.4121 &  $21.07 \pm 0.24$ & $1.6 \pm 0.5$ & $2.0 \pm 0.4$ & $21.42 \pm 0.25$ & $2.7 \pm 0.9$ & $15.6 \pm 7.2$ & 8.0 & 4\_4\\
HW22 & 11.6878 & -72.0632 &  $19.74 \pm 0.10$ & $4.0 \pm 0.5$ & $2.8 \pm 0.2$ & $19.66 \pm 0.13$ & $3.0 \pm 0.3$ & $57.3 \pm 12.8$ & 25.0 & 4\_5\\
HW26 & 12.3886 & -73.7053 &  $15.17 \pm 0.39$ & $0.8 \pm 0.1$ & $3.2 \pm 0.2$ & - & - & - & 8.0 & 3\_5\\
HW34 & 14.4662 & -73.5455 &  - & - & - & $21.37 \pm 0.15$ & $6.2 \pm 4.3$ & $9.3 \pm 3.2$ & 6.0 & 3\_5\\
HW35 & 14.6820 & -73.5834 &  $20.60 \pm 0.08$ & $3.6 \pm 0.6$ & $2.7 \pm 0.3$ & $20.58 \pm 0.08$ & $3.0 \pm 0.3$ & $36.7 \pm 10.9$ & 12.0 & 3\_5\\
HW36 & 14.7667 & -73.8414 &  $20.95 \pm 0.27$ & $5.6 \pm 3.1$ & $5.5 \pm 3.6$ & $21.00 \pm 0.29$ & $4.0 \pm 1.5$ & $13.1 \pm 2.7$ & 10.0 & 3\_5\\
HW37 & 13.8725 & -71.8852 &  - & - & - & $19.96 \pm 0.42$ & $11.3 \pm 8.7$ & $19.8 \pm 3.0$ & 20.0 & 5\_6\\
HW38 & 14.8576 & -73.8170 &  $22.45 \pm 0.18$ & $9.8 \pm 7.6$ & $2.7 \pm 2.4$ & $22.48 \pm 0.18$ & $10.8 \pm 5.7$ & $41.0 \pm 36.0$ & 20.0 & 3\_5\\
HW40 & 15.1060 & -71.2949 &  $21.50 \pm 0.08$ & $7.1 \pm 0.9$ & $3.0 \pm 0.3$ & $21.47 \pm 0.08$ & $5.5 \pm 0.5$ & $64.1 \pm 10.6$ & 32.0 & 5\_6\\
HW41 & 15.1485 & -71.4601 &  $22.45 \pm 0.21$ & $9.3 \pm 4.6$ & $2.0 \pm 0.7$ & $22.80 \pm 0.17$ & $18.4 \pm 7.1$ & $107.1 \pm 86.5$ & 50.0 & 5\_6\\
HW43 & 15.2849 & -71.7537 &  $21.10 \pm 0.16$ & $5.4 \pm 2.3$ & $2.0 \pm 0.7$ & $21.22 \pm 0.15$ & $8.5 \pm 2.7$ & $35.3 \pm 15.0$ & 20.0 & 5\_6\\
HW44 & 15.3429 & -73.7883 &  $21.45 \pm 0.30$ & $4.7 \pm 3.7$ & $2.0 \pm 1.3$ & $21.91 \pm 0.14$ & $14.3 \pm 11.4$ & $25.3 \pm 14.3$ & 16.0 & 3\_6\\
HW48 & 16.2437 & -73.6375 &  $19.85 \pm 0.10$ & $3.7 \pm 0.5$ & $2.5 \pm 0.2$ & $19.80 \pm 0.11$ & $3.1 \pm 0.3$ & $72.9 \pm 27.3$ & 20.0 & 3\_6\\
HW50 & 16.5115 & -71.7110 &  $21.72 \pm 0.53$ & $3.7 \pm 3.4$ & $2.0 \pm 1.1$ & - & - & - & 20.0 & 5\_6\\
HW52 & 16.7359 & -73.2362 &  $20.88 \pm 0.18$ & $8.1 \pm 4.6$ & $3.0 \pm 1.7$ & $21.03 \pm 0.11$ & $14.0 \pm 4.7$ & $24.1 \pm 3.0$ & 20.0 & 3\_6\\
HW53 & 16.7452 & -73.5779 &  $21.99 \pm 0.08$ & $5.5 \pm 2.9$ & $2.9 \pm 2.0$ & $22.04 \pm 0.07$ & $7.3 \pm 2.8$ & $14.3 \pm 4.9$ & 8.0 & 3\_6\\
HW54 & 16.8157 & -72.1007 &  $21.94 \pm 0.30$ & $6.2 \pm 4.9$ & $2.0 \pm 1.2$ & - & - & - & 25.0 & 4\_6\\
HW55 & 16.8292 & -73.3783 &  $22.17 \pm 0.17$ & $6.9 \pm 4.2$ & $2.0 \pm 1.1$ & $22.42 \pm 0.08$ & $17.6 \pm 8.7$ & $30.7 \pm 10.2$ & 20.0 & 3\_6\\
HW59 & 17.2230 & -73.2419 &  - & - & - & $20.03 \pm 0.23$ & $2.1 \pm 0.6$ & $3.3 \pm 0.1$ & 3.0 & 3\_6\\
HW61 & 17.4270 & -72.2952 &  $21.02 \pm 0.19$ & $6.5 \pm 4.7$ & $2.3 \pm 1.7$ & $21.11 \pm 0.10$ & $12.2 \pm 4.7$ & $21.3 \pm 4.2$ & 16.0 & 4\_6\\
HW68 & 18.4701 & -73.4165 &  $21.19 \pm 0.46$ & $2.2 \pm 1.5$ & $2.0 \pm 0.8$ & $21.62 \pm 0.39$ & $4.1 \pm 2.4$ & $24.4 \pm 22.2$ & 12.0 & 3\_6\\
HW74 & 19.2002 & -73.1601 &  $22.02 \pm 0.24$ & $8.4 \pm 6.5$ & $2.3 \pm 1.7$ & $22.10 \pm 0.19$ & $12.3 \pm 5.7$ & $38.5 \pm 17.0$ & 25.0 & 3\_7\\
HW78 & 20.3360 & -73.0942 &  $19.49 \pm 0.24$ & $4.7 \pm 1.4$ & $4.8 \pm 1.2$ & $19.41 \pm 0.34$ & $2.6 \pm 0.7$ & $20.2 \pm 3.2$ & 16.0 & 3\_7\\
HW8 & 8.4446 & -73.6333 &  $19.51 \pm 0.28$ & $3.4 \pm 1.1$ & $2.0 \pm 0.3$ & $19.66 \pm 0.31$ & $4.5 \pm 1.2$ & $100.9 \pm 48.6$ & 50.0 & 3\_3\\
HW82 & 21.1158 & -73.1707 &  $19.25 \pm 0.30$ & $1.9 \pm 0.6$ & $2.0 \pm 0.2$ & $19.56 \pm 0.29$ & $2.7 \pm 0.6$ & $57.1 \pm 26.5$ & 25.0 & 3\_7\\
HW9 & 9.1051 & -73.0011 &  $22.03 \pm 0.15$ & $4.0 \pm 1.4$ & $2.0 \pm 0.5$ & $22.22 \pm 0.17$ & $6.8 \pm 2.4$ & $28.1 \pm 13.3$ & 16.0 & 3\_4\\
IC1611 & 14.9513 & -72.3340 &  $18.73 \pm 0.09$ & $6.5 \pm 1.0$ & $2.3 \pm 0.2$ & $18.73 \pm 0.10$ & $6.2 \pm 0.6$ & $184.5 \pm 89.8$ & 50.0 & 4\_6\\
IC1612 & 15.0079 & -72.3700 &  $20.74 \pm 0.18$ & $11.0 \pm 6.6$ & $2.0 \pm 1.0$ & $21.07 \pm 0.14$ & $28.0 \pm 19.4$ & $72.7 \pm 61.1$ & 40.0 & 4\_6\\
IC1624 & 16.3380 & -72.0434 &  $19.23 \pm 0.06$ & $12.1 \pm 1.5$ & $3.5 \pm 0.4$ & $19.21 \pm 0.07$ & $8.9 \pm 0.8$ & $70.1 \pm 9.2$ & 40.0 & 4\_6\\
IC1662 & 18.1363 & -73.4569 &  $20.88 \pm 0.15$ & $11.8 \pm 3.8$ & $3.4 \pm 1.0$ & $20.87 \pm 0.16$ & $9.0 \pm 1.9$ & $72.7 \pm 24.9$ & 40.0 & 3\_6\\
\hline
    \end{tabular}
    \caption{Parameters derived by fitting the EFF and King profiles.}
    \label{tab:results_fit_params}
\end{table*}
\begin{table*}
    \centering
    \hspace{2cm}
    \tiny
    \begin{tabular}{l|c|c|c|c|c|c|c|c|c|c|}
    \hline
     & & & \multicolumn{3}{c}{EFF} & \multicolumn{3}{c}{KING} & & \\
     \hline
    ID & R.A. & Dec & $\mu_0$ & $\alpha$ & $\gamma$ & $\mu_0$ & $r_{\rm c}$ & $r_{\rm t}$ & $r_{\rm f}$ & Tile\\ 
     & (J2000) & (J2000) & (mag/arcsec$^2$) & (arcsec) & & (mag/arcsec$^2$) & (arcsec) & (arcsec) & (arcsec) & \\
    \hline
K1 & 5.3578 & -73.7486 &  $22.65 \pm 0.08$ & $38.3 \pm 12.7$ & $3.7 \pm 1.4$ & $22.65 \pm 0.08$ & $29.9 \pm 6.0$ & $160.6 \pm 61.5$ & 80.0 & 3\_3\\
K11 & 9.1131 & -72.4786 &  $22.06 \pm 0.10$ & $18.6 \pm 4.9$ & $5.0 \pm 1.5$ & $22.03 \pm 0.11$ & $11.5 \pm 1.9$ & $58.6 \pm 9.8$ & 40.0 & 4\_4\\
K13 & 8.9196 & -73.5981 &  $22.60 \pm 0.17$ & $19.9 \pm 14.0$ & $2.0 \pm 1.3$ & $22.84 \pm 0.10$ & $63.1 \pm 50.0$ & $90.0 \pm 35.5$ & 64.0 & 3\_4\\
K15 & 10.0556 & -72.6988 &  $20.16 \pm 0.05$ & $6.3 \pm 0.6$ & $2.2 \pm 0.1$ & $20.16 \pm 0.05$ & $5.8 \pm 0.4$ & $223.6 \pm 109.6$ & 40.0 & 4\_4\\
K16 & 10.1389 & -72.7400 &  $21.81 \pm 0.07$ & $12.5 \pm 5.4$ & $5.2 \pm 3.2$ & $21.81 \pm 0.08$ & $9.5 \pm 2.3$ & $26.8 \pm 7.0$ & 16.0 & 4\_4\\
K17 & 10.2544 & -72.5728 &  $19.54 \pm 0.07$ & $7.7 \pm 0.8$ & $2.6 \pm 0.1$ & $19.50 \pm 0.08$ & $6.1 \pm 0.5$ & $169.4 \pm 39.7$ & 64.0 & 4\_4\\
K21 & 10.3532 & -72.8890 &  $21.77 \pm 0.08$ & $27.3 \pm 6.9$ & $2.0 \pm 0.4$ & $21.81 \pm 0.09$ & $35.5 \pm 6.5$ & $292.0 \pm 138.4$ & 127.0 & 4\_4\\
K25 & 12.0063 & -73.4861 &  $20.20 \pm 0.22$ & $6.0 \pm 2.9$ & $2.0 \pm 0.7$ & $20.35 \pm 0.25$ & $9.1 \pm 3.6$ & $59.4 \pm 40.2$ & 32.0 & 3\_4\\
K27 & 12.0579 & -73.8614 &  $21.16 \pm 0.08$ & $12.0 \pm 2.5$ & $2.0 \pm 0.3$ & $21.27 \pm 0.12$ & $17.0 \pm 3.5$ & $132.5 \pm 59.1$ & 64.0 & 3\_4\\
K28 & 12.9233 & -71.9998 &  $20.97 \pm 0.05$ & $20.1 \pm 2.5$ & $2.6 \pm 0.3$ & $20.96 \pm 0.05$ & $17.2 \pm 1.4$ & $272.6 \pm 100.6$ & 80.0 & 4\_5\\
K30 & 13.1499 & -72.1920 &  $21.43 \pm 0.14$ & $21.9 \pm 9.7$ & $2.0 \pm 0.7$ & $21.51 \pm 0.15$ & $32.9 \pm 10.7$ & $176.5 \pm 92.7$ & 100.0 & 4\_5\\
K31 & 13.2541 & -72.8978 &  $21.56 \pm 0.09$ & $43.7 \pm 21.1$ & $2.0 \pm 1.0$ & $21.61 \pm 0.10$ & $68.5 \pm 26.8$ & $230.5 \pm 130.3$ & 127.0 & 4\_5\\
K34 & 13.8875 & -72.8330 &  $19.38 \pm 0.04$ & $16.9 \pm 2.4$ & $5.1 \pm 0.9$ & $19.36 \pm 0.04$ & $11.1 \pm 0.8$ & $47.5 \pm 3.3$ & 32.0 & 4\_5\\
K38 & 14.4478 & -73.4204 &  $22.36 \pm 0.10$ & $41.6 \pm 20.5$ & $2.0 \pm 1.0$ & $22.42 \pm 0.10$ & $65.0 \pm 24.5$ & $228.3 \pm 124.9$ & 127.0 & 3\_5\\
K4 & 5.7610 & -73.6702 &  $22.56 \pm 0.07$ & $14.6 \pm 3.1$ & $2.1 \pm 0.3$ & $22.60 \pm 0.07$ & $17.6 \pm 2.5$ & $181.8 \pm 88.3$ & 64.0 & 3\_3\\
K42 & 15.1420 & -72.3658 &  $17.75 \pm 0.13$ & $2.6 \pm 0.4$ & $2.3 \pm 0.2$ & $17.73 \pm 0.13$ & $2.4 \pm 0.3$ & $80.4 \pm 37.0$ & 20.0 & 4\_6\\
K43 & 15.2033 & -73.3491 &  $20.73 \pm 0.12$ & $7.8 \pm 2.2$ & $2.0 \pm 0.4$ & $20.87 \pm 0.16$ & $11.6 \pm 3.1$ & $76.5 \pm 35.6$ & 40.0 & 3\_5\\
K44 & 15.5264 & -73.9253 &  $22.73 \pm 0.09$ & $52.7 \pm 20.3$ & $3.3 \pm 1.4$ & $22.73 \pm 0.09$ & $43.8 \pm 10.5$ & $254.2 \pm 127.1$ & 127.0 & 3\_6\\
K45w & 15.6973 & -73.7384 &  $20.63 \pm 0.17$ & $8.3 \pm 3.3$ & $4.1 \pm 1.7$ & $20.68 \pm 0.15$ & $7.4 \pm 1.7$ & $26.4 \pm 4.0$ & 20.0 & 3\_6\\
K47 & 15.7982 & -72.2721 &  $18.88 \pm 0.13$ & $9.6 \pm 2.5$ & $4.6 \pm 1.2$ & $18.86 \pm 0.16$ & $6.0 \pm 1.2$ & $36.2 \pm 6.4$ & 25.0 & 4\_6\\
K5 & 6.1770 & -73.7546 &  $20.85 \pm 0.06$ & $14.2 \pm 1.8$ & $2.6 \pm 0.3$ & $20.84 \pm 0.06$ & $12.3 \pm 1.0$ & $187.6 \pm 55.1$ & 64.0 & 3\_3\\
K50 & 16.1512 & -72.1611 &  $19.73 \pm 0.08$ & $11.2 \pm 4.8$ & $3.1 \pm 1.6$ & $19.75 \pm 0.08$ & $11.3 \pm 3.2$ & $41.1 \pm 19.4$ & 20.0 & 4\_6\\
K53 & 16.5568 & -73.2973 &  $20.62 \pm 0.23$ & $7.5 \pm 4.6$ & $2.0 \pm 1.0$ & $20.90 \pm 0.07$ & $27.5 \pm 7.0$ & $33.9 \pm 1.6$ & 32.0 & 3\_6\\
K54 & 16.6988 & -72.2725 &  $19.67 \pm 0.16$ & $17.0 \pm 6.2$ & $5.1 \pm 2.0$ & $19.65 \pm 0.19$ & $10.2 \pm 2.5$ & $56.1 \pm 12.7$ & 40.0 & 4\_6\\
K55 & 16.8846 & -73.1217 &  $20.89 \pm 0.16$ & $8.9 \pm 2.8$ & $2.3 \pm 0.5$ & $20.93 \pm 0.17$ & $9.6 \pm 2.1$ & $110.9 \pm 55.1$ & 50.0 & 3\_6\\
K56 & 16.8670 & -72.4927 &  $20.61 \pm 0.14$ & $7.3 \pm 2.5$ & $2.1 \pm 0.5$ & $20.70 \pm 0.13$ & $10.3 \pm 2.4$ & $61.9 \pm 24.4$ & 32.0 & 4\_6\\
K57 & 17.0552 & -73.2583 &  $21.41 \pm 0.14$ & $8.0 \pm 4.7$ & $2.0 \pm 1.2$ & $21.54 \pm 0.12$ & $13.8 \pm 7.0$ & $39.5 \pm 27.7$ & 20.0 & 3\_6\\
K61 & 17.2663 & -73.0870 &  $20.86 \pm 0.18$ & $6.5 \pm 3.2$ & $2.0 \pm 0.8$ & $21.14 \pm 0.12$ & $16.6 \pm 7.0$ & $33.2 \pm 8.4$ & 25.0 & 3\_6\\
K63 & 17.6987 & -72.7935 &  $20.39 \pm 0.21$ & $5.1 \pm 2.2$ & $2.0 \pm 0.6$ & $20.56 \pm 0.20$ & $8.0 \pm 2.6$ & $45.1 \pm 21.1$ & 25.0 & 4\_6\\
K8 & 7.0067 & -73.3039 &  $21.97 \pm 0.11$ & $8.1 \pm 2.2$ & $2.0 \pm 0.4$ & $22.08 \pm 0.13$ & $11.3 \pm 2.6$ & $86.3 \pm 42.7$ & 40.0 & 3\_3\\
K9 & 7.5033 & -73.3790 &  $22.44 \pm 0.19$ & $9.6 \pm 4.6$ & $2.0 \pm 0.7$ & $22.80 \pm 0.13$ & $20.6 \pm 6.9$ & $91.7 \pm 48.6$ & 50.0 & 3\_3\\
L14 & 8.1683 & -72.5803 &  $22.43 \pm 0.13$ & $10.2 \pm 3.5$ & $2.3 \pm 0.6$ & $22.48 \pm 0.13$ & $12.3 \pm 3.0$ & $78.9 \pm 35.4$ & 40.0 & 4\_4\\
L19 & 9.4240 & -73.9055 &  $22.72 \pm 0.20$ & $14.8 \pm 13.7$ & $2.0 \pm 1.9$ & - & - & - & 40.0 & 3\_4\\
L28 & 10.7481 & -72.5890 &  $19.35 \pm 0.12$ & $5.4 \pm 0.8$ & $2.7 \pm 0.2$ & $19.32 \pm 0.10$ & $4.4 \pm 0.4$ & $73.3 \pm 11.1$ & 40.0 & 4\_4\\
L31 & 11.6503 & -72.7419 &  $20.37 \pm 0.08$ & $5.7 \pm 0.8$ & $2.2 \pm 0.2$ & $20.38 \pm 0.08$ & $5.6 \pm 0.5$ & $124.3 \pm 59.1$ & 32.0 & 4\_5\\
L33 & 11.8541 & -72.8415 &  $20.19 \pm 0.17$ & $6.5 \pm 2.0$ & $2.6 \pm 0.6$ & $20.21 \pm 0.18$ & $6.3 \pm 1.4$ & $64.7 \pm 27.5$ & 32.0 & 4\_5\\
L48 & 13.3672 & -71.3989 &  $18.90 \pm 0.14$ & $9.6 \pm 1.8$ & $5.0 \pm 0.8$ & $18.85 \pm 0.18$ & $5.2 \pm 0.8$ & $38.7 \pm 3.6$ & 32.0 & 5\_5\\
L51 & 13.7271 & -72.1141 &  $18.92 \pm 0.17$ & $4.8 \pm 1.5$ & $2.2 \pm 0.5$ & $18.99 \pm 0.17$ & $5.5 \pm 1.2$ & $56.3 \pm 27.7$ & 26.0 & 4\_5\\
L52 & 13.8220 & -73.5071 &  $19.38 \pm 0.07$ & $4.6 \pm 0.6$ & $2.7 \pm 0.2$ & $19.35 \pm 0.08$ & $3.7 \pm 0.3$ & $54.3 \pm 14.1$ & 20.0 & 3\_5\\
L56 & 14.3753 & -72.2648 &  $16.58 \pm 0.11$ & $5.7 \pm 0.7$ & $3.9 \pm 0.3$ & $16.46 \pm 0.20$ & $3.2 \pm 0.5$ & $41.6 \pm 4.1$ & 32.0 & 4\_5\\
L65 & 15.2521 & -72.7498 &  $20.65 \pm 0.31$ & $6.7 \pm 4.3$ & $2.0 \pm 0.8$ & $21.11 \pm 0.28$ & $15.2 \pm 9.2$ & $72.4 \pm 66.7$ & 40.0 & 4\_6\\
L66 & 15.4374 & -72.5643 &  $17.31 \pm 0.12$ & $7.1 \pm 1.4$ & $4.7 \pm 0.9$ & $17.24 \pm 0.11$ & $4.3 \pm 0.5$ & $26.3 \pm 2.0$ & 20.0 & 4\_6\\
L80 & 16.8675 & -72.7694 &  $20.87 \pm 0.11$ & $10.6 \pm 3.1$ & $2.0 \pm 0.4$ & $20.97 \pm 0.13$ & $15.5 \pm 3.6$ & $94.0 \pm 37.3$ & 50.0 & 4\_6\\
L91 & 18.2130 & -73.1198 &  $21.89 \pm 0.09$ & $15.1 \pm 6.3$ & $2.8 \pm 1.3$ & $21.92 \pm 0.09$ & $17.2 \pm 5.1$ & $57.1 \pm 22.5$ & 32.0 & 3\_6\\
L93 & 18.2029 & -73.4744 &  $22.12 \pm 0.06$ & $15.5 \pm 3.7$ & $3.3 \pm 0.9$ & $22.11 \pm 0.06$ & $12.6 \pm 1.8$ & $77.6 \pm 28.6$ & 32.0 & 3\_6\\
NGC152 & 8.2329 & -73.1159 &  $21.20 \pm 0.07$ & $24.4 \pm 4.4$ & $2.5 \pm 0.4$ & $21.20 \pm 0.07$ & $22.1 \pm 2.6$ & $320.0 \pm 157.9$ & 100.0 & 3\_3\\
NGC176 & 8.9929 & -73.1659 &  $20.18 \pm 0.11$ & $16.4 \pm 4.9$ & $4.1 \pm 1.3$ & $20.17 \pm 0.12$ & $11.4 \pm 2.2$ & $67.9 \pm 18.5$ & 40.0 & 3\_4\\
NGC220 & 10.1280 & -73.4027 &  $19.53 \pm 0.09$ & $11.4 \pm 2.0$ & $2.5 \pm 0.3$ & $19.52 \pm 0.10$ & $9.7 \pm 1.2$ & $200.0 \pm 96.4$ & 70.0 & 3\_4\\
NGC222 & 10.1838 & -73.3839 &  $19.54 \pm 0.17$ & $8.0 \pm 2.5$ & $2.0 \pm 0.4$ & $19.62 \pm 0.17$ & $10.5 \pm 2.2$ & $132.2 \pm 60.7$ & 70.0 & 3\_4\\
NGC231 & 10.2770 & -73.3514 &  $20.08 \pm 0.16$ & $9.4 \pm 2.9$ & $2.0 \pm 0.4$ & $20.25 \pm 0.20$ & $14.0 \pm 3.6$ & $152.9 \pm 74.0$ & 80.0 & 3\_4\\
NGC241 & 10.8827 & -73.4397 &  $19.22 \pm 0.16$ & $7.4 \pm 1.9$ & $2.7 \pm 0.5$ & $19.19 \pm 0.16$ & $6.0 \pm 1.1$ & $91.8 \pm 38.7$ & 40.0 & 3\_4\\
NGC242 & 10.9026 & -73.4436 &  $19.10 \pm 0.19$ & $4.6 \pm 1.3$ & $2.0 \pm 0.3$ & $19.27 \pm 0.19$ & $6.3 \pm 1.3$ & $86.7 \pm 39.6$ & 40.0 & 3\_4\\
NGC256 & 11.4753 & -73.5069 &  $18.49 \pm 0.11$ & $14.1 \pm 4.7$ & $8.3 \pm 3.7$ & $18.57 \pm 0.09$ & $9.9 \pm 1.4$ & $21.9 \pm 0.9$ & 20.0 & 3\_4\\
NGC265 & 11.7984 & -73.4775 &  $19.22 \pm 0.06$ & $12.8 \pm 1.7$ & $3.2 \pm 0.4$ & $19.20 \pm 0.07$ & $9.8 \pm 0.9$ & $96.7 \pm 17.4$ & 50.0 & 3\_4\\
NGC269 & 12.0892 & -73.5303 &  $19.09 \pm 0.04$ & $7.0 \pm 0.5$ & $2.6 \pm 0.1$ & $19.07 \pm 0.04$ & $5.9 \pm 0.3$ & $119.4 \pm 20.5$ & 40.0 & 3\_5\\
NGC290 & 12.8113 & -73.1615 &  $17.27 \pm 0.22$ & $2.3 \pm 0.6$ & $2.0 \pm 0.2$ & $17.49 \pm 0.25$ & $3.1 \pm 0.7$ & $77.7 \pm 37.6$ & 32.0 & 3\_5\\
NGC294 & 13.2736 & -73.3803 &  $18.74 \pm 0.03$ & $11.4 \pm 0.7$ & $3.8 \pm 0.2$ & $18.71 \pm 0.05$ & $7.8 \pm 0.4$ & $62.3 \pm 4.0$ & 40.0 & 3\_5\\
NGC299 & 13.3533 & -72.1970 &  $17.12 \pm 0.12$ & $6.5 \pm 1.4$ & $4.0 \pm 0.8$ & $17.06 \pm 0.12$ & $4.4 \pm 0.6$ & $30.7 \pm 4.0$ & 20.0 & 4\_5\\
NGC306 & 13.5617 & -72.2418 &  $18.62 \pm 0.17$ & $8.3 \pm 3.5$ & $3.7 \pm 1.7$ & $18.75 \pm 0.13$ & $9.6 \pm 2.3$ & $24.7 \pm 3.0$ & 20.0 & 4\_5\\
NGC330 & 14.0764 & -72.4636 &  $17.10 \pm 0.10$ & $13.3 \pm 1.9$ & $2.9 \pm 0.3$ & $17.07 \pm 0.12$ & $9.8 \pm 1.1$ & $185.6 \pm 37.4$ & 100.0 & 4\_5\\
NGC361 & 15.5451 & -71.6060 &  $20.38 \pm 0.05$ & $19.2 \pm 2.2$ & $2.7 \pm 0.3$ & $20.38 \pm 0.04$ & $16.6 \pm 1.1$ & $196.7 \pm 38.4$ & 80.0 & 5\_6\\
NGC376 & 15.9732 & -72.8260 &  $17.88 \pm 0.19$ & $12.5 \pm 3.5$ & $4.0 \pm 0.9$ & $17.85 \pm 0.22$ & $7.9 \pm 1.7$ & $69.5 \pm 13.0$ & 50.0 & 4\_6\\
NGC416 & 16.9963 & -72.3552 &  $18.57 \pm 0.09$ & $9.2 \pm 1.3$ & $2.4 \pm 0.2$ & $18.54 \pm 0.10$ & $7.6 \pm 0.8$ & $286.0 \pm 136.7$ & 80.0 & 4\_6\\
NGC419 & 17.0730 & -72.8844 &  $18.16 \pm 0.04$ & $15.1 \pm 0.9$ & $2.8 \pm 0.1$ & $18.11 \pm 0.06$ & $10.9 \pm 0.6$ & $240.3 \pm 25.0$ & 127.0 & 4\_6\\
OGLE132 & 15.5551 & -72.9679 &  - & - & - & $21.29 \pm 0.25$ & $8.2 \pm 6.2$ & $20.8 \pm 15.1$ & 12.0 & 3\_6\\
OGLE172 & 10.4508 & -73.3919 &  $20.87 \pm 0.35$ & $3.3 \pm 2.4$ & $5.3 \pm 4.7$ & $21.22 \pm 0.08$ & $5.5 \pm 1.5$ & $5.5 \pm 0.2$ & 5.0 & 3\_4\\
OGLE28 & 11.3667 & -72.8199 &  $20.80 \pm 0.30$ & $1.9 \pm 0.9$ & $2.0 \pm 0.6$ & $21.19 \pm 0.29$ & $3.7 \pm 1.9$ & $13.6 \pm 8.0$ & 8.0 & 4\_4\\
OGLE5 & 9.8419 & -73.2580 &  $22.16 \pm 0.22$ & $3.1 \pm 2.0$ & $4.2 \pm 3.1$ & $22.38 \pm 0.09$ & $5.5 \pm 2.0$ & $6.0 \pm 0.5$ & 5.0 & 3\_4\\
OGLE53 & 12.3182 & -73.2116 &  $20.33 \pm 0.17$ & $3.0 \pm 2.1$ & $2.0 \pm 1.5$ & $20.56 \pm 0.08$ & $8.0 \pm 6.0$ & $10.2 \pm 3.9$ & 6.0 & 3\_5\\
OGLE6 & 9.8885 & -73.1815 &  $19.62 \pm 0.65$ & $0.6 \pm 0.3$ & $2.0 \pm 0.2$ & $20.34 \pm 0.42$ & $1.1 \pm 0.4$ & $14.3 \pm 6.2$ & 6.0 & 3\_4\\
RZ140 & 15.6803 & -71.4808 &  $22.04 \pm 0.27$ & $4.1 \pm 2.2$ & $2.0 \pm 0.7$ & $22.23 \pm 0.26$ & $6.5 \pm 2.5$ & $36.3 \pm 20.3$ & 20.0 & 5\_6\\
RZ82 & 13.2911 & -71.9957 &  $21.75 \pm 0.12$ & $4.3 \pm 1.2$ & $2.6 \pm 0.7$ & $21.80 \pm 0.12$ & $4.4 \pm 0.9$ & $26.7 \pm 10.0$ & 12.0 & 4\_5\\
\hline
\end{tabular}
    \contcaption{}
\end{table*}

\section{Profiles - figure}
In Figure~\ref{fig:best_fit_both_profiles} we display all SBPs with the best fit obtained through the EFF (red line) and King (blue line) profiles.

\begin{figure*}
    \centering
    \hspace{-1.cm}
    \includegraphics[width = 0.2\textwidth]{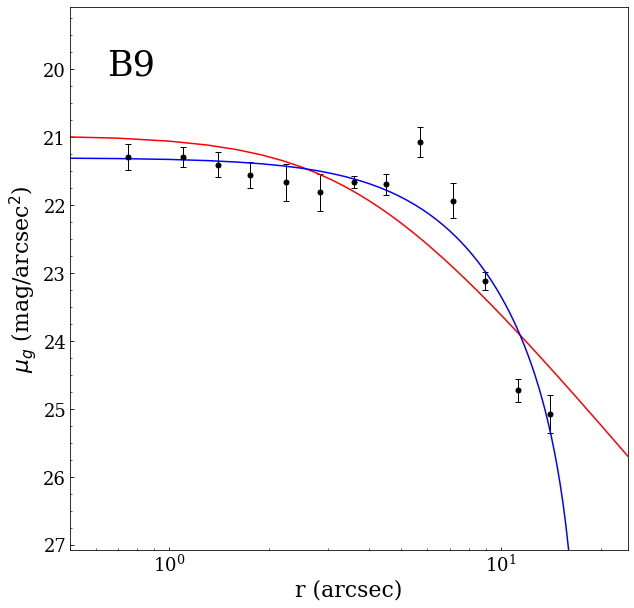}
    \includegraphics[width = 0.2\textwidth]{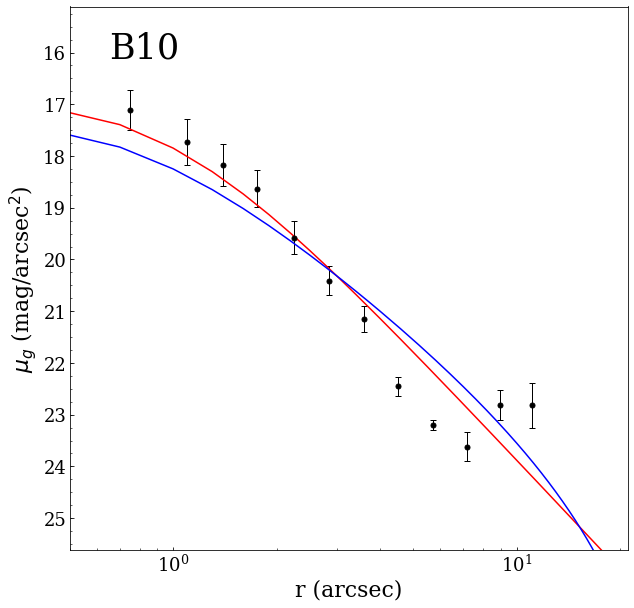}
    \includegraphics[width = 0.2\textwidth]{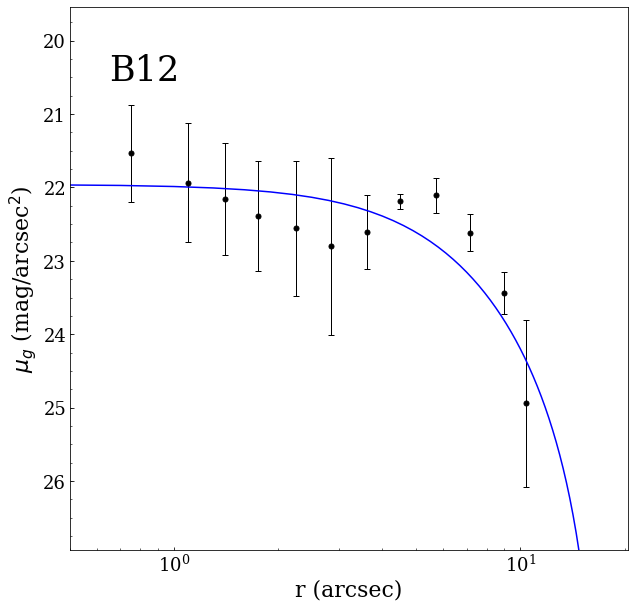}
    \includegraphics[width = 0.2\textwidth]{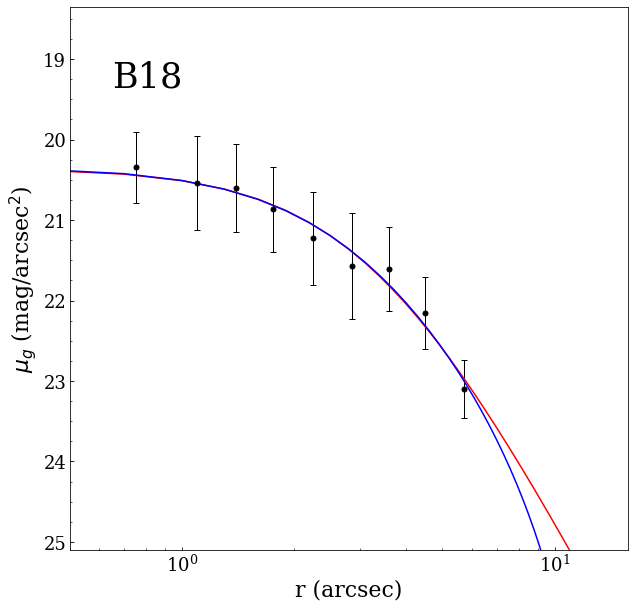}
    \includegraphics[width = 0.2\textwidth]{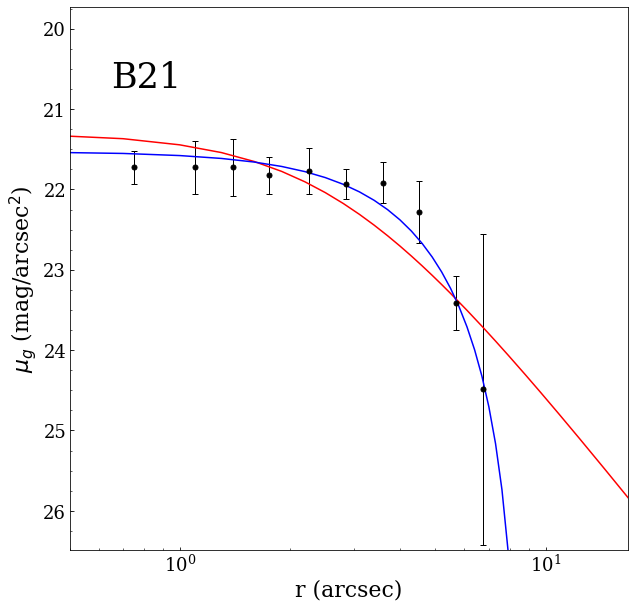}\\
    \hspace{-1.cm}
    \includegraphics[width = 0.2\textwidth]{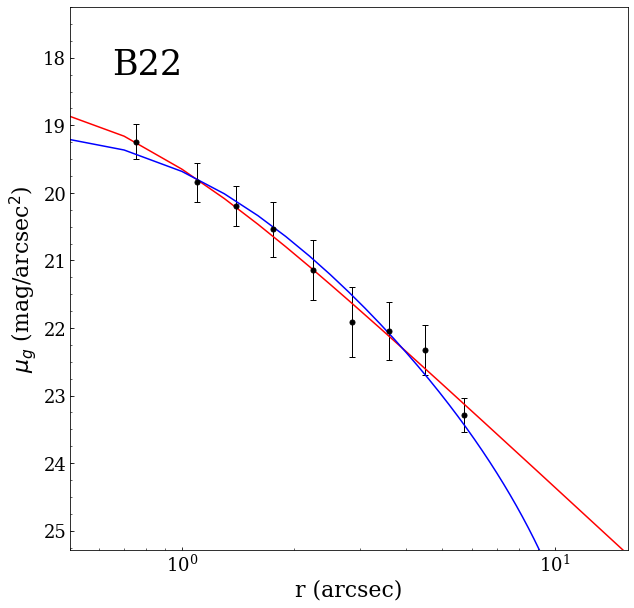}
    \includegraphics[width = 0.2\textwidth]{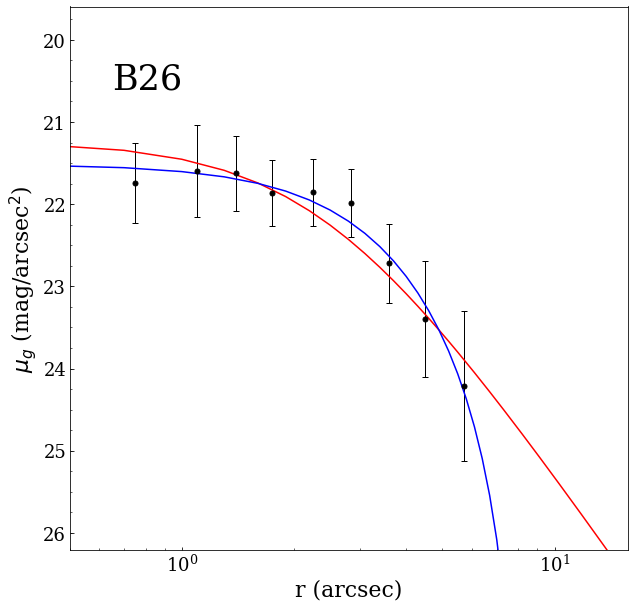}
    \includegraphics[width = 0.2\textwidth]{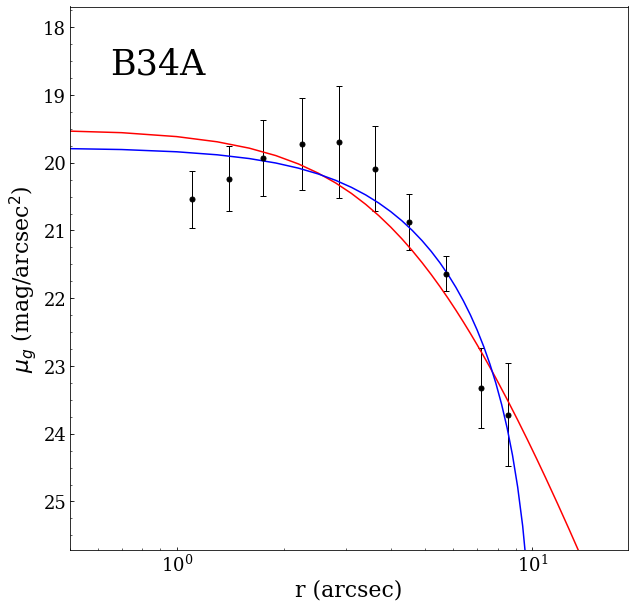}
    \includegraphics[width = 0.2\textwidth]{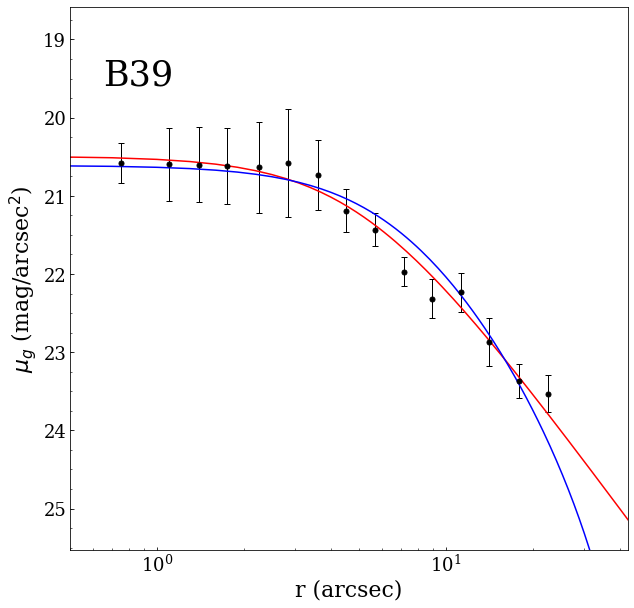}
    \includegraphics[width = 0.2\textwidth]{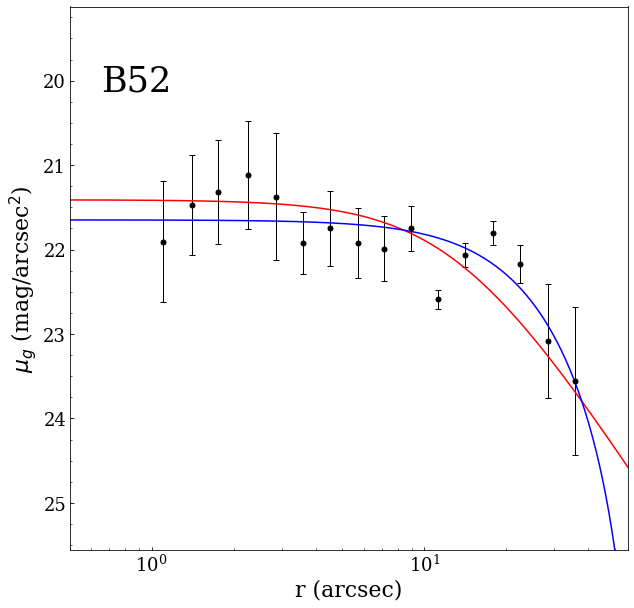}\\
    \hspace{-1.cm}
    \includegraphics[width = 0.2\textwidth]{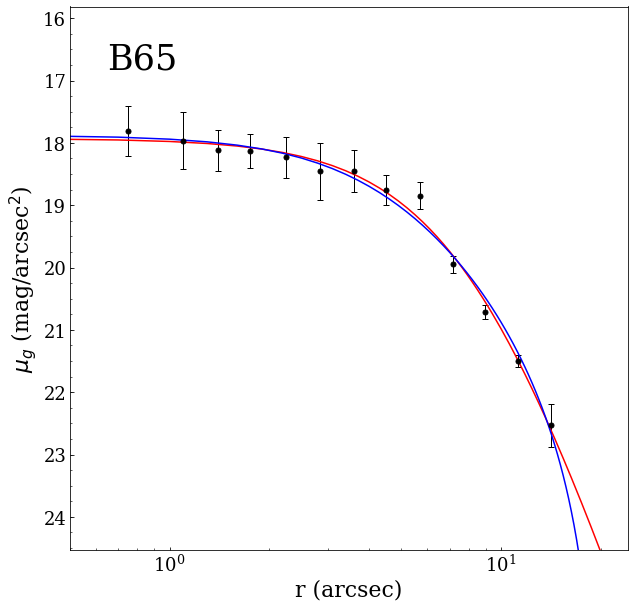}
    \includegraphics[width = 0.2\textwidth]{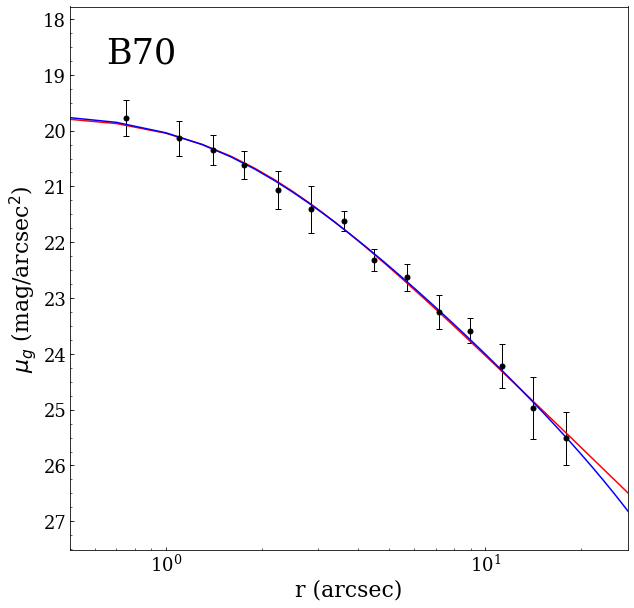}
    \includegraphics[width = 0.2\textwidth]{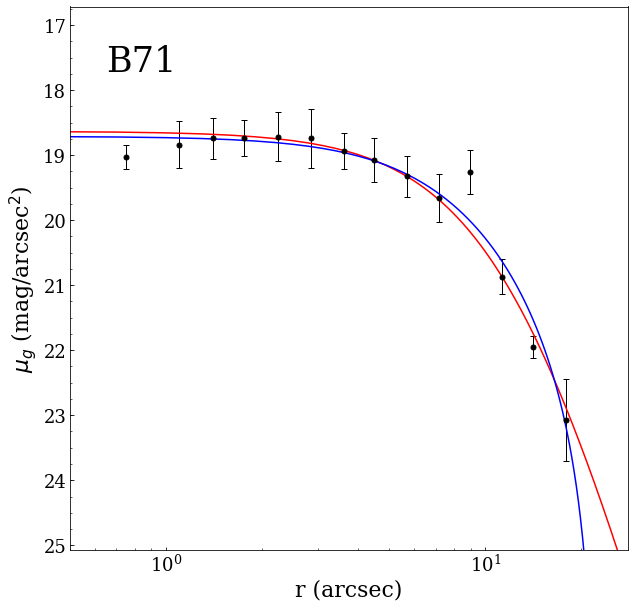}
    \includegraphics[width = 0.2\textwidth]{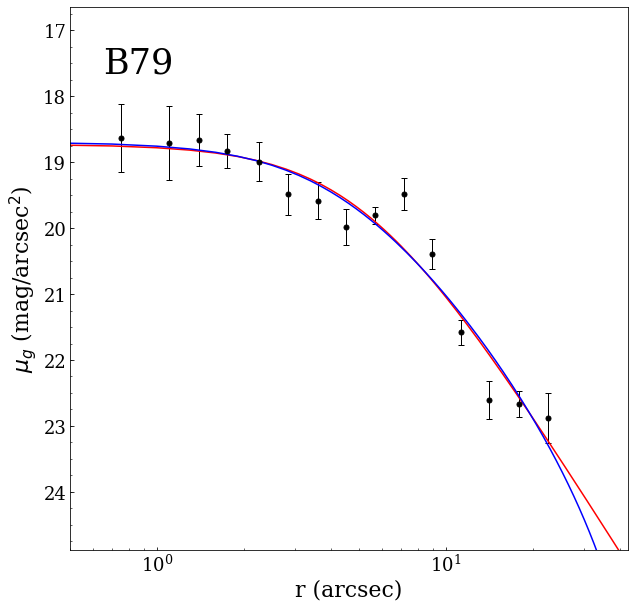}
    \includegraphics[width = 0.2\textwidth]{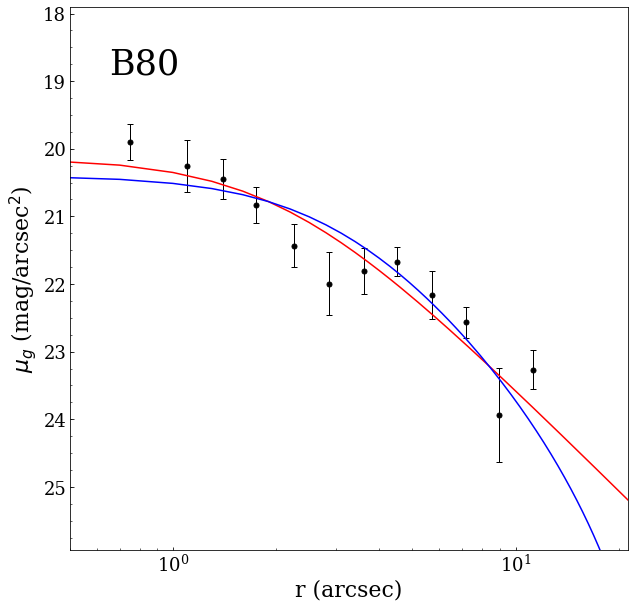}\\
    \hspace{-1.cm}
    \includegraphics[width = 0.2\textwidth]{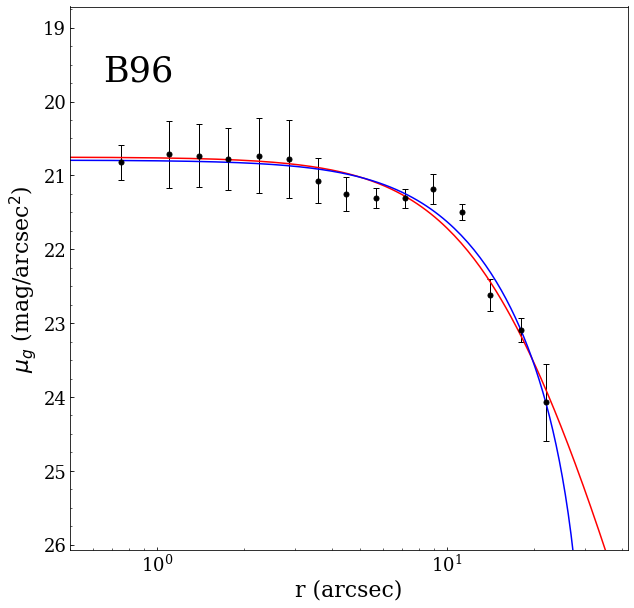}
    \includegraphics[width = 0.2\textwidth]{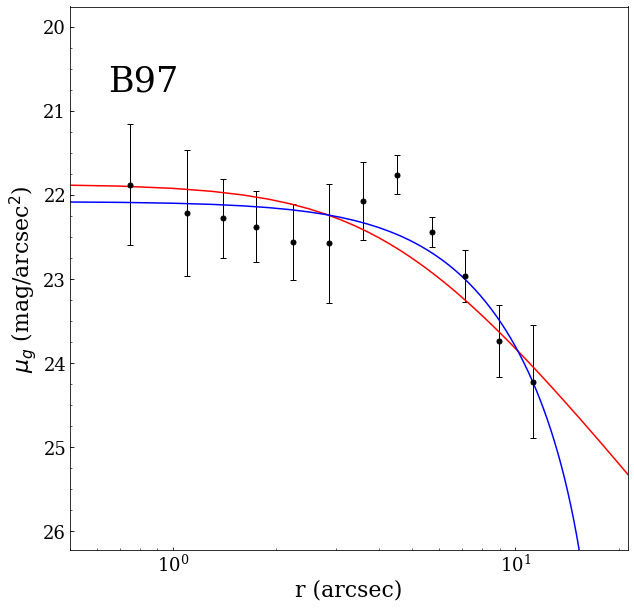}
    \includegraphics[width = 0.2\textwidth]{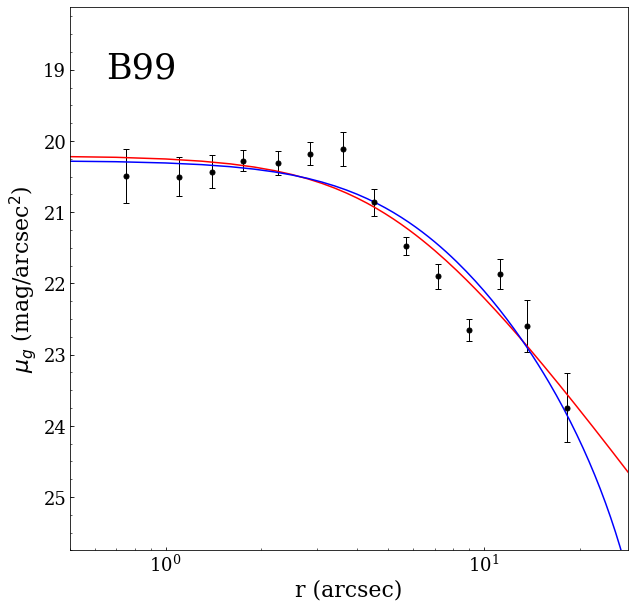}
    \includegraphics[width = 0.2\textwidth]{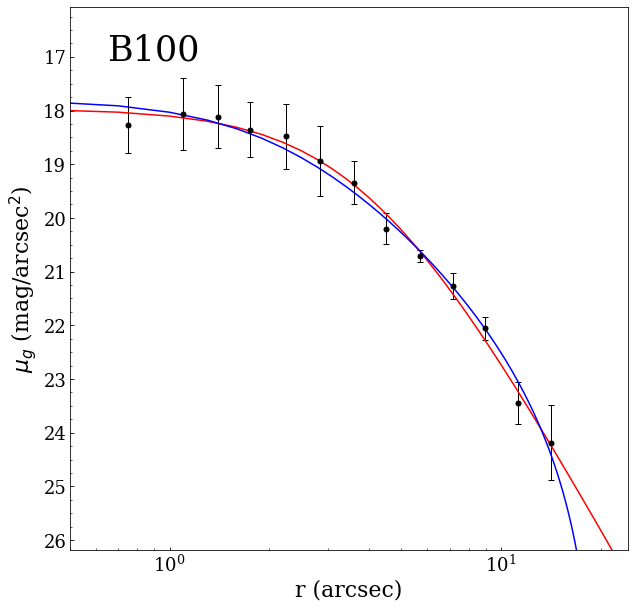}
    \includegraphics[width = 0.2\textwidth]{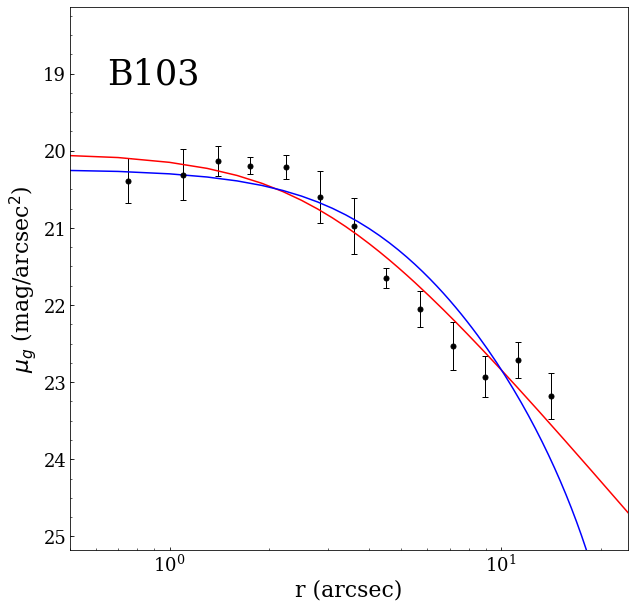}\\
    \hspace{-1.cm}
    \includegraphics[width = 0.2\textwidth]{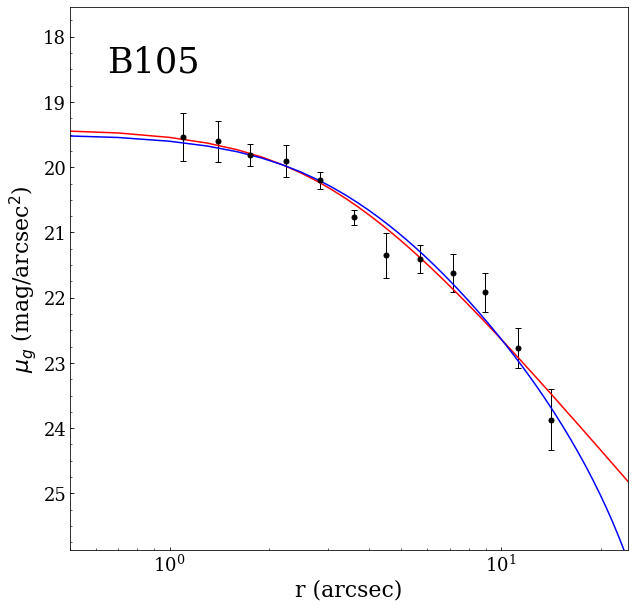}
    \includegraphics[width = 0.2\textwidth]{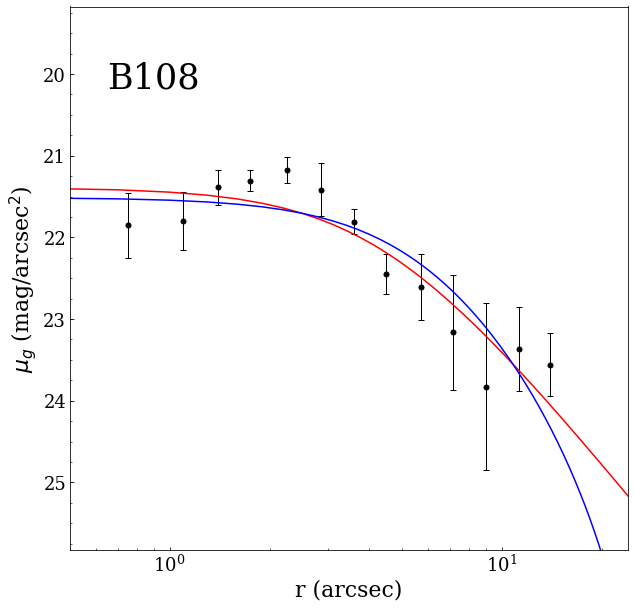}
    \includegraphics[width = 0.2\textwidth]{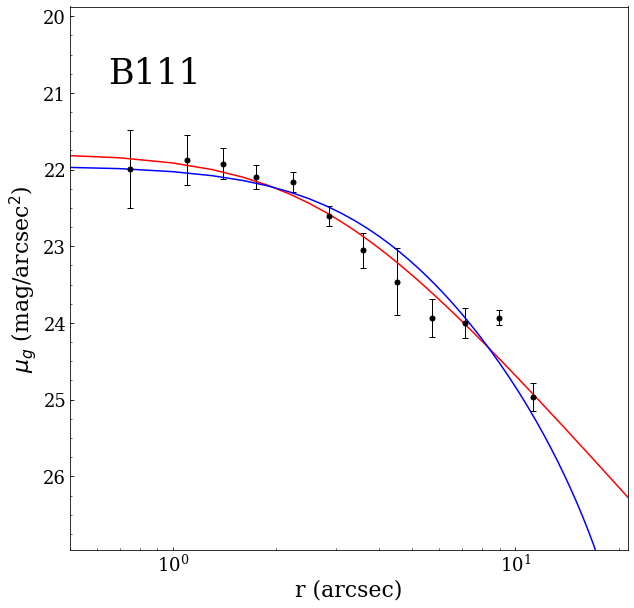}
    \includegraphics[width = 0.2\textwidth]{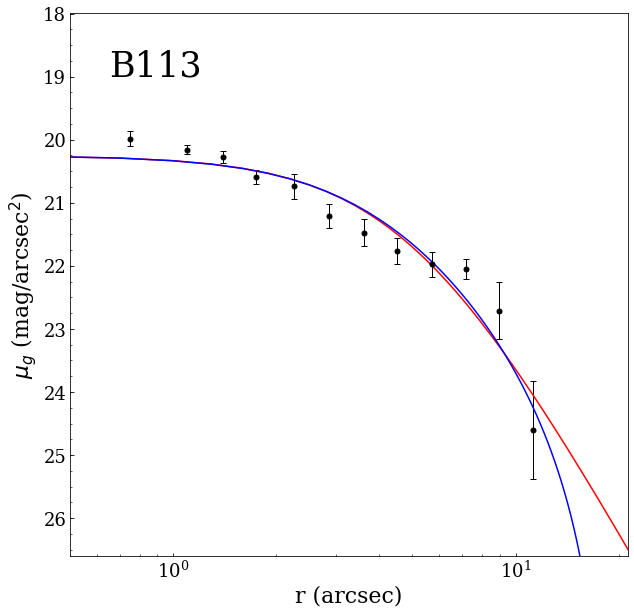}
    \includegraphics[width = 0.2\textwidth]{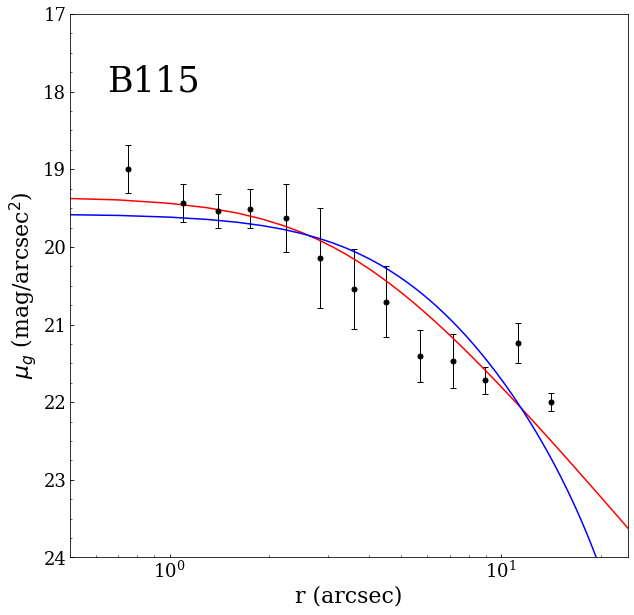}\\
    \hspace{-1.cm}
    \includegraphics[width = 0.2\textwidth]{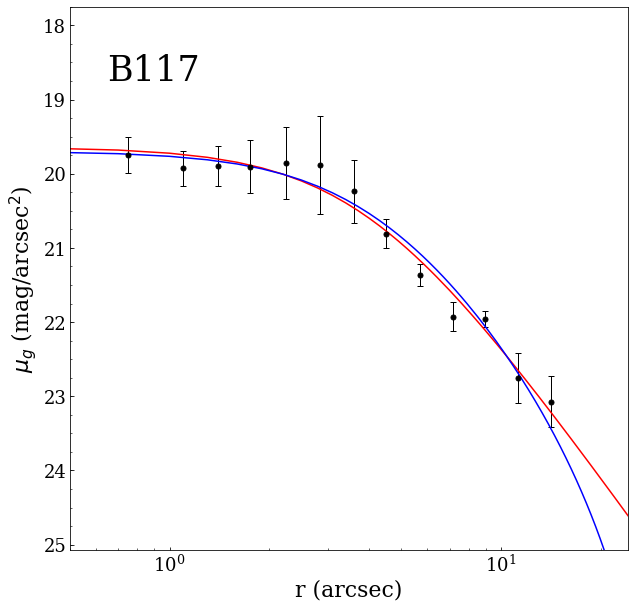}
    \includegraphics[width = 0.2\textwidth]{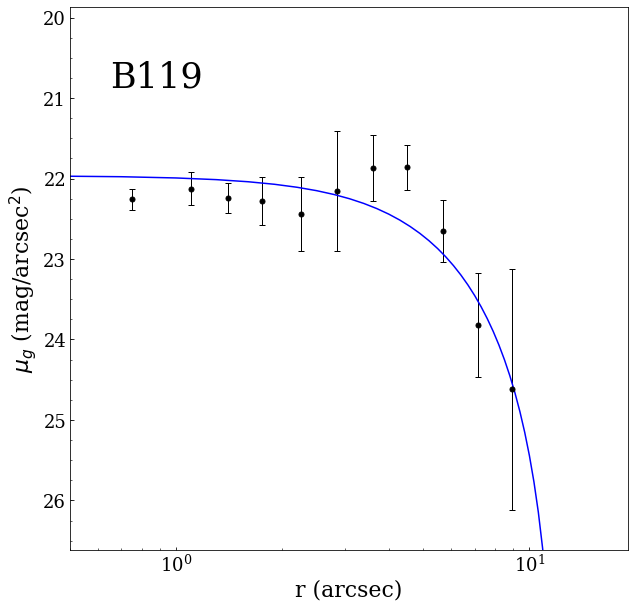}
    \includegraphics[width = 0.2\textwidth]{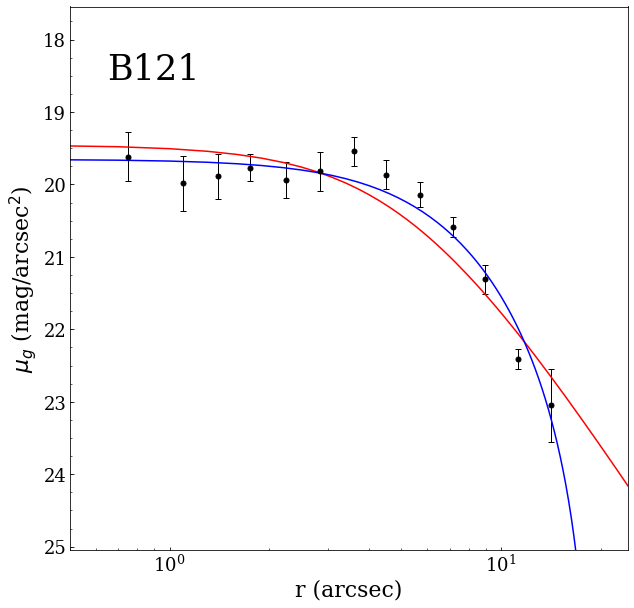}
    \includegraphics[width = 0.2\textwidth]{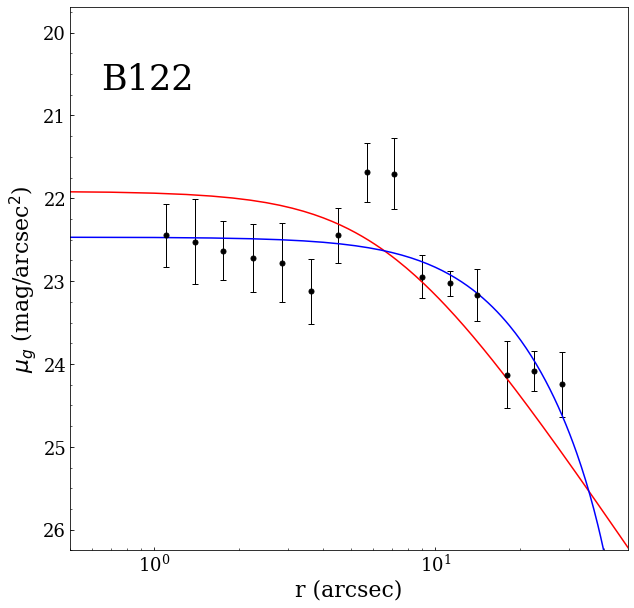}
    \includegraphics[width = 0.2\textwidth]{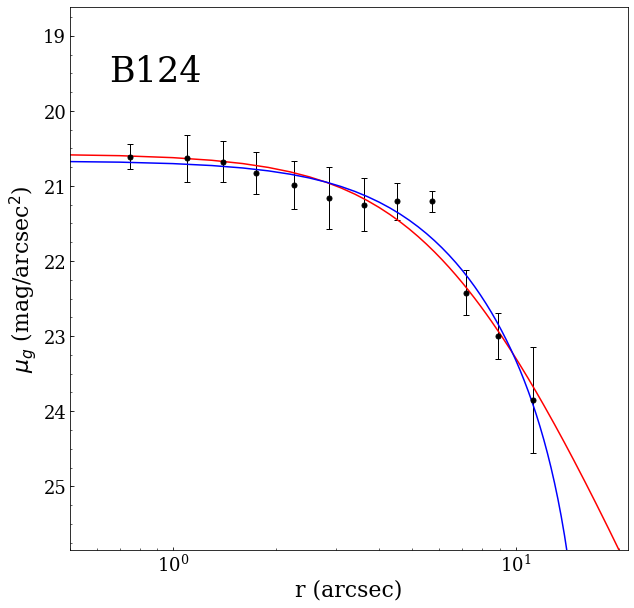}\\
    \caption{SBPs derived as described in Sec.\ref{sec:method}. The red and blue lines represent the best fit obtained through the EFF and King's profiles, respectively.}
    \label{fig:best_fit_both_profiles}
\end{figure*}

\begin{figure*}
    \centering
    \hspace{-1.cm}
    \includegraphics[width = 0.2\textwidth]{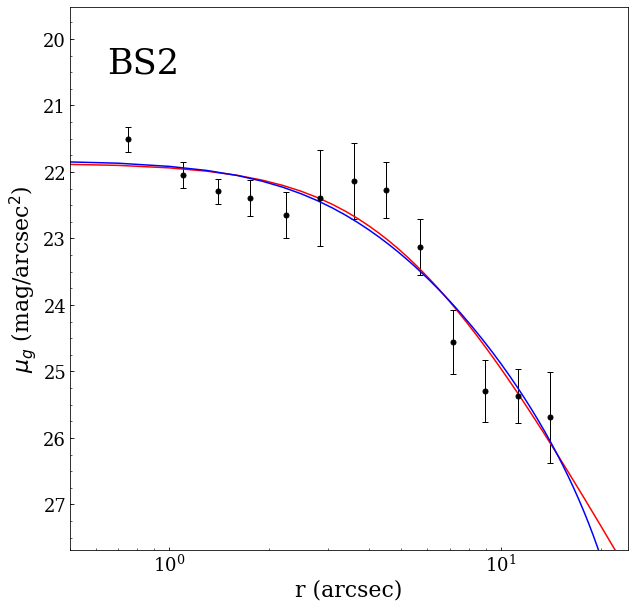}
    \includegraphics[width = 0.2\textwidth]{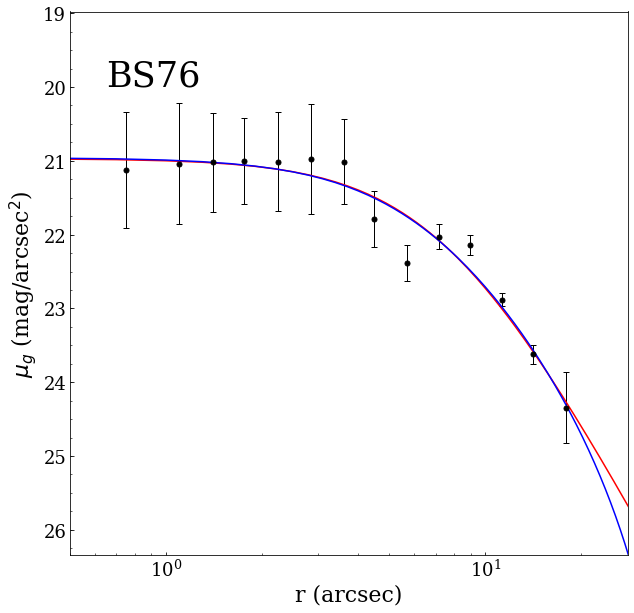}
    \includegraphics[width = 0.2\textwidth]{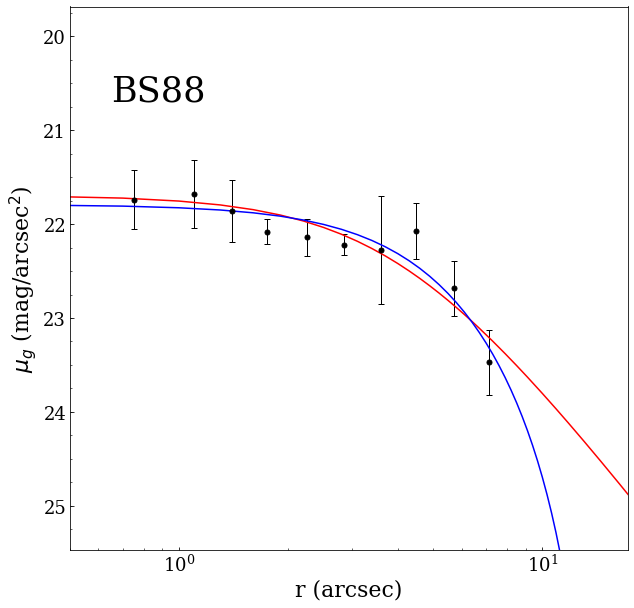}
    \includegraphics[width = 0.2\textwidth]{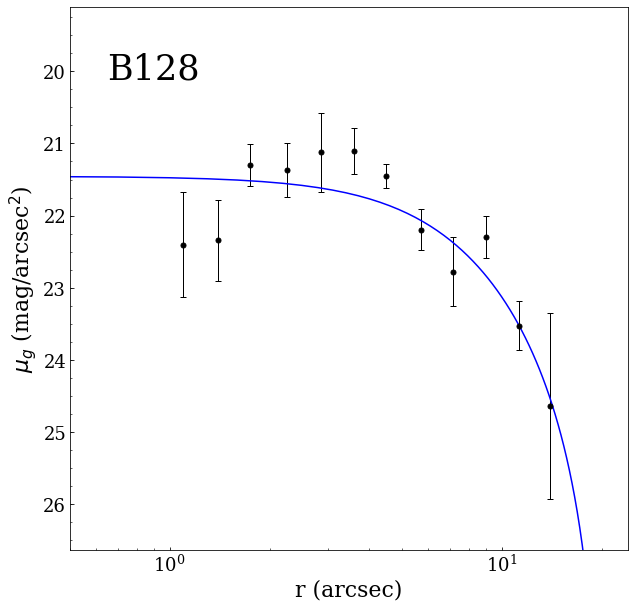}
    \includegraphics[width = 0.2\textwidth]{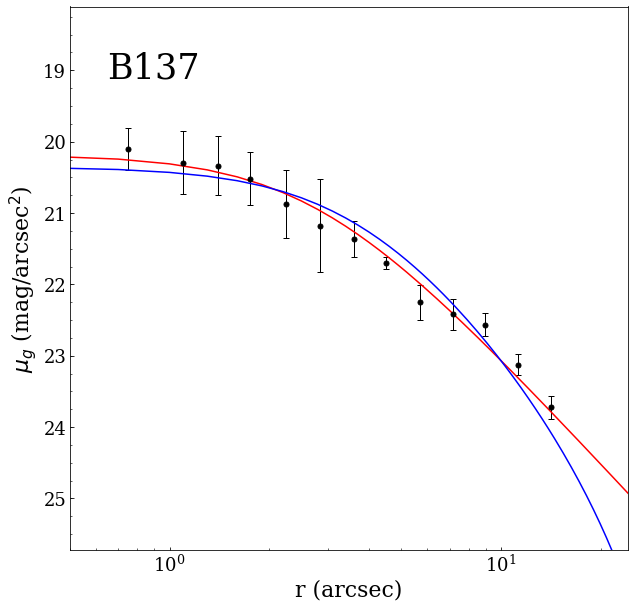}\\
    \hspace{-1.cm}
    \includegraphics[width = 0.2\textwidth]{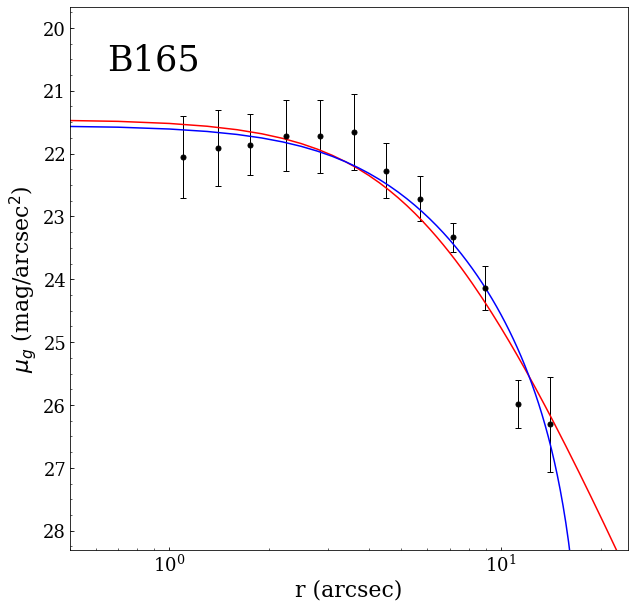}
    \includegraphics[width = 0.2\textwidth]{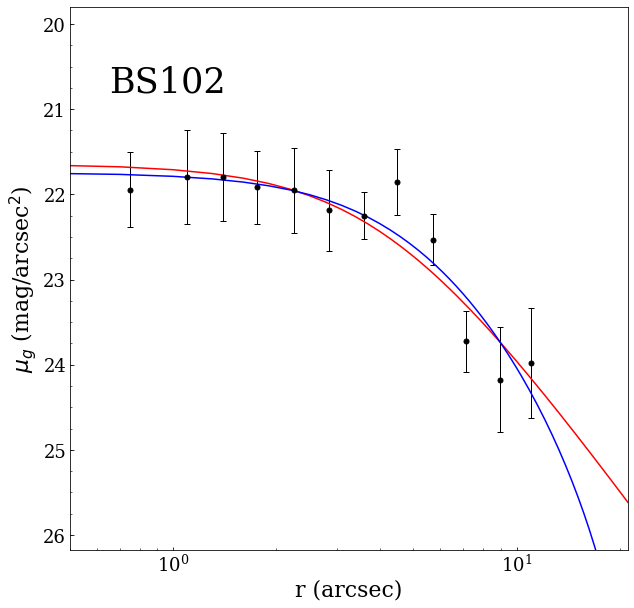}
    \includegraphics[width = 0.2\textwidth]{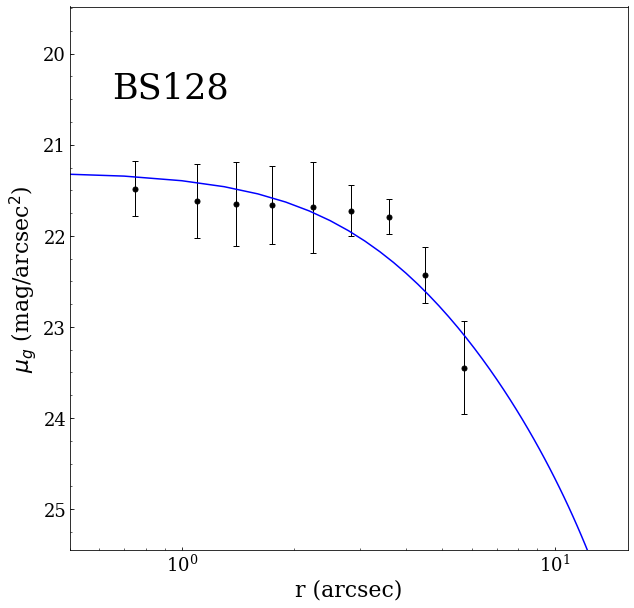}
    \includegraphics[width = 0.2\textwidth]{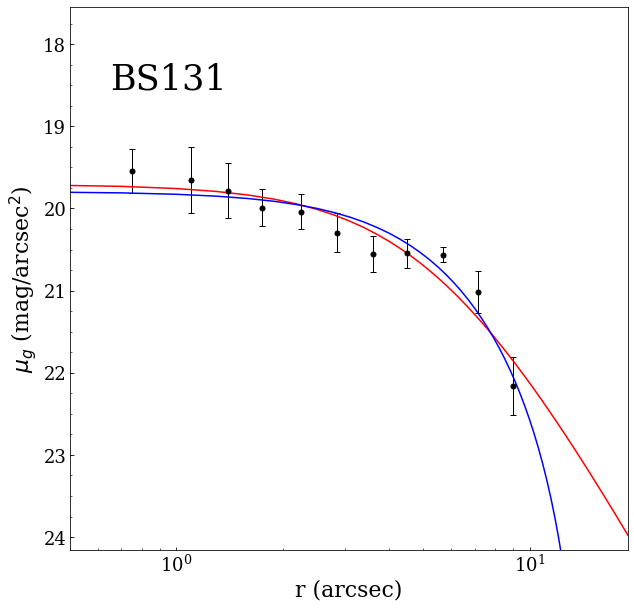}
    \includegraphics[width = 0.2\textwidth]{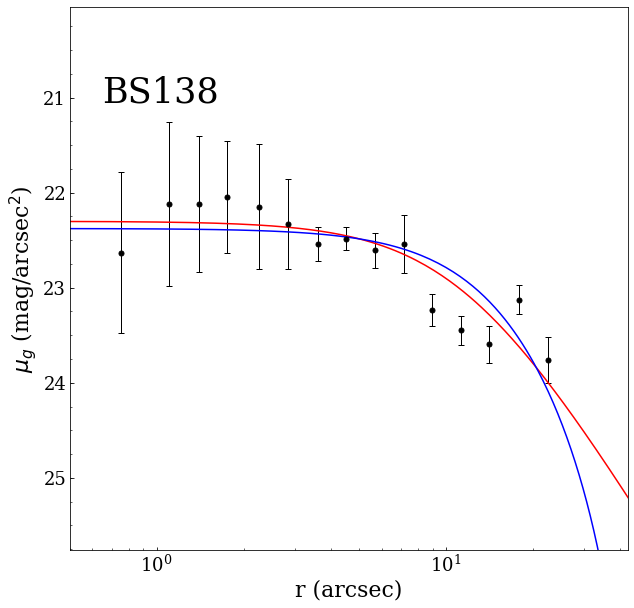}\\
    \hspace{-1.cm}
    \includegraphics[width = 0.2\textwidth]{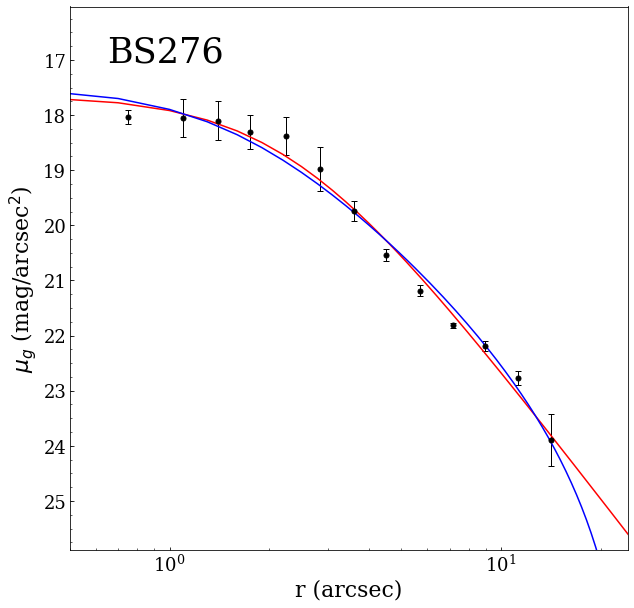}
    \includegraphics[width = 0.2\textwidth]{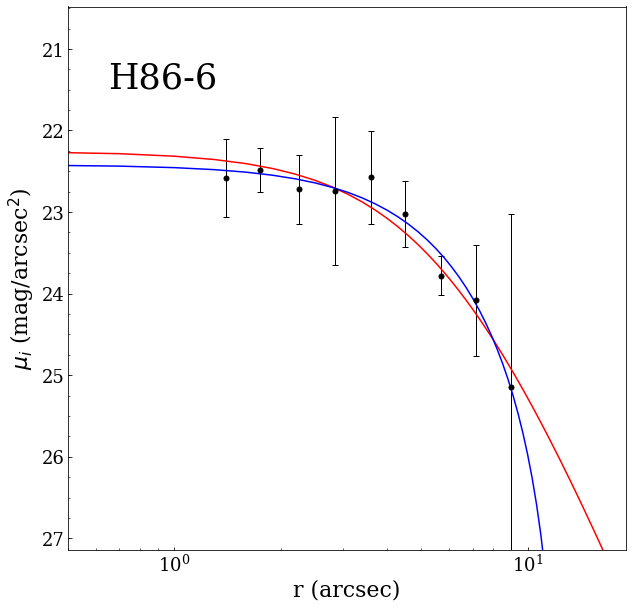}
    \includegraphics[width = 0.2\textwidth]{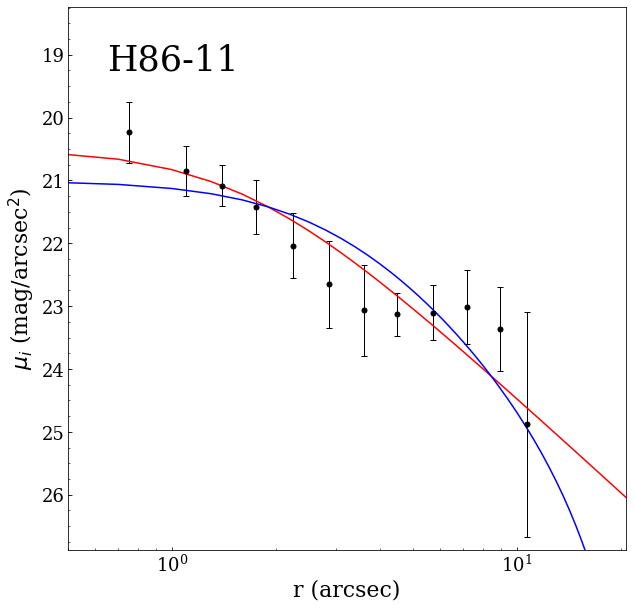}
    \includegraphics[width = 0.2\textwidth]{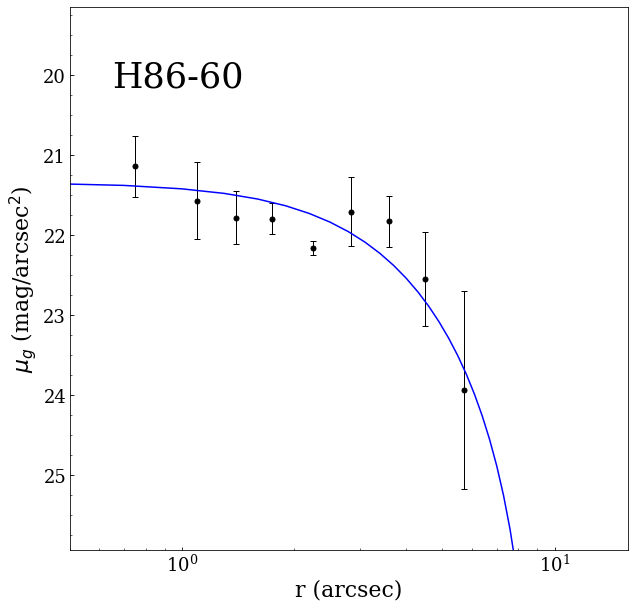}
    \includegraphics[width = 0.2\textwidth]{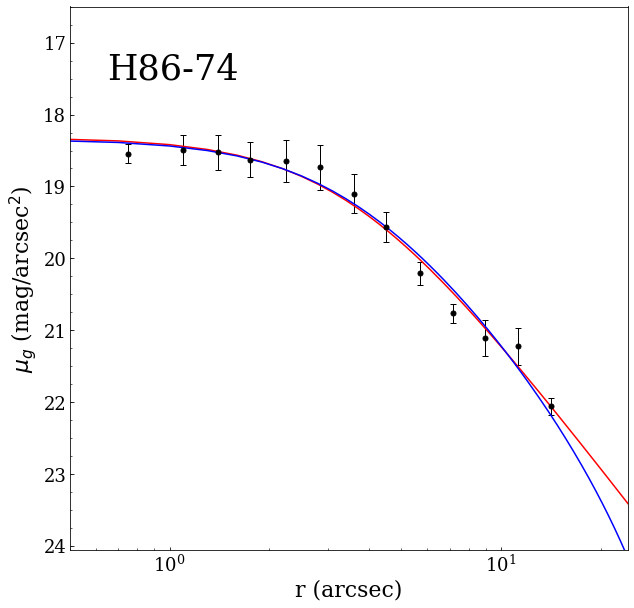}\\
    \hspace{-1.cm}
    \includegraphics[width = 0.2\textwidth]{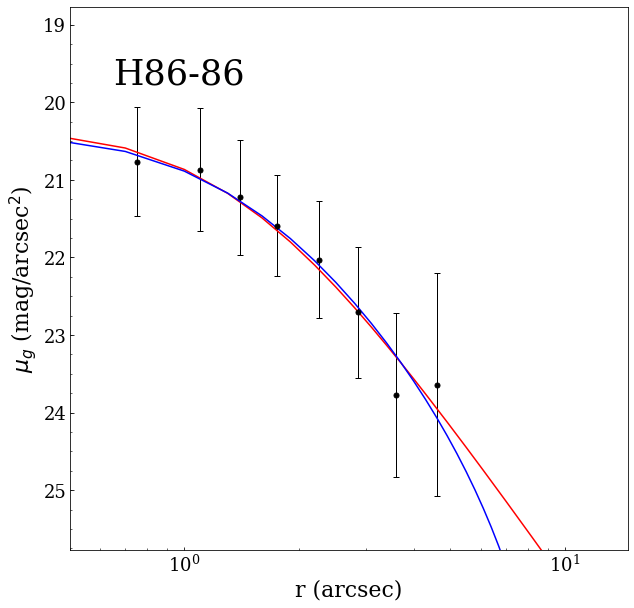}
    \includegraphics[width = 0.2\textwidth]{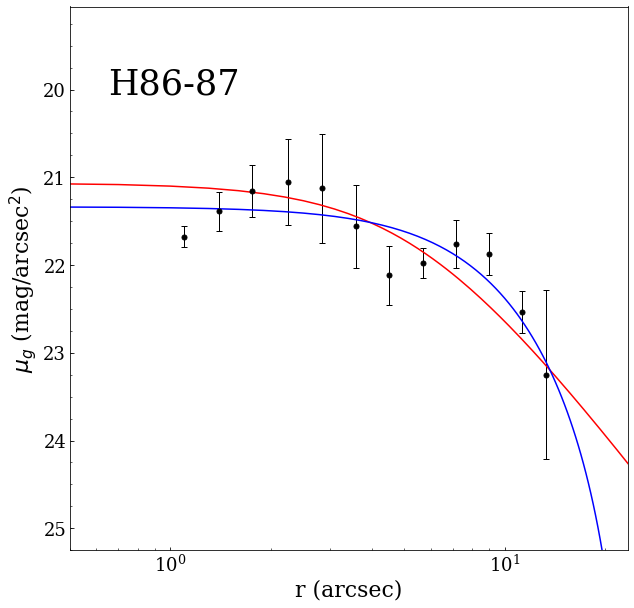}
    \includegraphics[width = 0.2\textwidth]{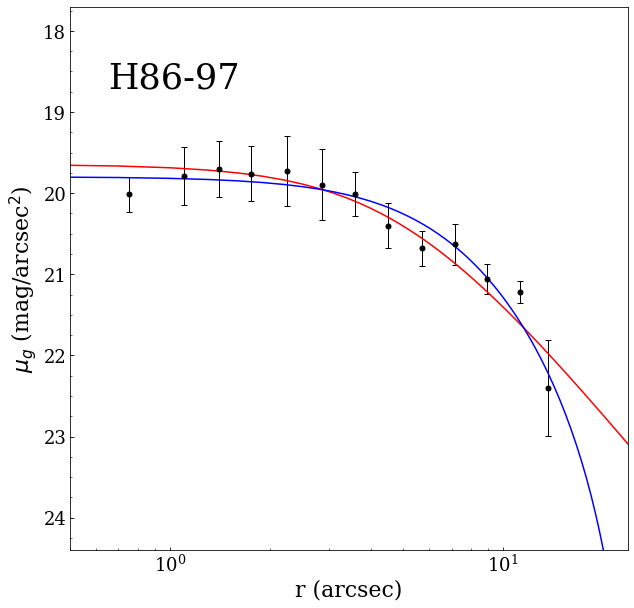}
    \includegraphics[width = 0.2\textwidth]{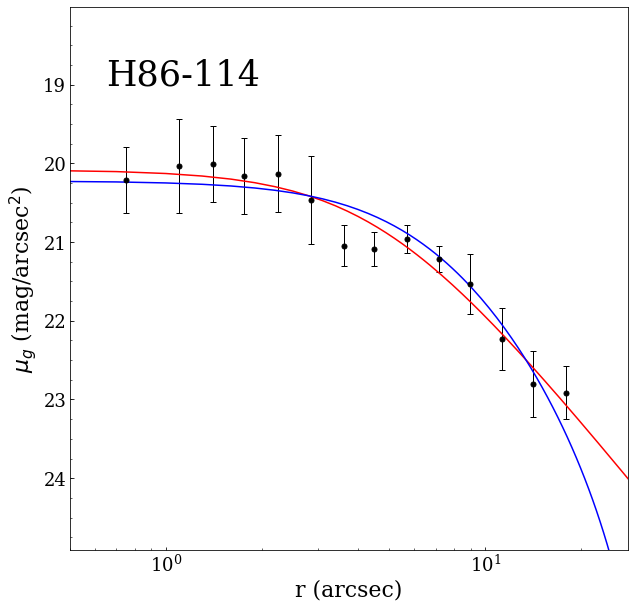}
    \includegraphics[width = 0.2\textwidth]{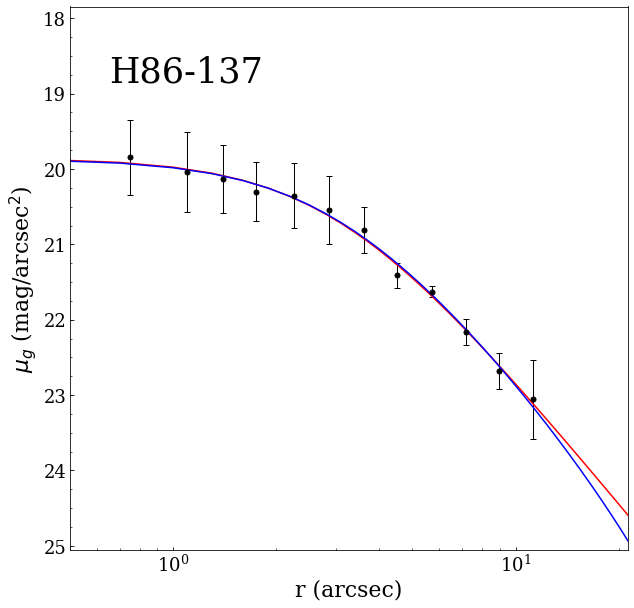}\\
    \hspace{-1.cm}
    \includegraphics[width = 0.2\textwidth]{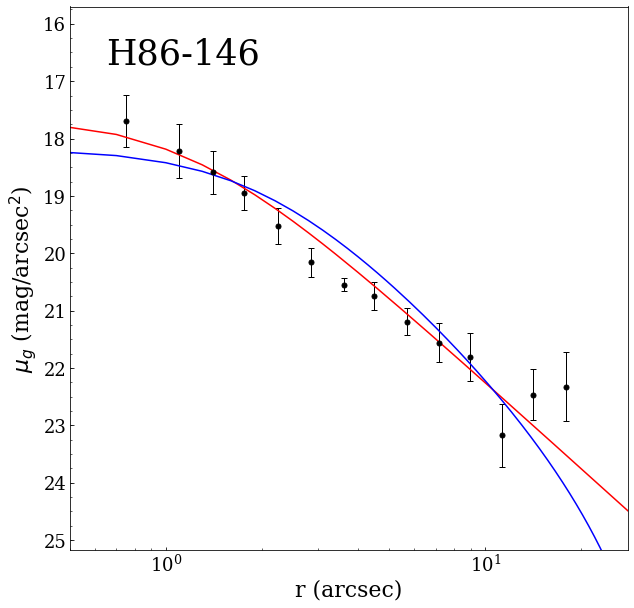}
    \includegraphics[width = 0.2\textwidth]{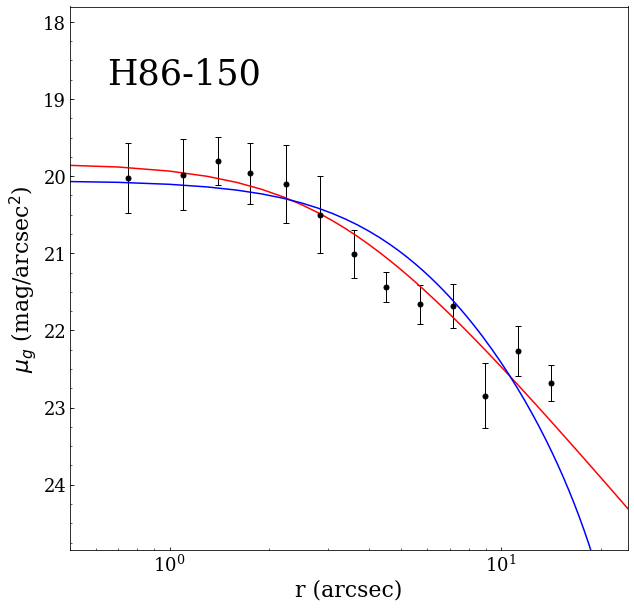}
    \includegraphics[width = 0.2\textwidth]{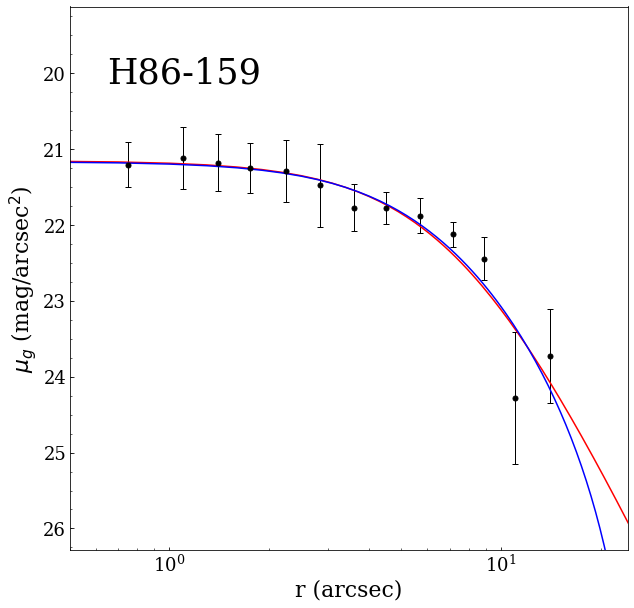}
    \includegraphics[width = 0.2\textwidth]{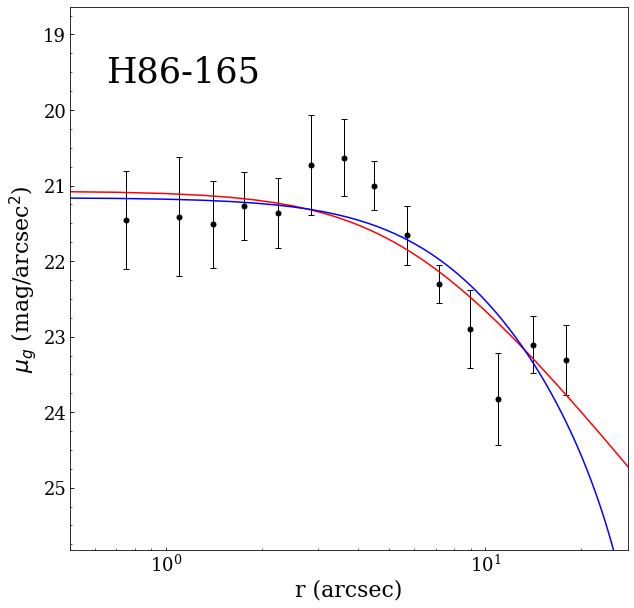}
    \includegraphics[width = 0.2\textwidth]{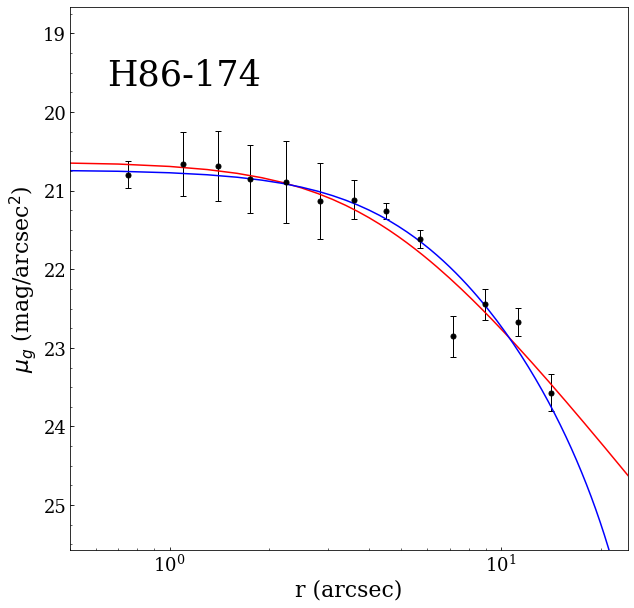}\\
    \hspace{-1.cm}
    \includegraphics[width = 0.2\textwidth]{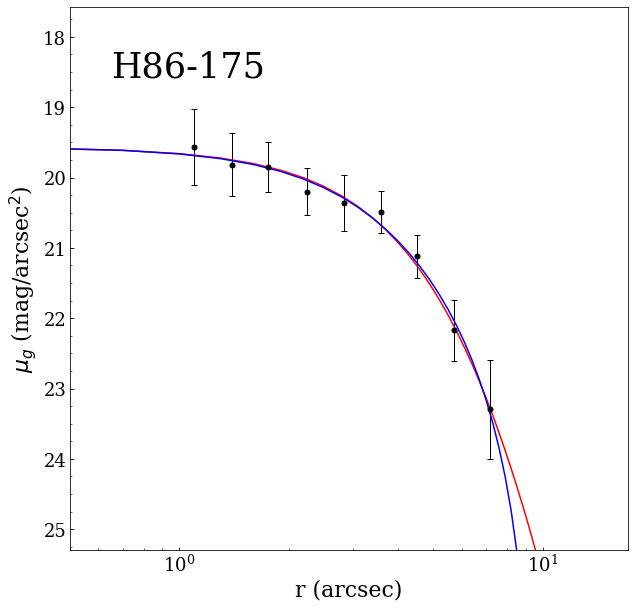}
    \includegraphics[width = 0.2\textwidth]{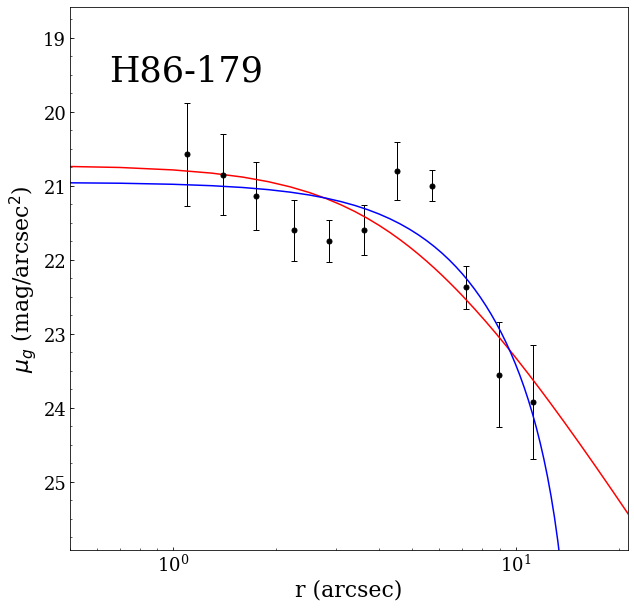}
    \includegraphics[width = 0.2\textwidth]{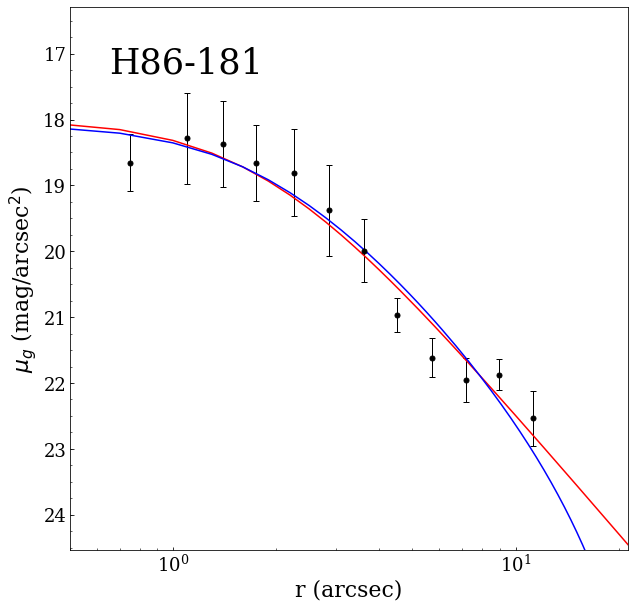}
    \includegraphics[width = 0.2\textwidth]{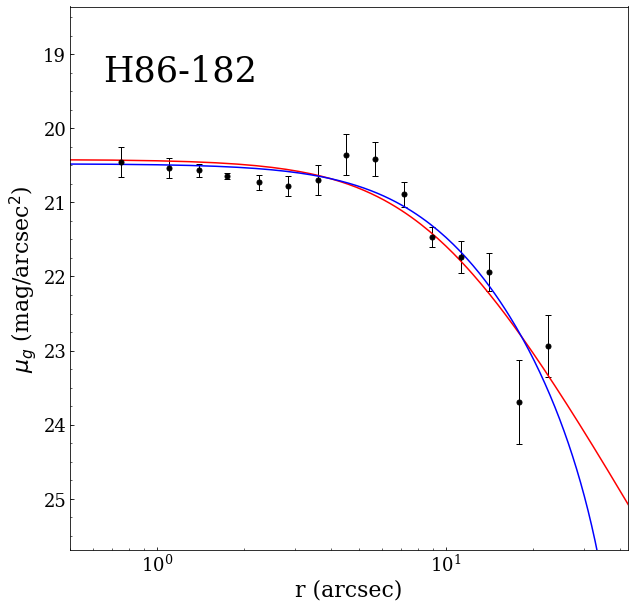}
    \includegraphics[width = 0.2\textwidth]{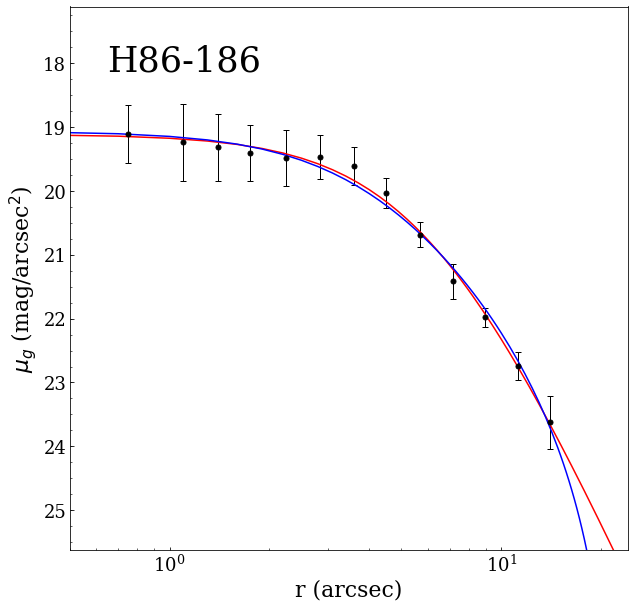}\\
    \contcaption{}
\end{figure*}

\begin{figure*}
    \centering
    \hspace{-1.cm}
    \includegraphics[width = 0.2\textwidth]{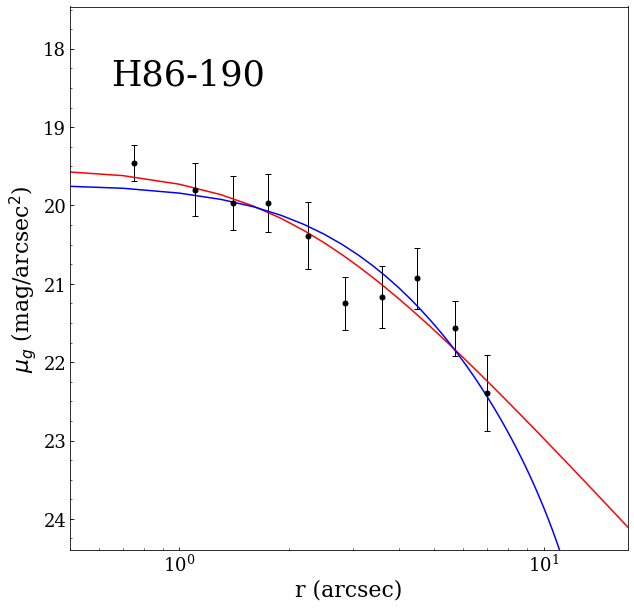}
    \includegraphics[width = 0.2\textwidth]{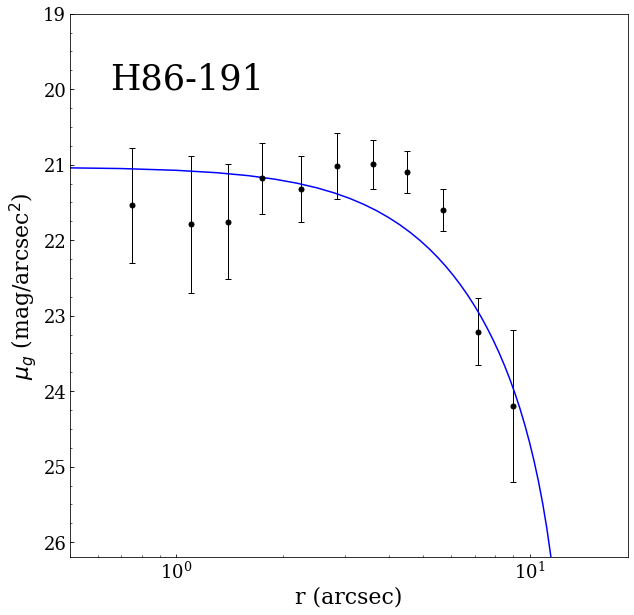}
    \includegraphics[width = 0.2\textwidth]{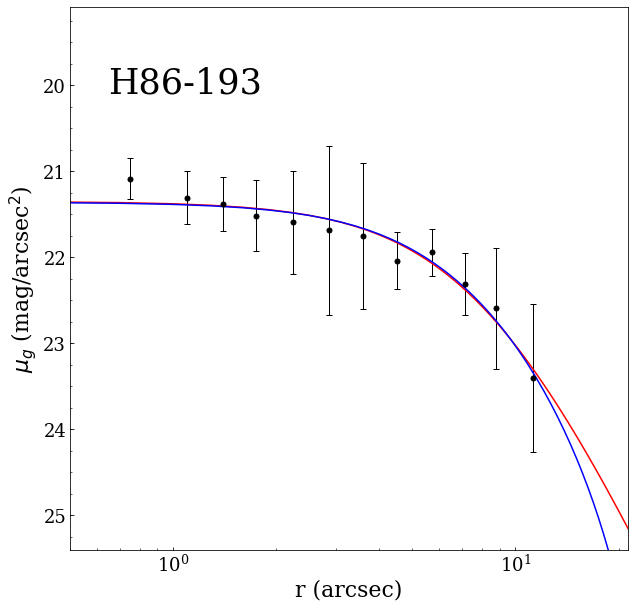}
    \includegraphics[width = 0.2\textwidth]{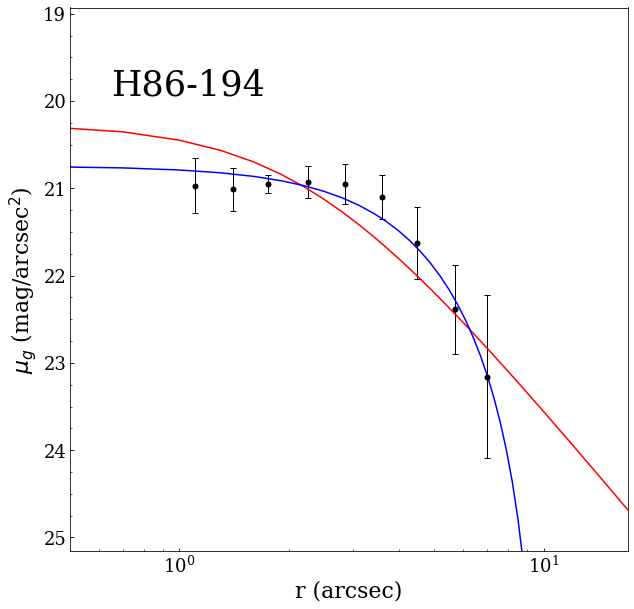}
    \includegraphics[width = 0.2\textwidth]{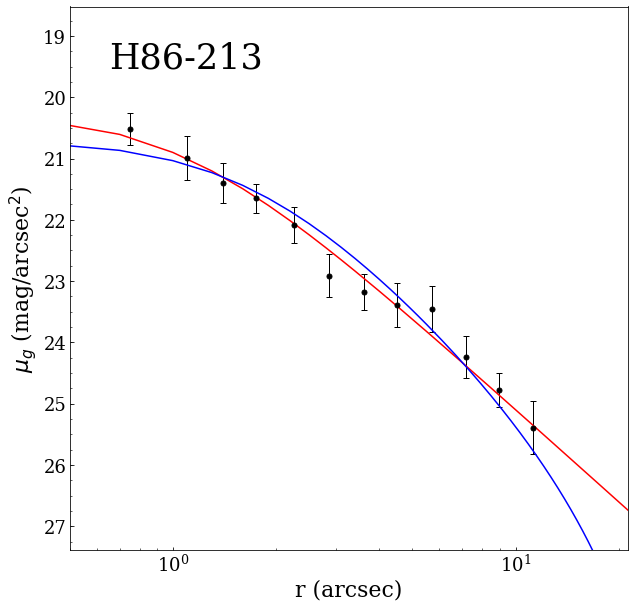}\\
    \hspace{-1.cm}
    \includegraphics[width = 0.2\textwidth]{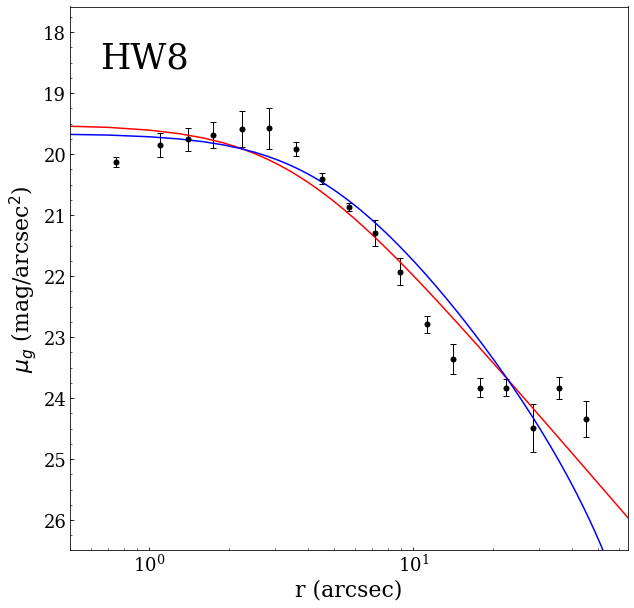}
    \includegraphics[width = 0.2\textwidth]{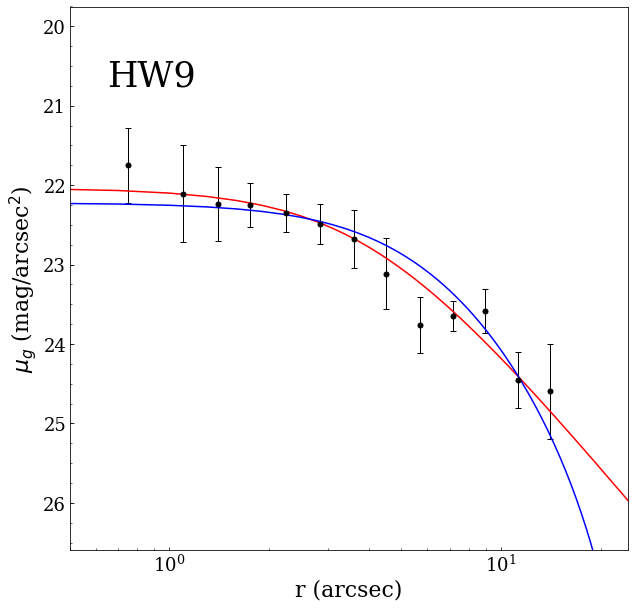}
    \includegraphics[width = 0.2\textwidth]{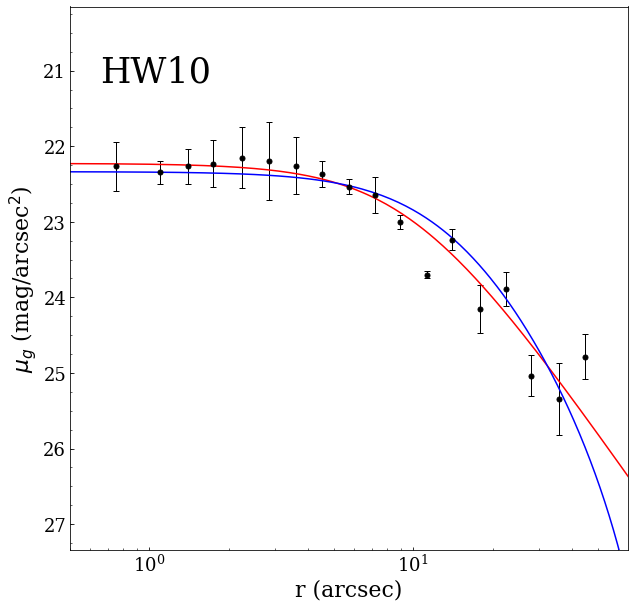}
    \includegraphics[width = 0.2\textwidth]{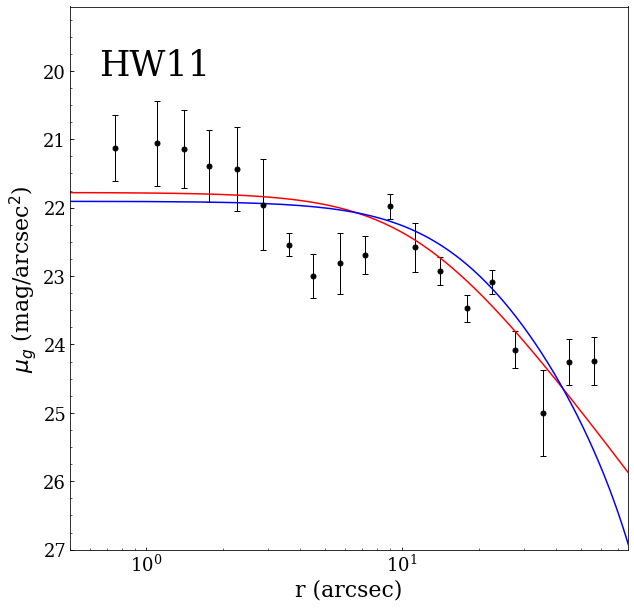}
    \includegraphics[width = 0.2\textwidth]{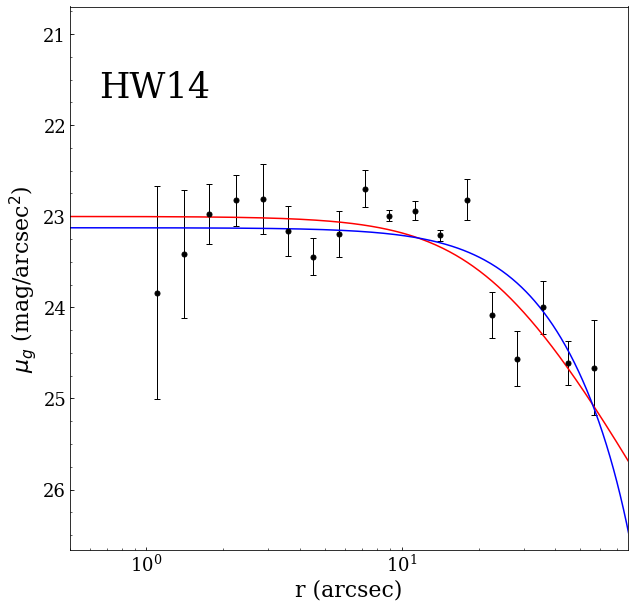}\\
    \hspace{-1.cm}
    \includegraphics[width = 0.2\textwidth]{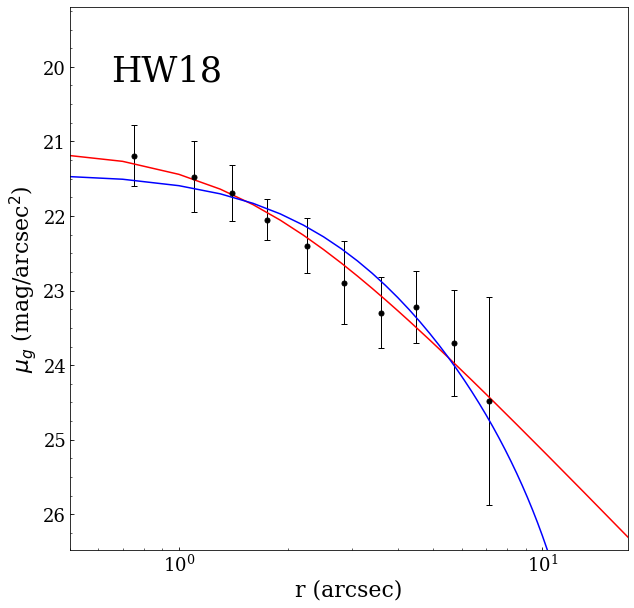}
    \includegraphics[width = 0.2\textwidth]{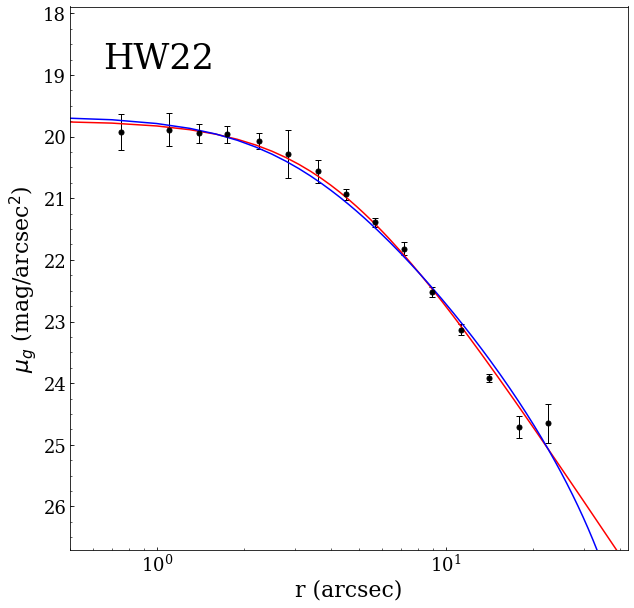}
    \includegraphics[width = 0.2\textwidth]{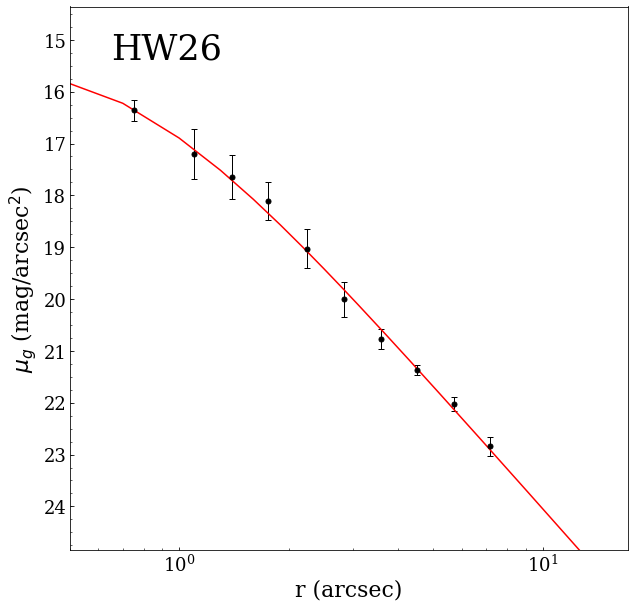}
    \includegraphics[width = 0.2\textwidth]{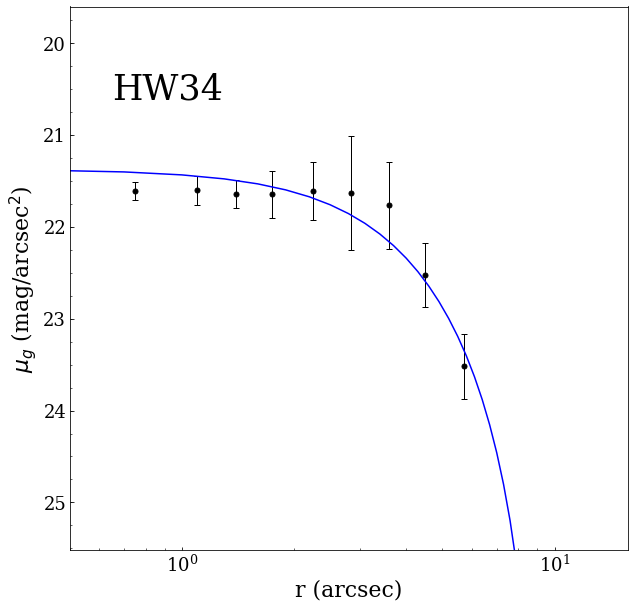}
    \includegraphics[width = 0.2\textwidth]{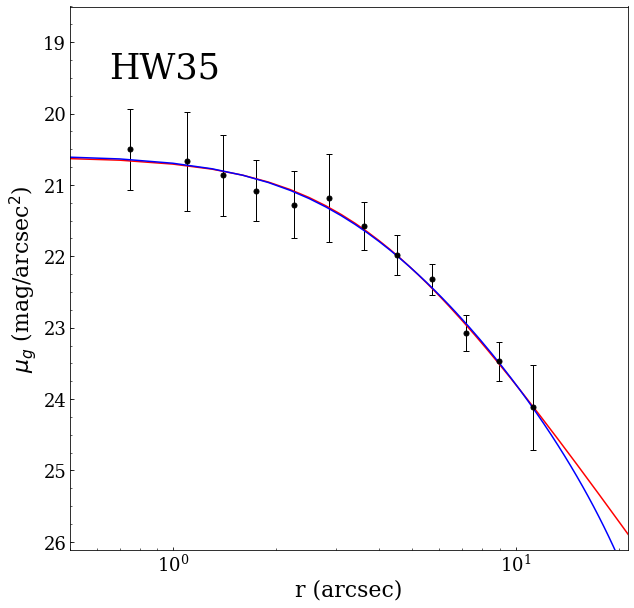}\\
    \hspace{-1.cm}
    \includegraphics[width = 0.2\textwidth]{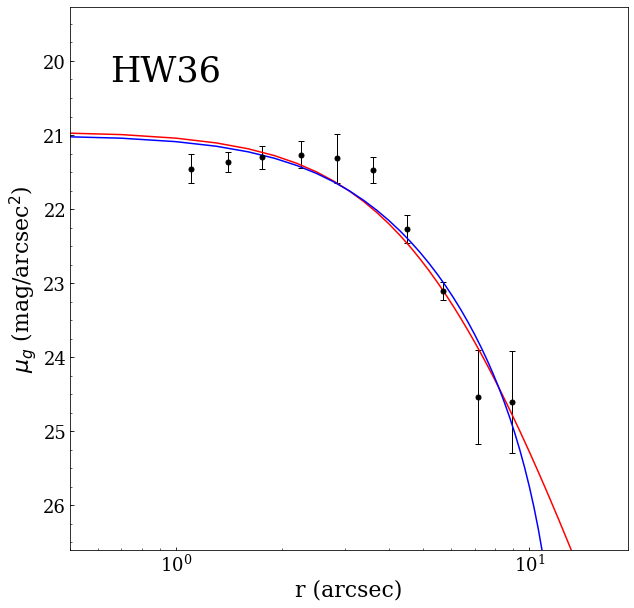}
    \includegraphics[width = 0.2\textwidth]{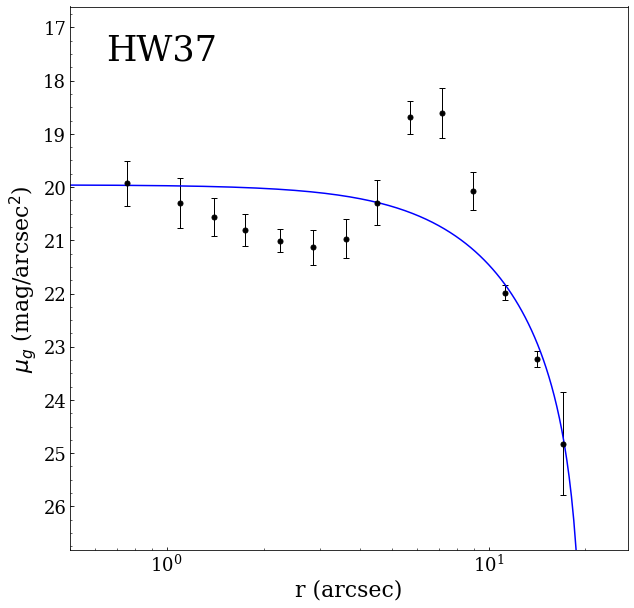}
    \includegraphics[width = 0.2\textwidth]{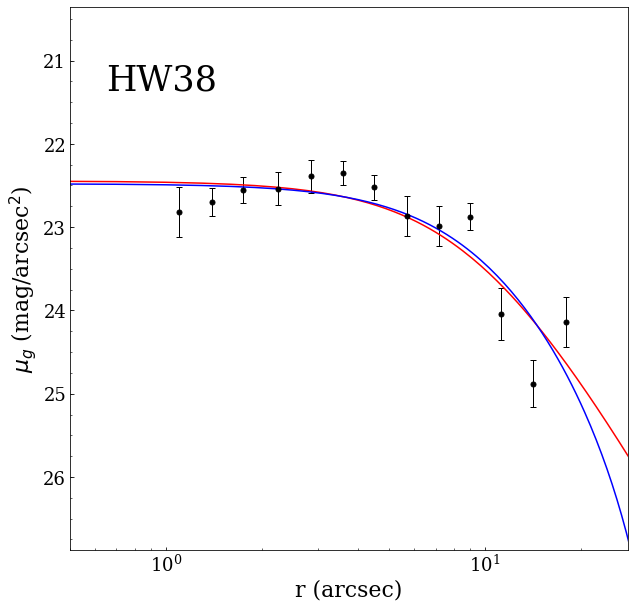}
    \includegraphics[width = 0.2\textwidth]{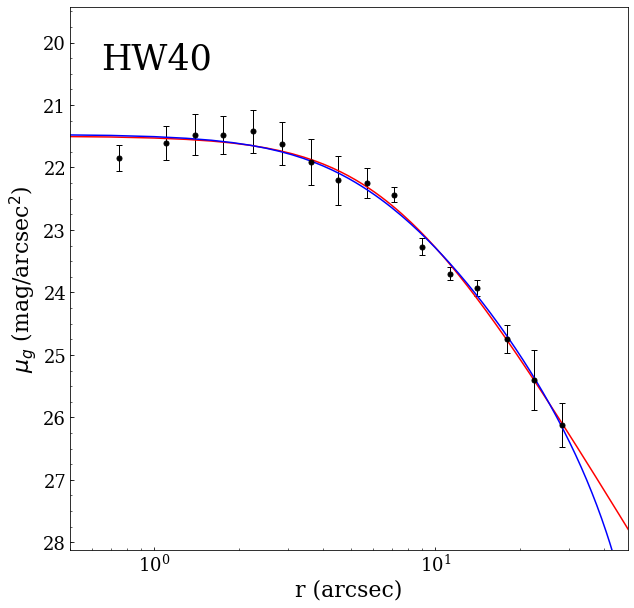}
    \includegraphics[width = 0.2\textwidth]{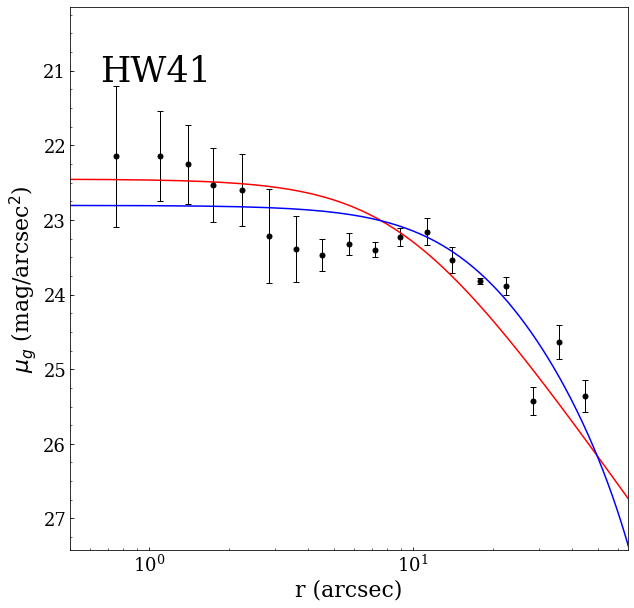}\\
    \hspace{-1.cm}
    \includegraphics[width = 0.2\textwidth]{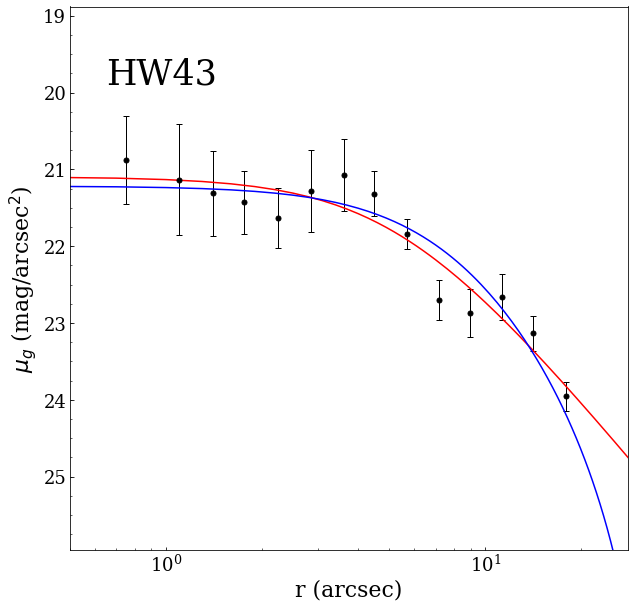}
    \includegraphics[width = 0.2\textwidth]{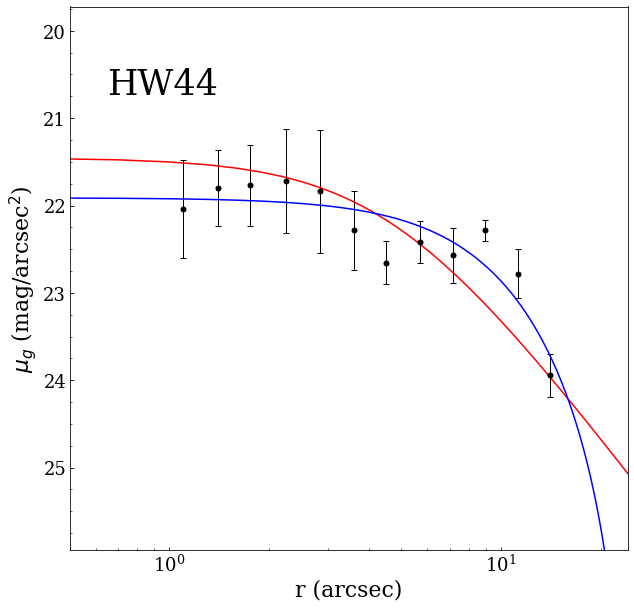}
    \includegraphics[width = 0.2\textwidth]{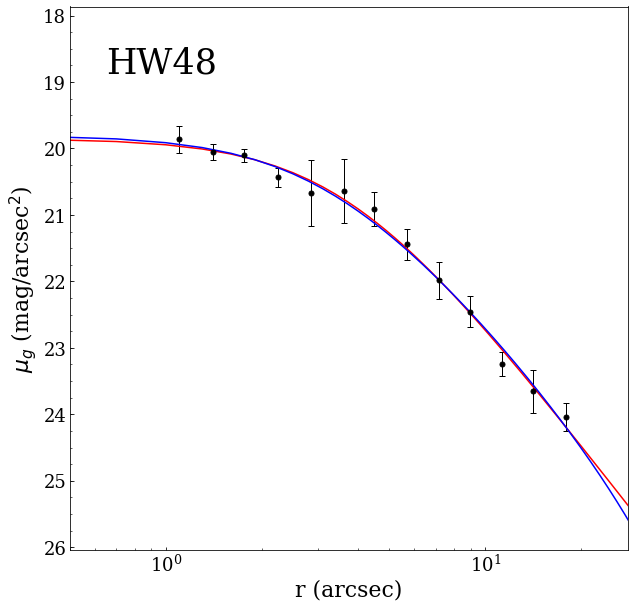}
    \includegraphics[width = 0.2\textwidth]{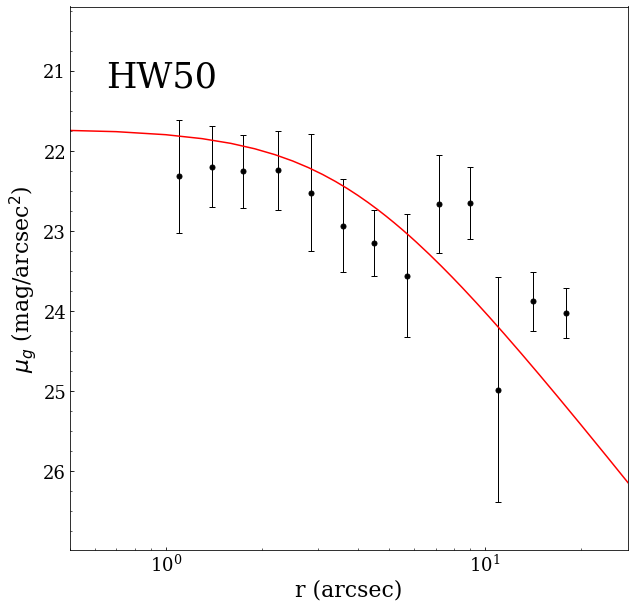}
    \includegraphics[width = 0.2\textwidth]{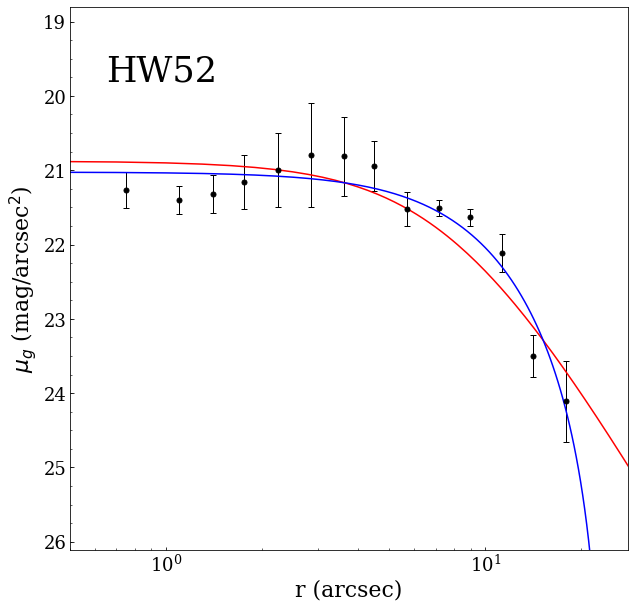}\\
    \hspace{-1.cm}
    \includegraphics[width = 0.2\textwidth]{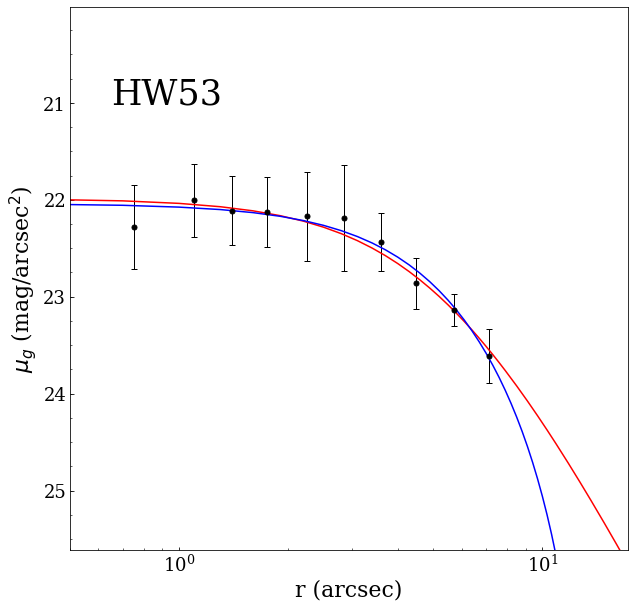}
    \includegraphics[width = 0.2\textwidth]{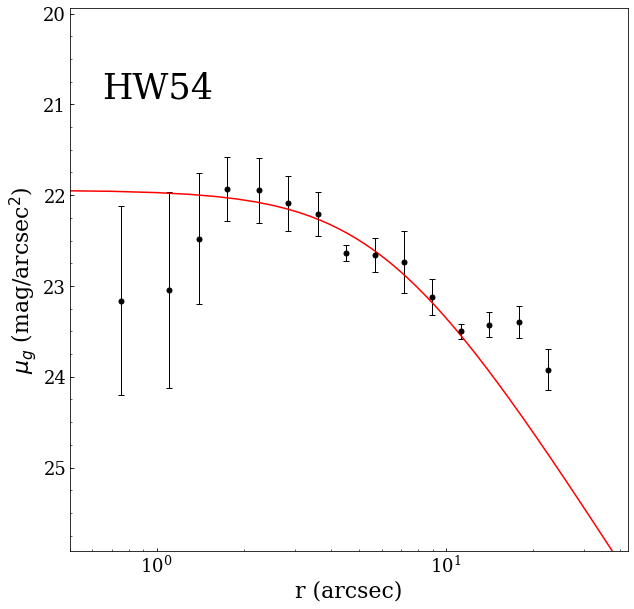}
    \includegraphics[width = 0.2\textwidth]{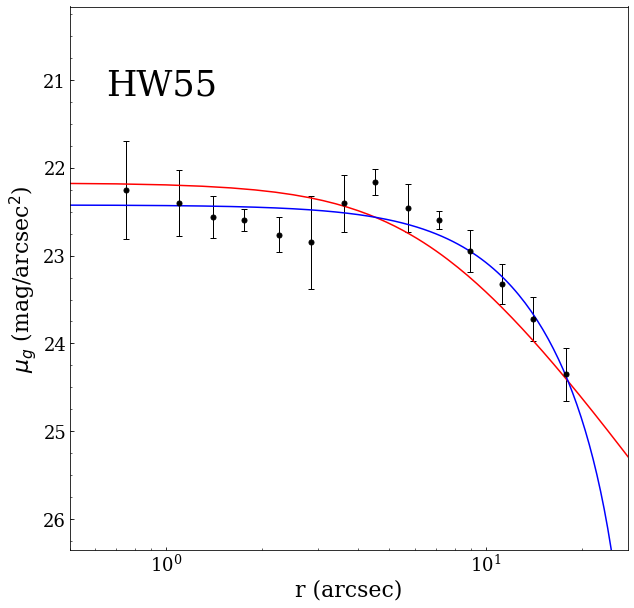}
    \includegraphics[width = 0.2\textwidth]{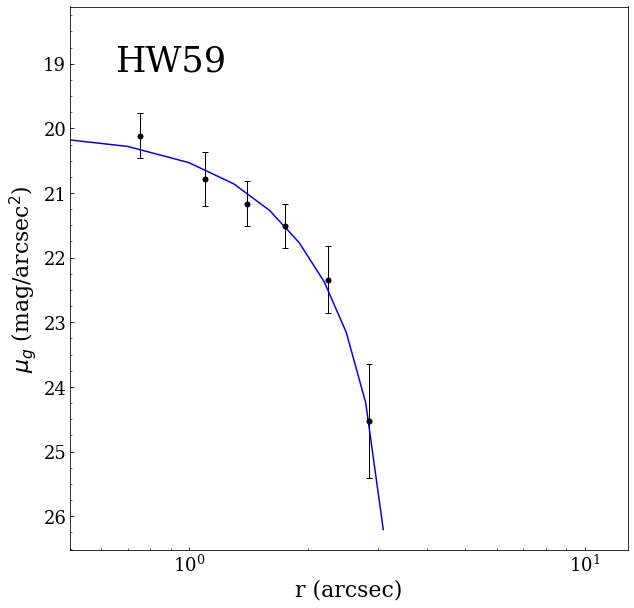}
    \includegraphics[width = 0.2\textwidth]{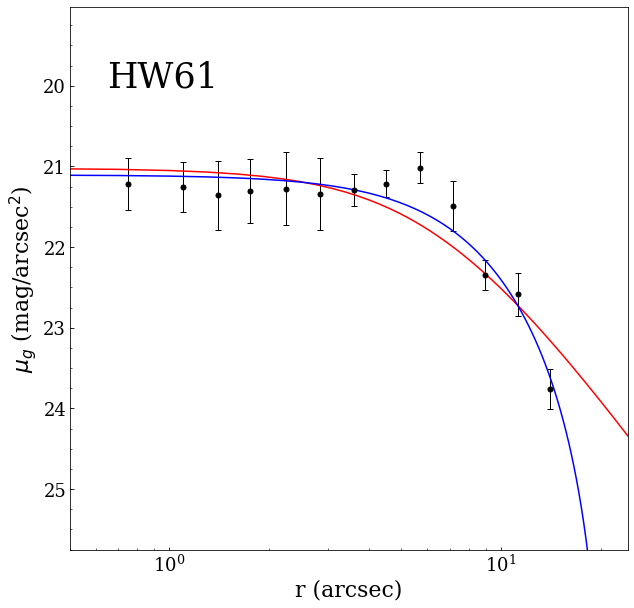}\\
    \contcaption{}
\end{figure*}

\begin{figure*}
    \centering
    \hspace{-1.cm}
    \includegraphics[width = 0.2\textwidth]{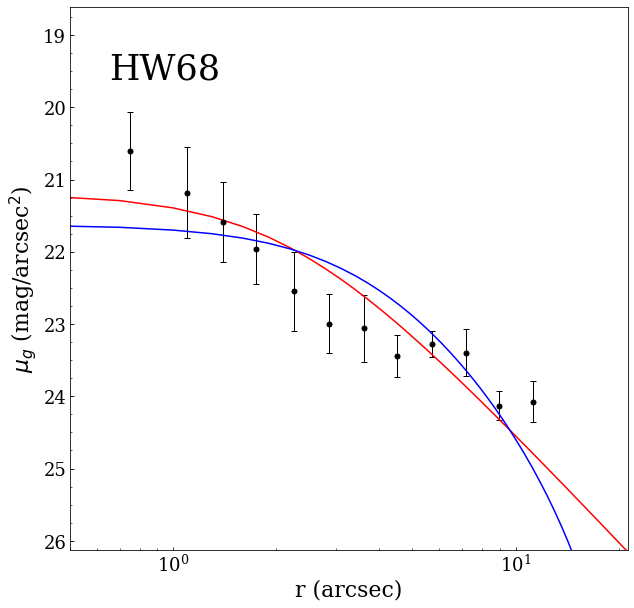}
    \includegraphics[width = 0.2\textwidth]{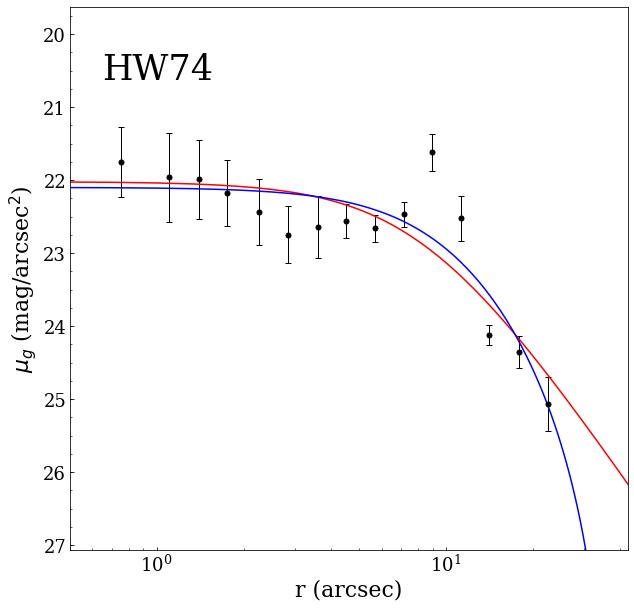}
    \includegraphics[width = 0.2\textwidth]{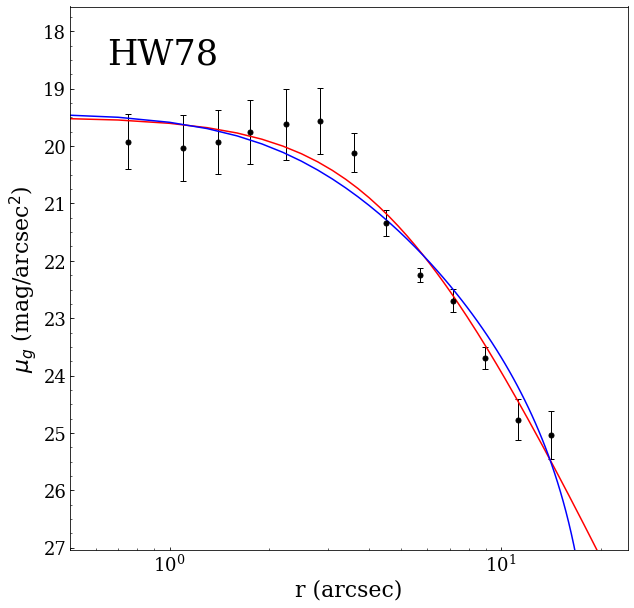}
    \includegraphics[width = 0.2\textwidth]{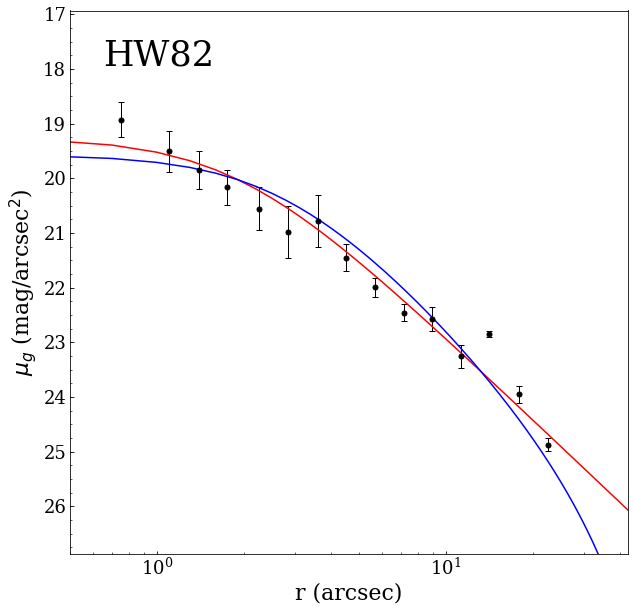}
    \includegraphics[width = 0.2\textwidth]{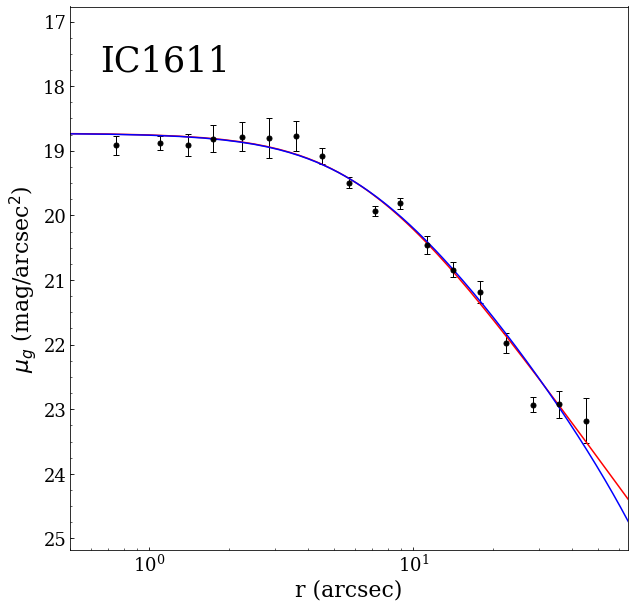}\\
    \hspace{-1.cm}
    \includegraphics[width = 0.2\textwidth]{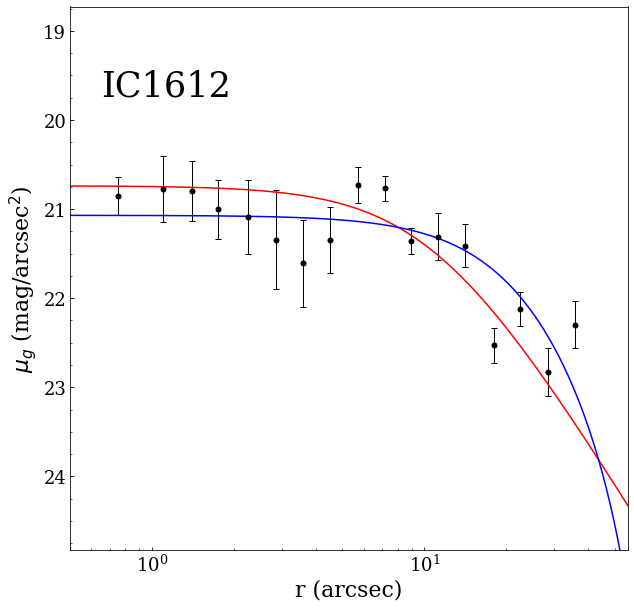}
    \includegraphics[width = 0.2\textwidth]{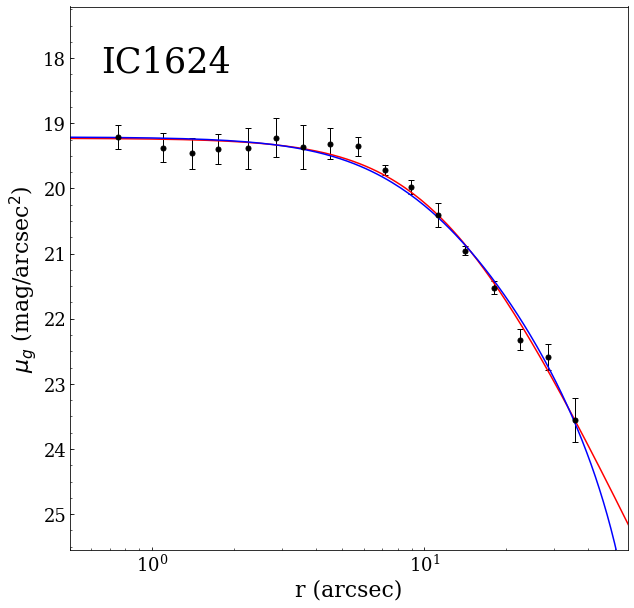}
    \includegraphics[width = 0.2\textwidth]{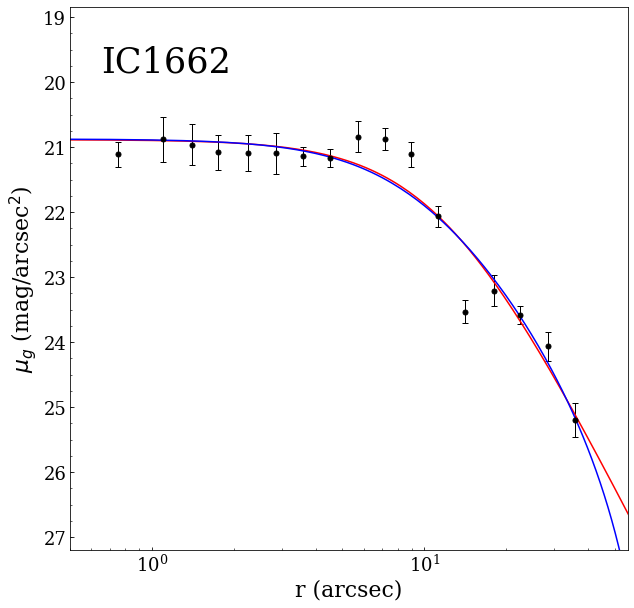}
    \includegraphics[width = 0.2\textwidth]{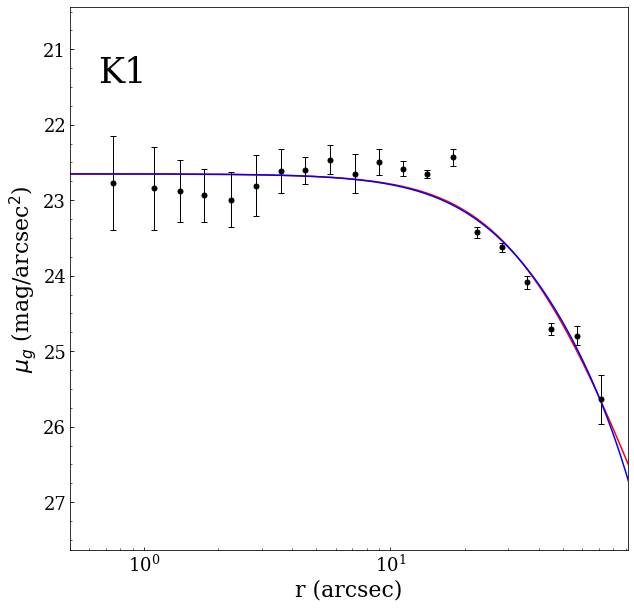}
    \includegraphics[width = 0.2\textwidth]{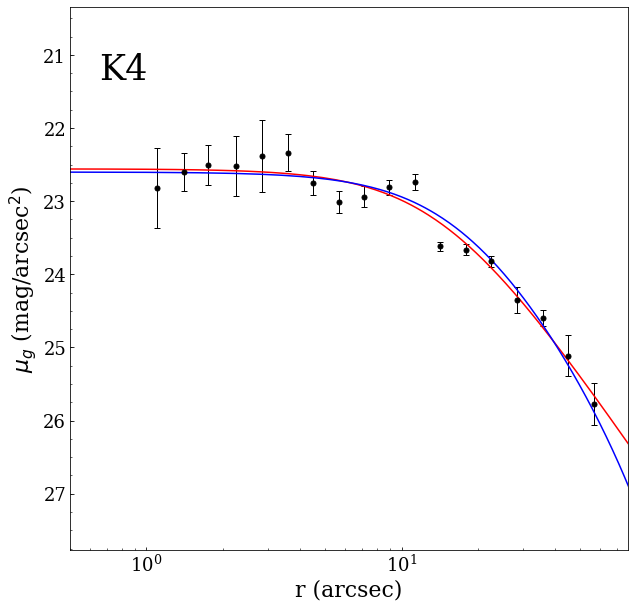}\\
    \hspace{-1.cm}
    \includegraphics[width = 0.2\textwidth]{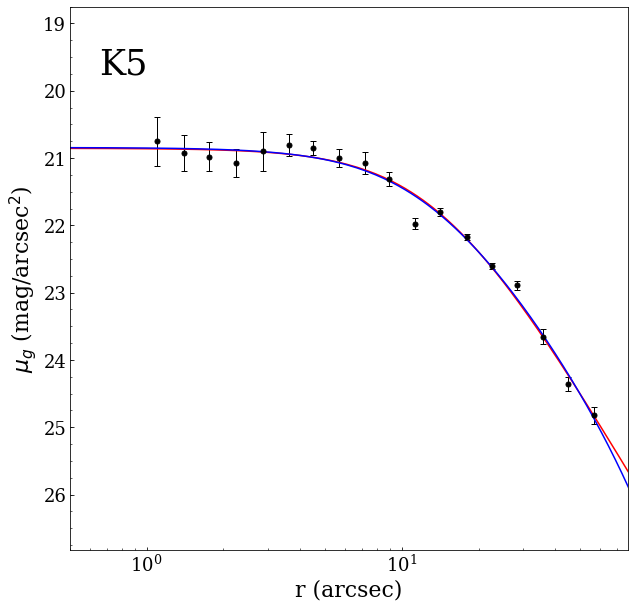}
    \includegraphics[width = 0.2\textwidth]{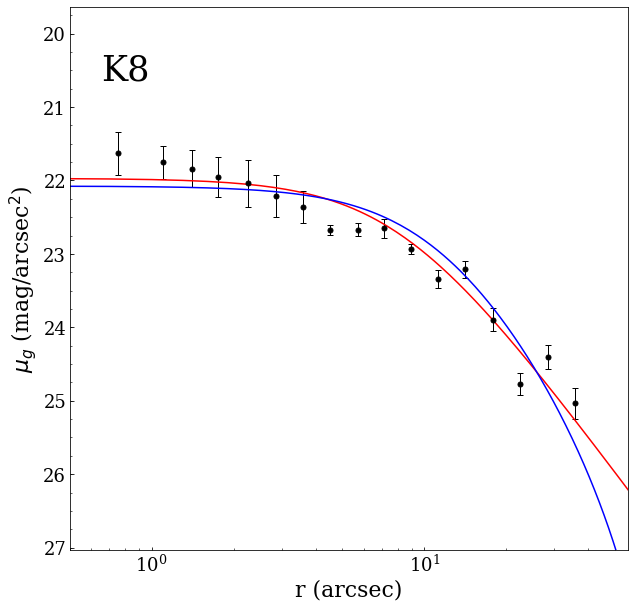}
    \includegraphics[width = 0.2\textwidth]{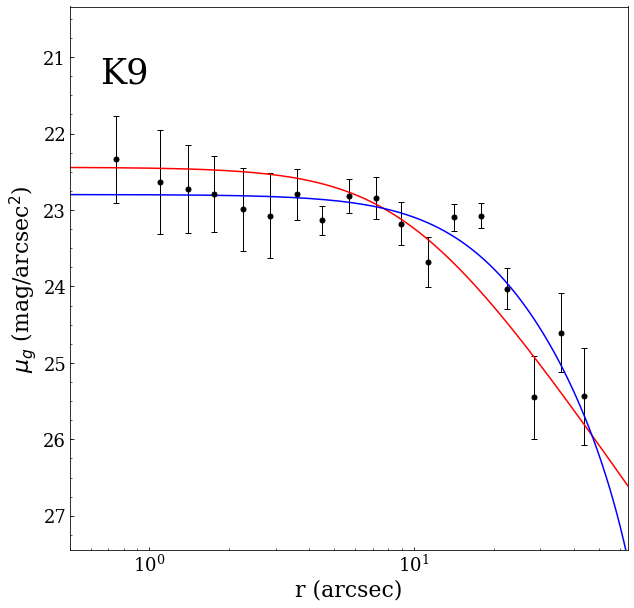}
    \includegraphics[width = 0.2\textwidth]{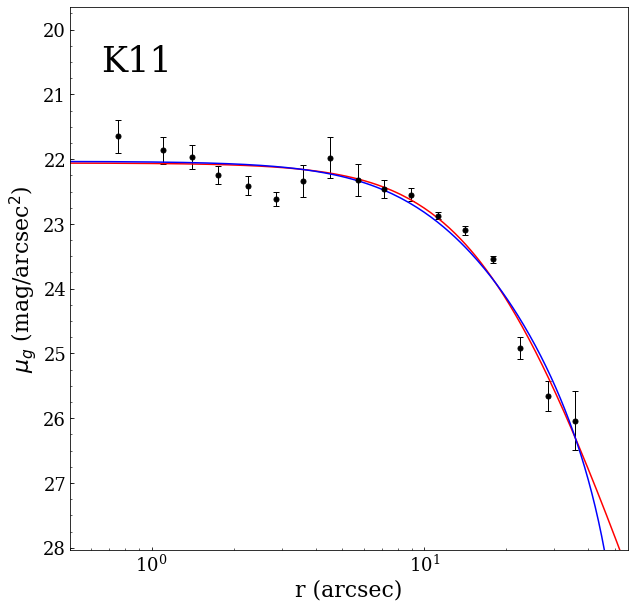}
    \includegraphics[width = 0.2\textwidth]{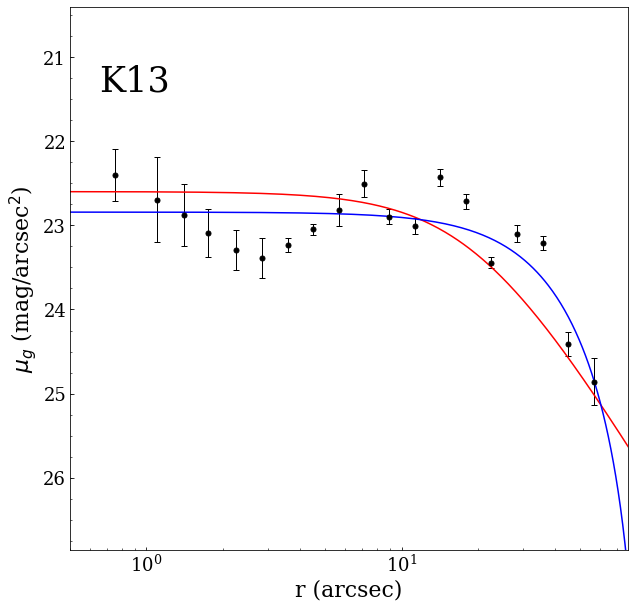}\\
    \hspace{-1.cm}
    \includegraphics[width = 0.2\textwidth]{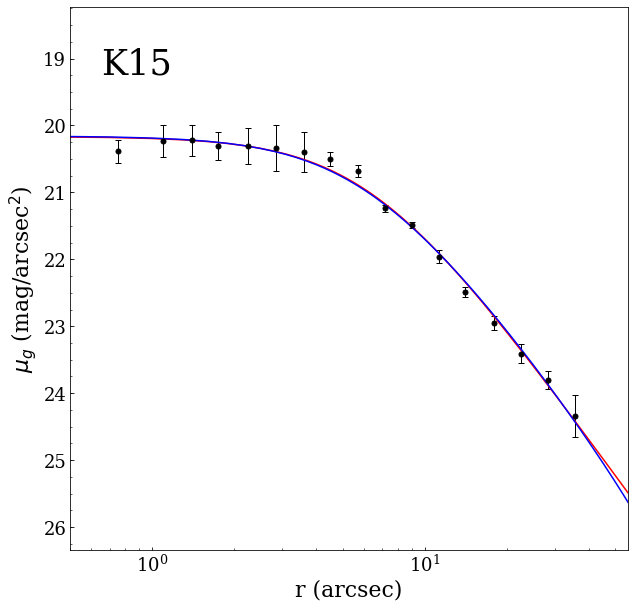}
    \includegraphics[width = 0.2\textwidth]{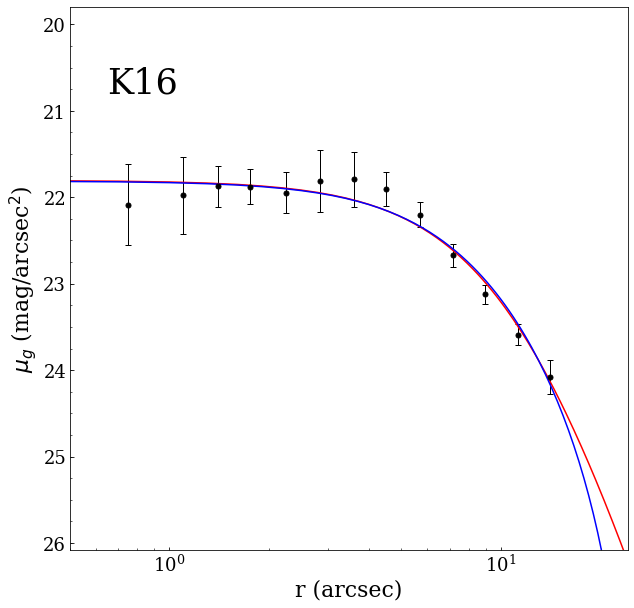}
    \includegraphics[width = 0.2\textwidth]{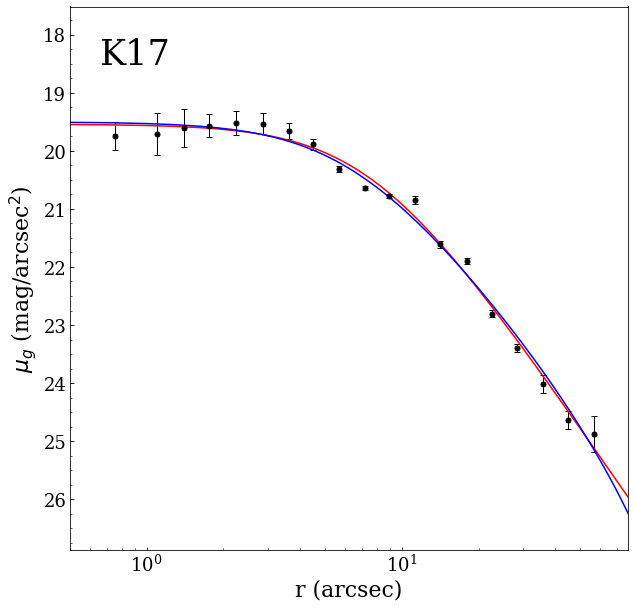}
    \includegraphics[width = 0.2\textwidth]{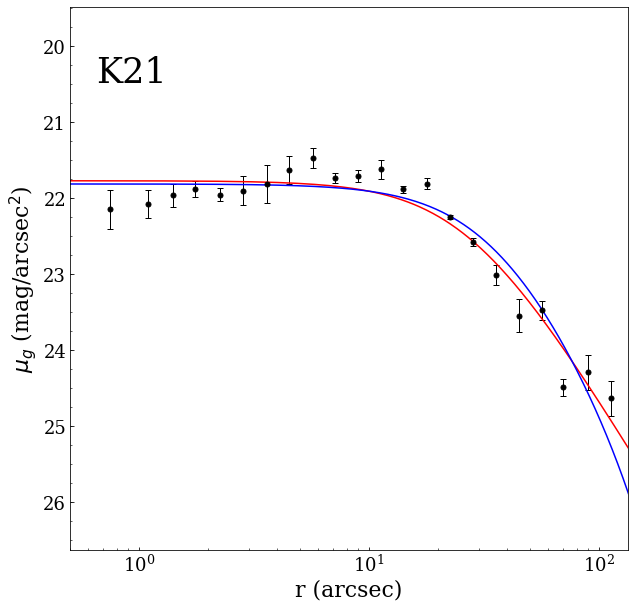}
    \includegraphics[width = 0.2\textwidth]{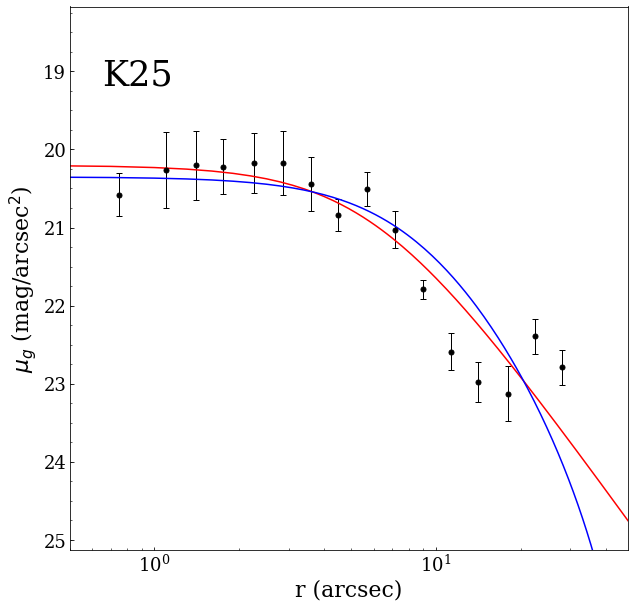}\\
    \hspace{-1.cm}
    \includegraphics[width = 0.2\textwidth]{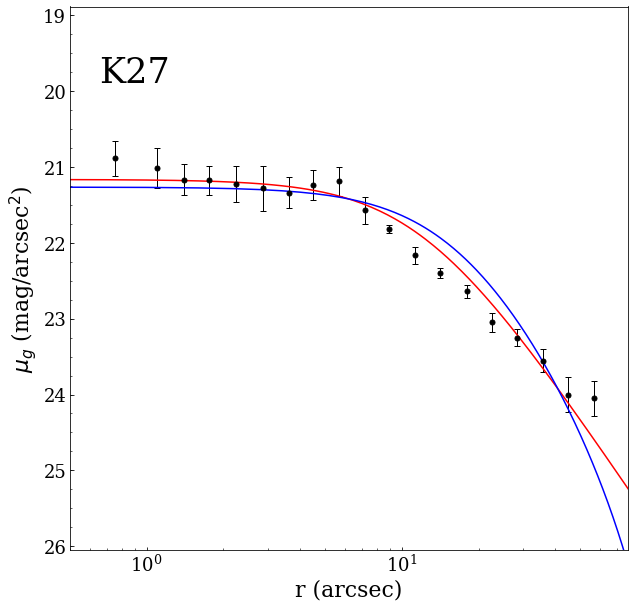}
    \includegraphics[width = 0.2\textwidth]{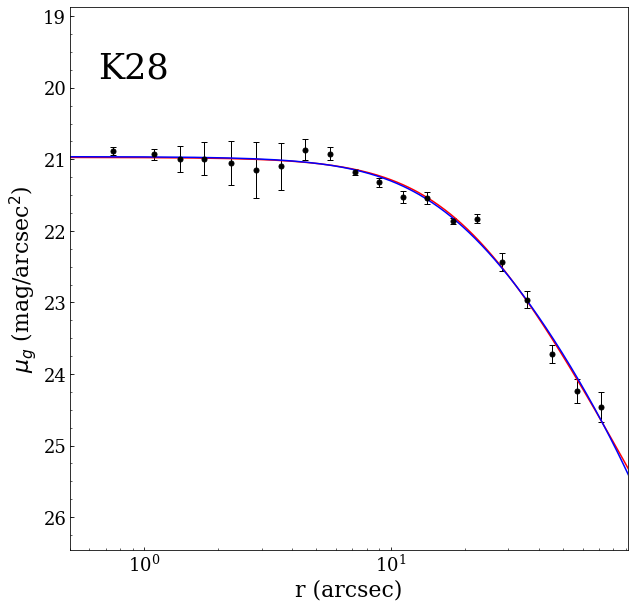}
    \includegraphics[width = 0.2\textwidth]{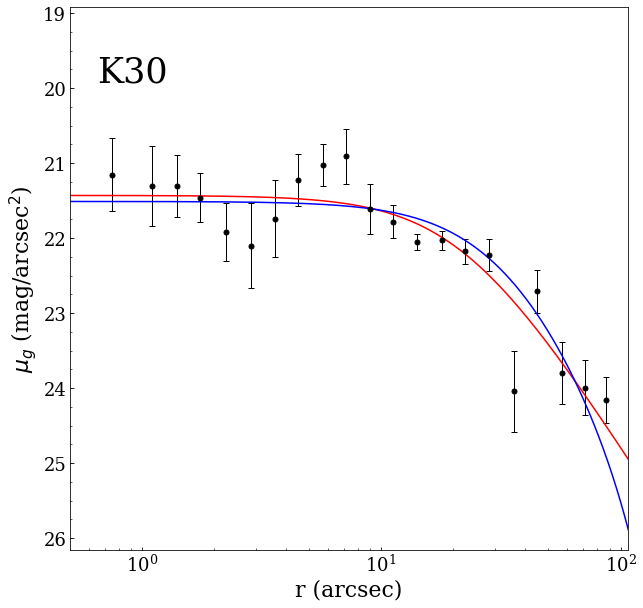}
    \includegraphics[width = 0.2\textwidth]{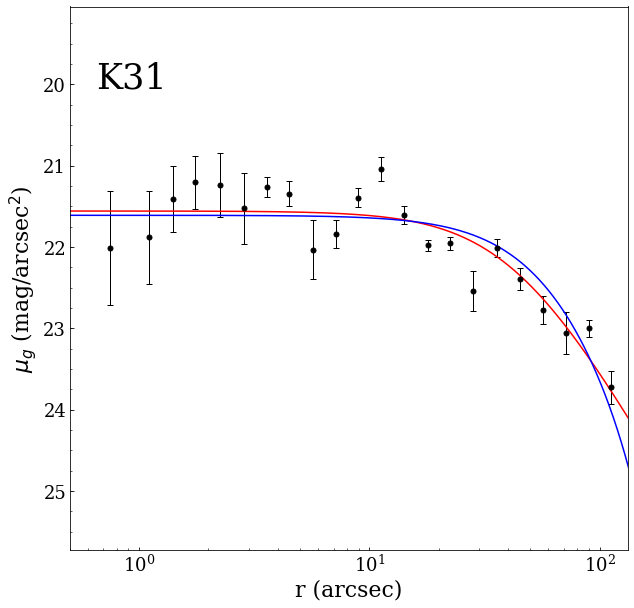}
    \includegraphics[width = 0.2\textwidth]{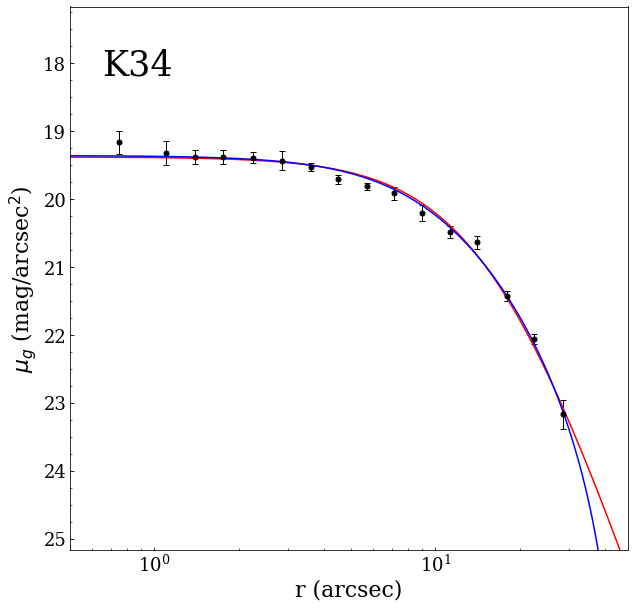}\\
    \hspace{-1.cm}
    \includegraphics[width = 0.2\textwidth]{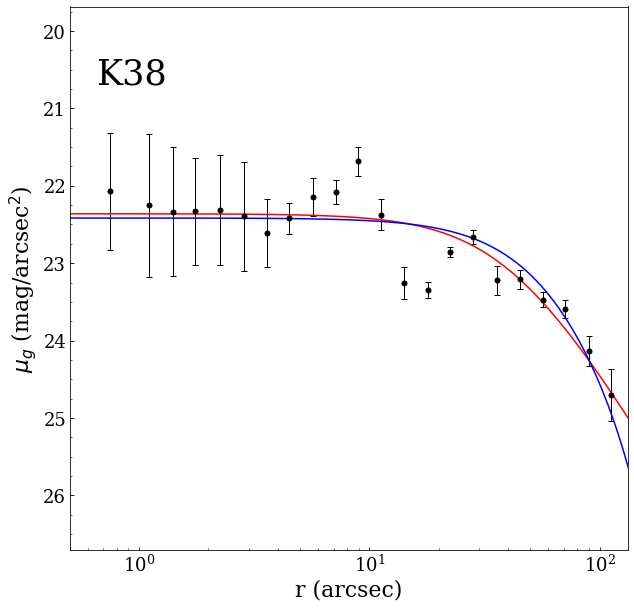}
    \includegraphics[width = 0.2\textwidth]{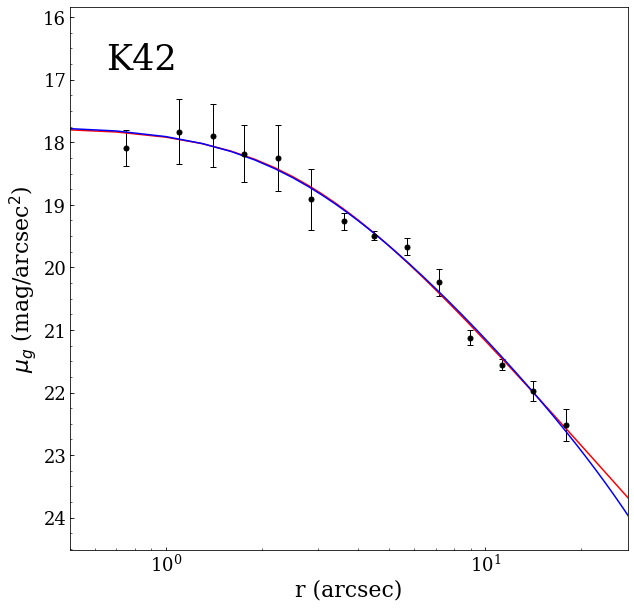}
    \includegraphics[width = 0.2\textwidth]{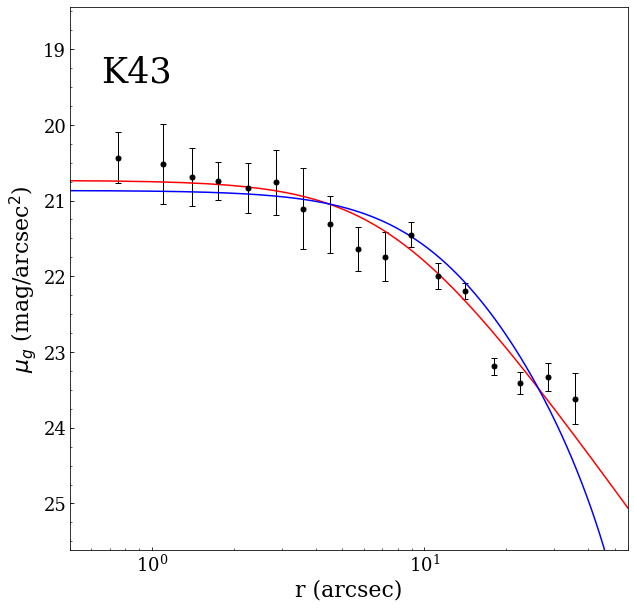}
    \includegraphics[width =0.21\textwidth]{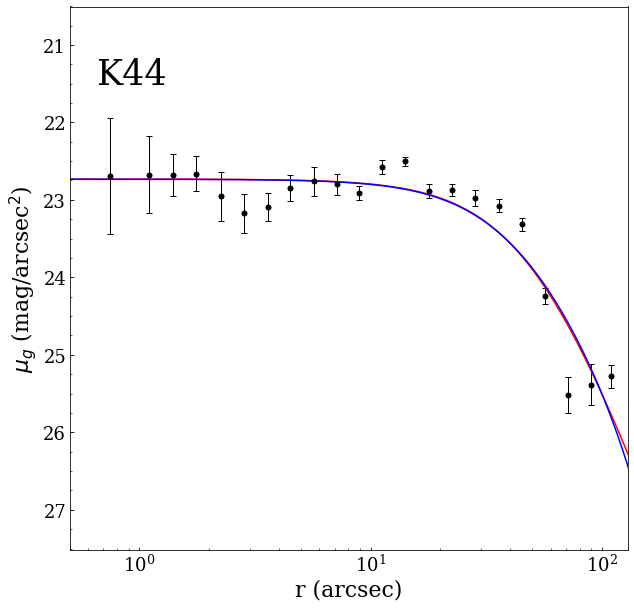}
    \includegraphics[width = 0.2\textwidth]{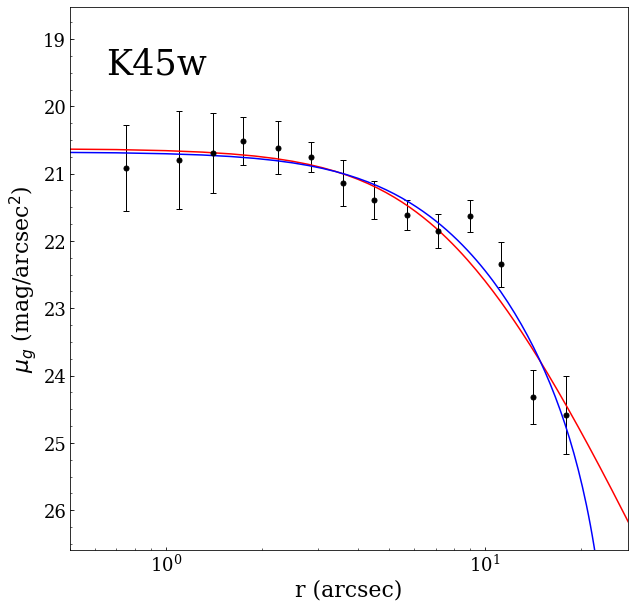}\\
    \contcaption{}
\end{figure*}

\begin{figure*}
    \centering
    \hspace{-1.cm}
    \includegraphics[width = 0.2\textwidth]{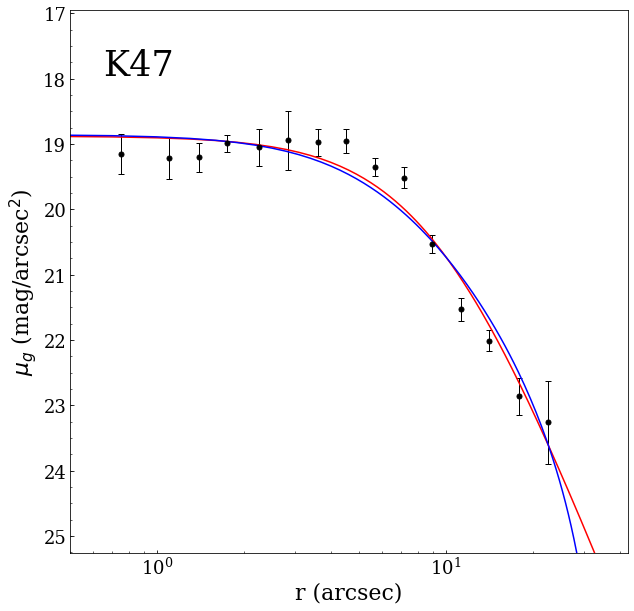}
    \includegraphics[width = 0.2\textwidth]{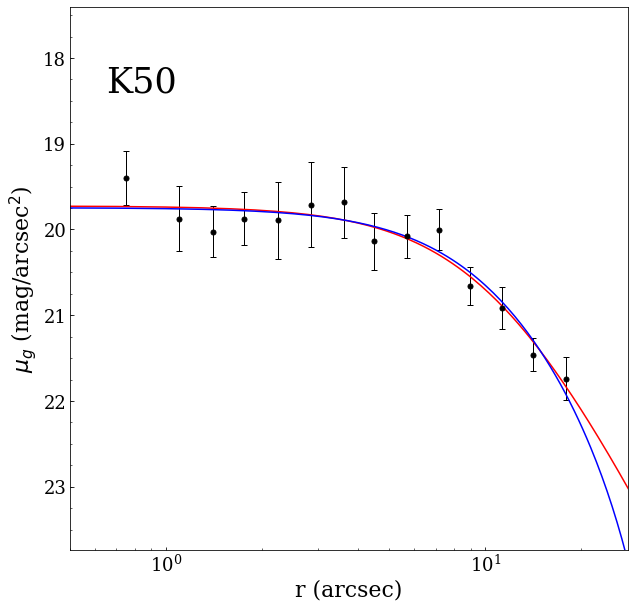}
    \includegraphics[width = 0.2\textwidth]{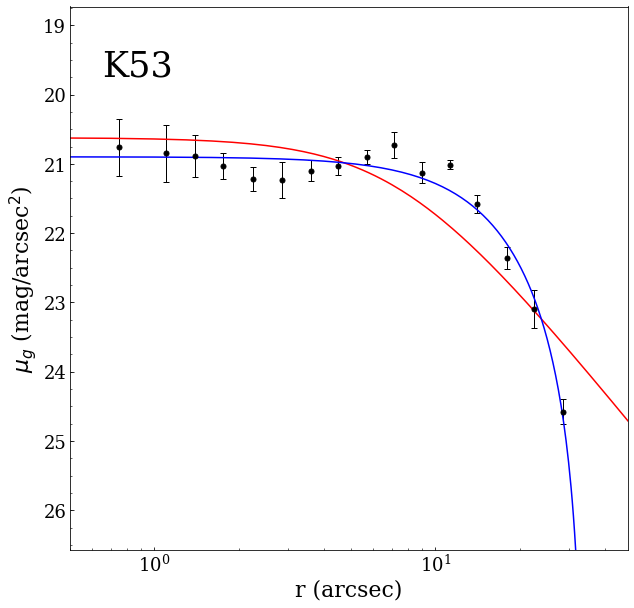}
    \includegraphics[width = 0.2\textwidth]{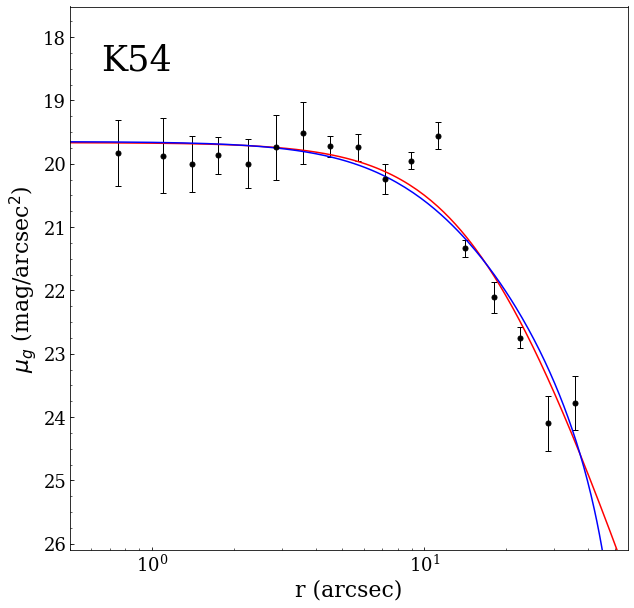}
    \includegraphics[width = 0.2\textwidth]{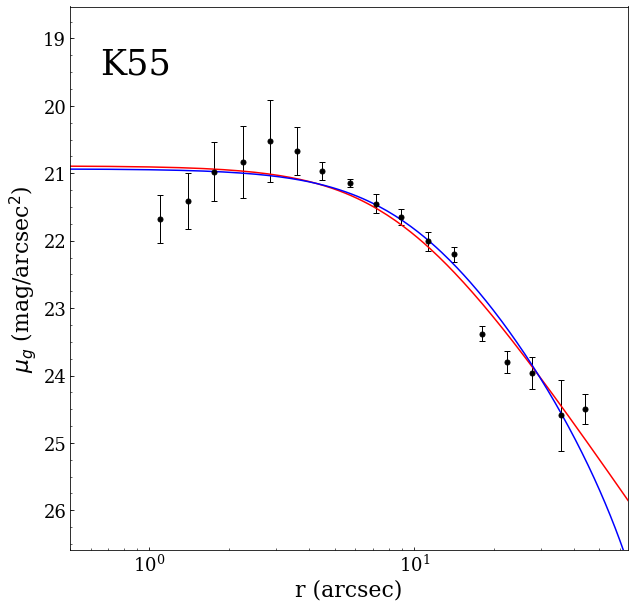}\\
    \hspace{-1.cm}
    \includegraphics[width = 0.2\textwidth]{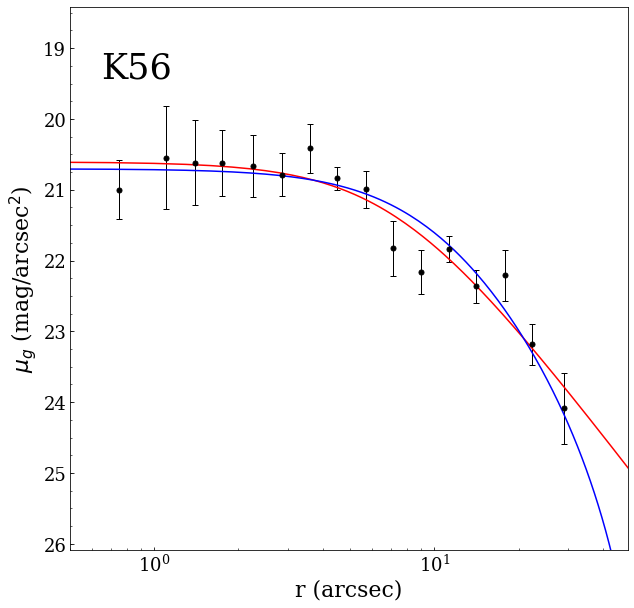}
    \includegraphics[width = 0.2\textwidth]{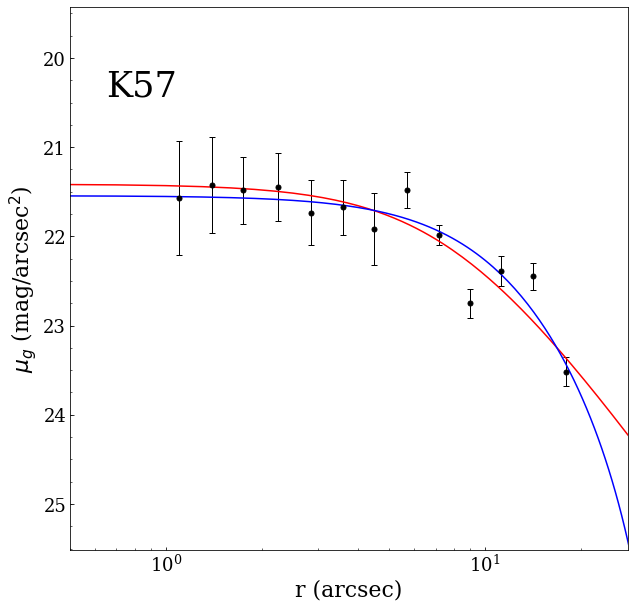}
    \includegraphics[width = 0.2\textwidth]{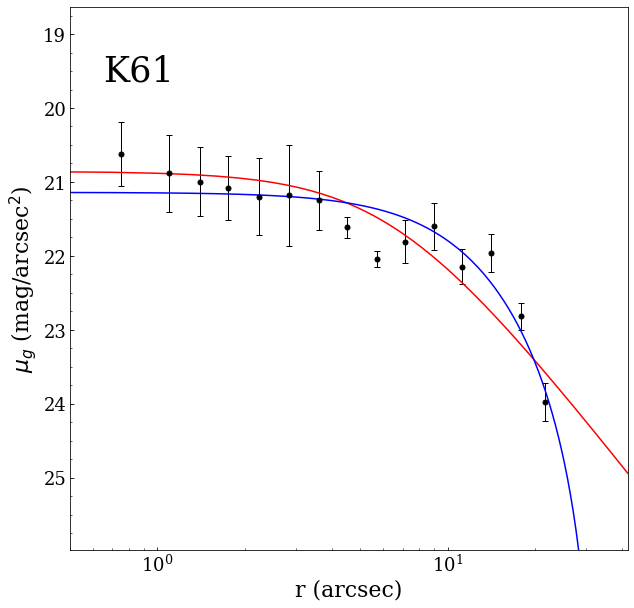}
    \includegraphics[width = 0.2\textwidth]{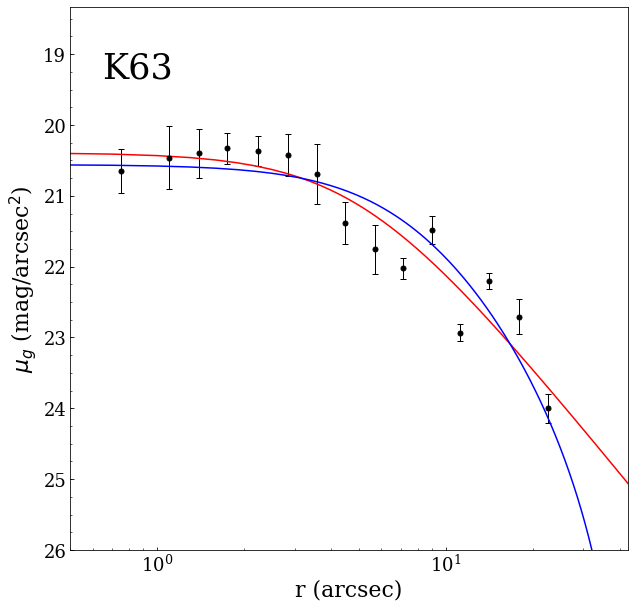}
    \includegraphics[width = 0.2\textwidth]{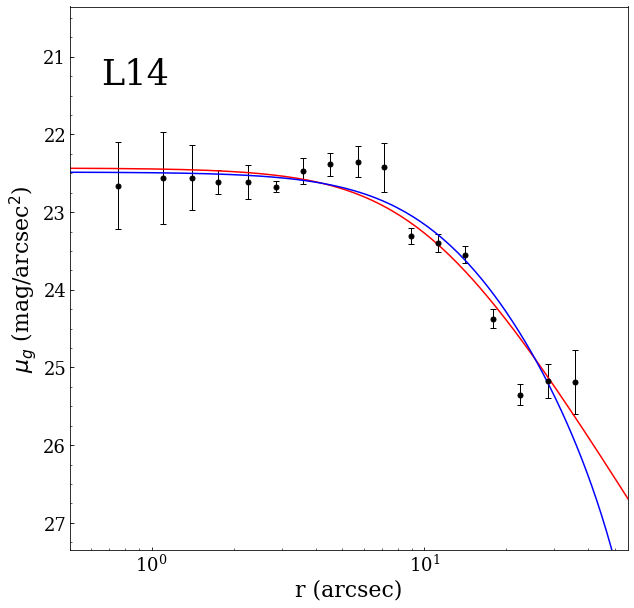}\\
    \hspace{-1.cm}
    \includegraphics[width = 0.2\textwidth]{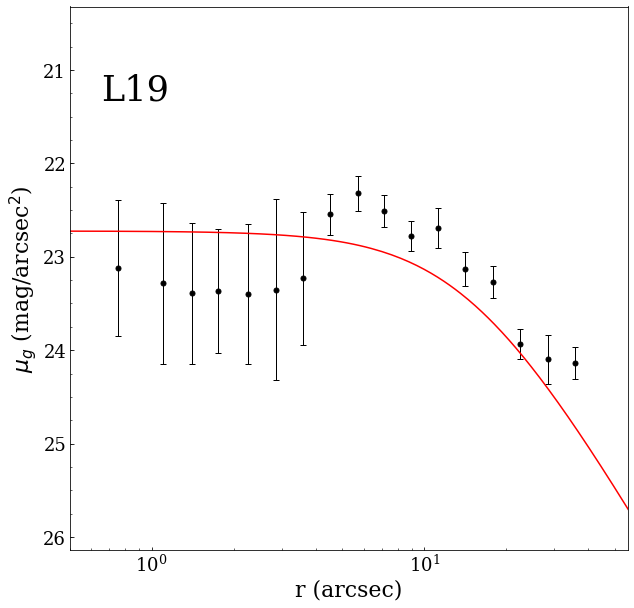}
    \includegraphics[width = 0.2\textwidth]{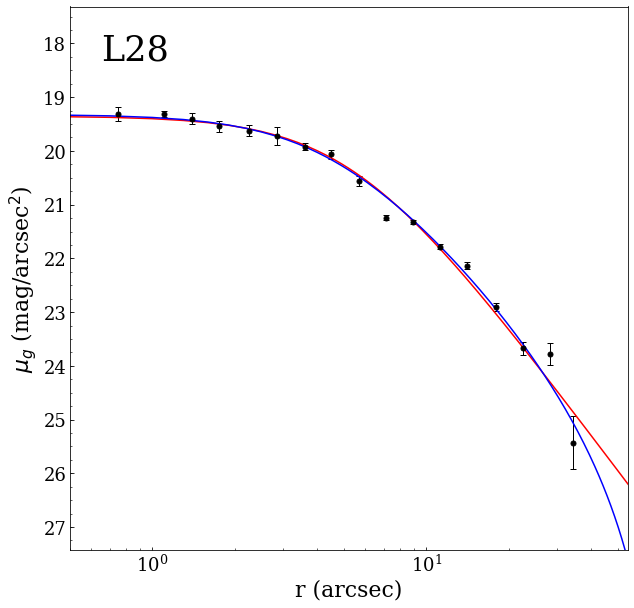}
    \includegraphics[width = 0.2\textwidth]{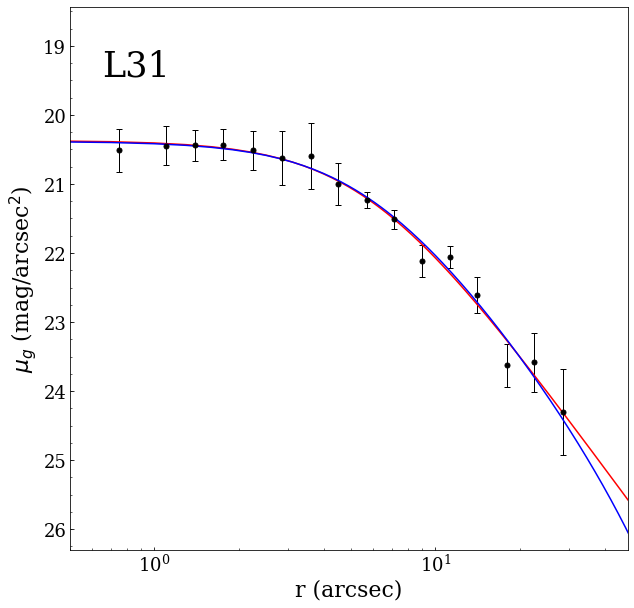}
    \includegraphics[width = 0.2\textwidth]{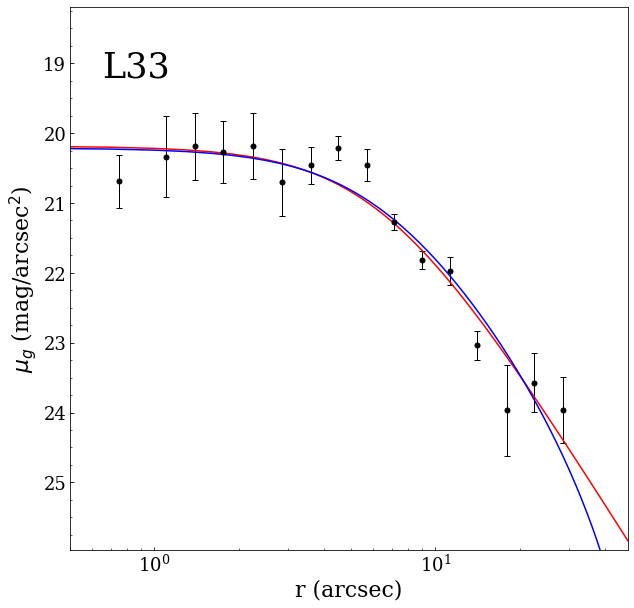}
    \includegraphics[width = 0.2\textwidth]{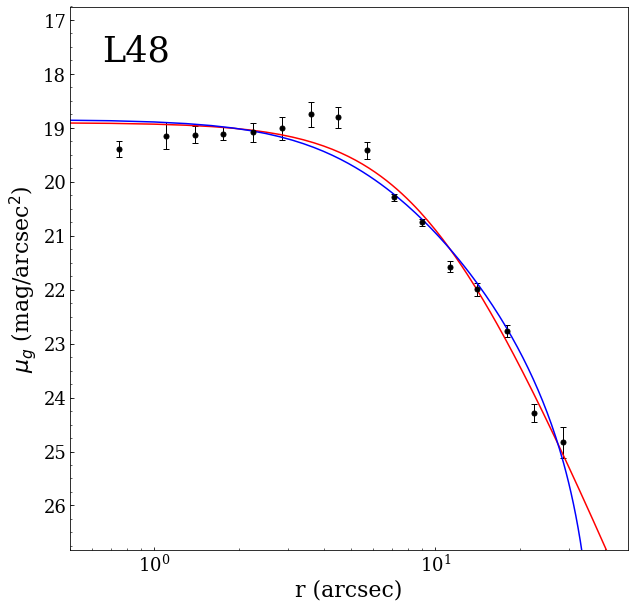}\\
    \hspace{-1.cm}
    \includegraphics[width = 0.2\textwidth]{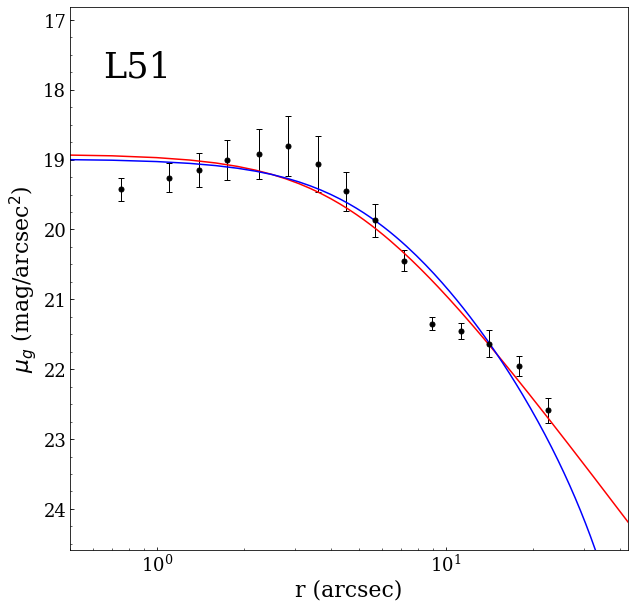}
    \includegraphics[width = 0.2\textwidth]{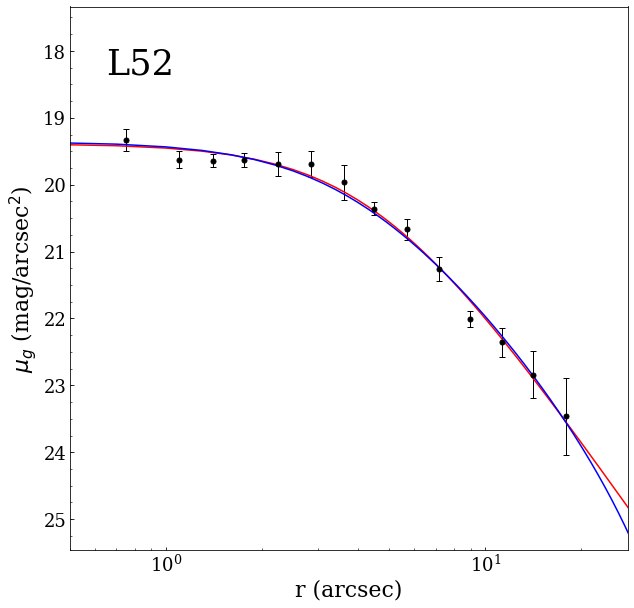}
    \includegraphics[width = 0.2\textwidth]{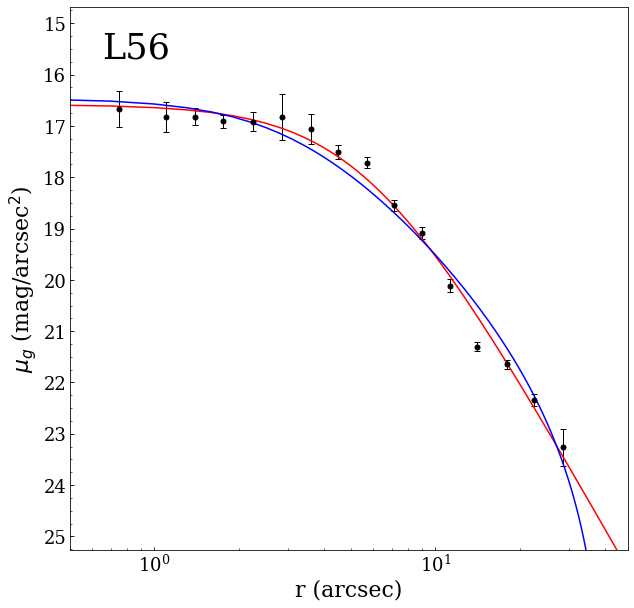}
    \includegraphics[width = 0.2\textwidth]{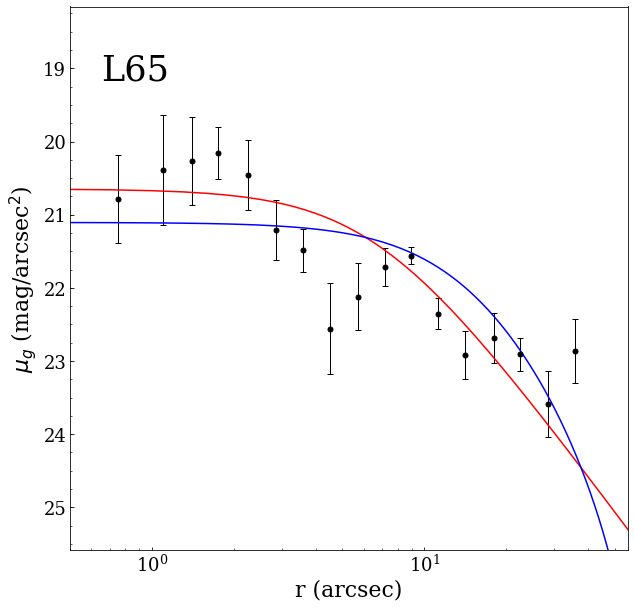}
    \includegraphics[width = 0.2\textwidth]{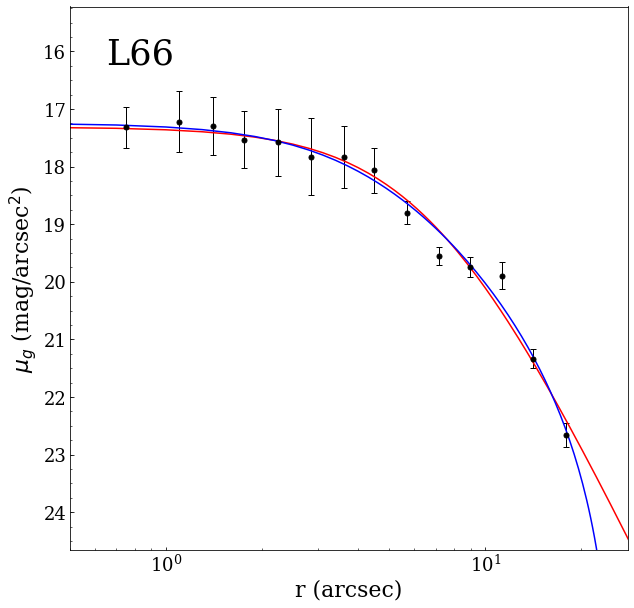}\\
    \hspace{-1.cm}
    \includegraphics[width = 0.2\textwidth]{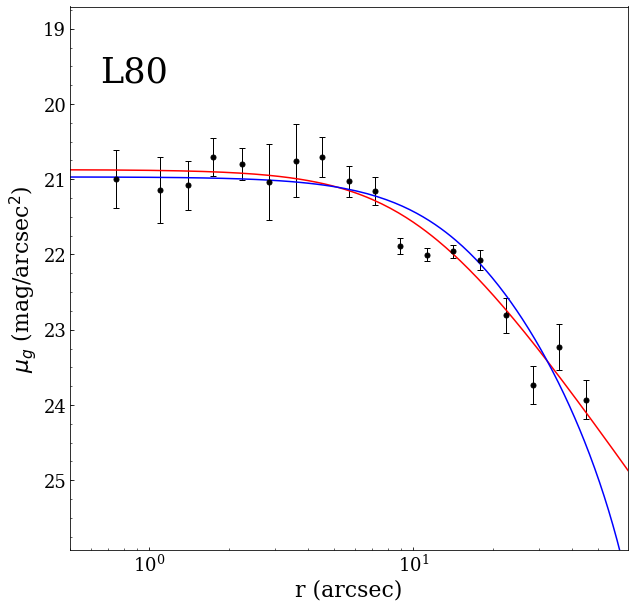}
    \includegraphics[width = 0.2\textwidth]{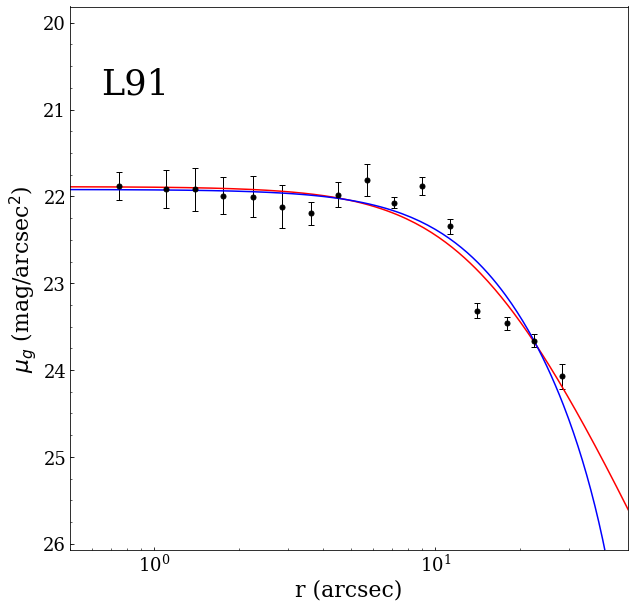}
    \includegraphics[width = 0.2\textwidth]{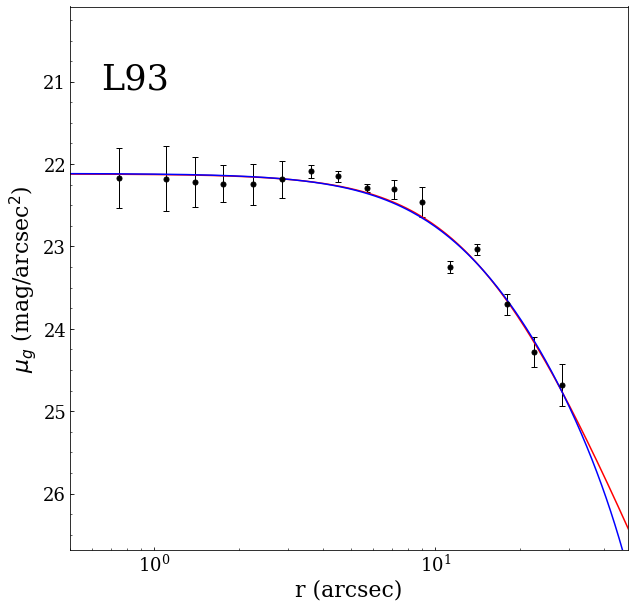}
    \includegraphics[width = 0.2\textwidth]{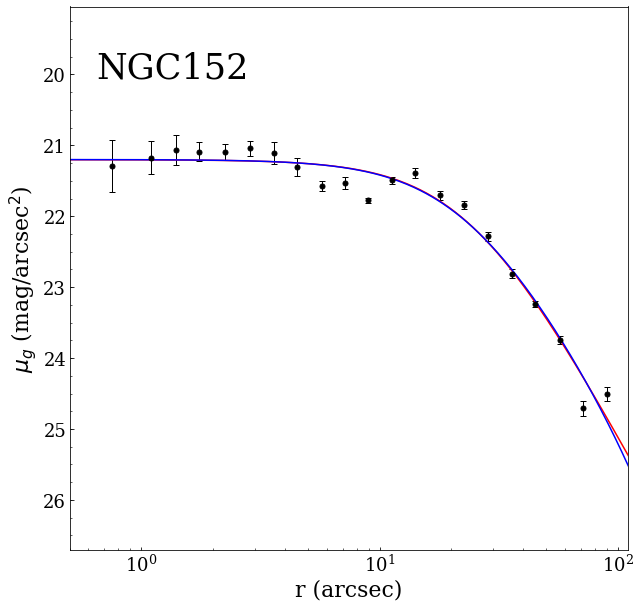}
    \includegraphics[width = 0.2\textwidth]{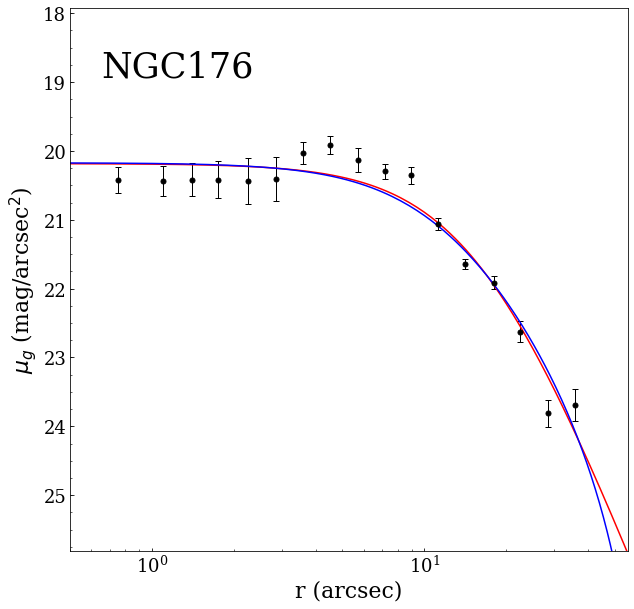}\\
    \hspace{-1.cm}
    \includegraphics[width = 0.2\textwidth]{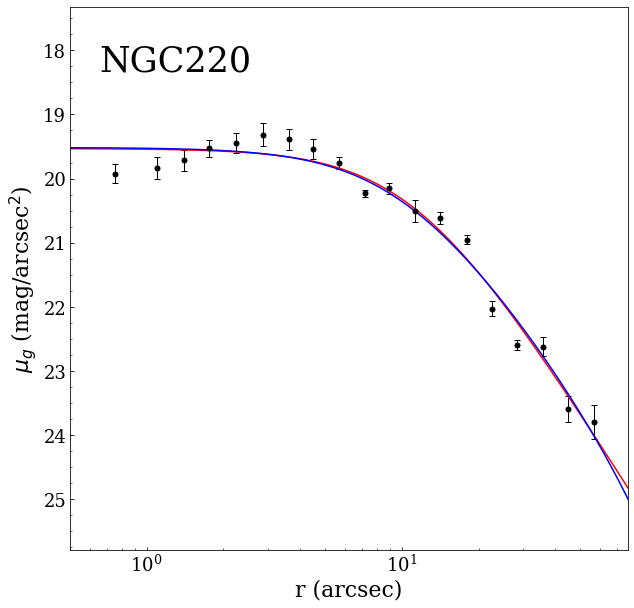}
    \includegraphics[width = 0.2\textwidth]{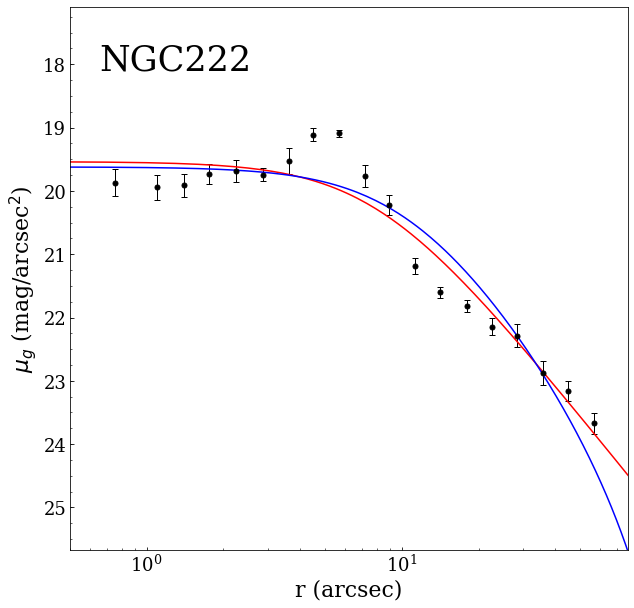}
    \includegraphics[width = 0.2\textwidth]{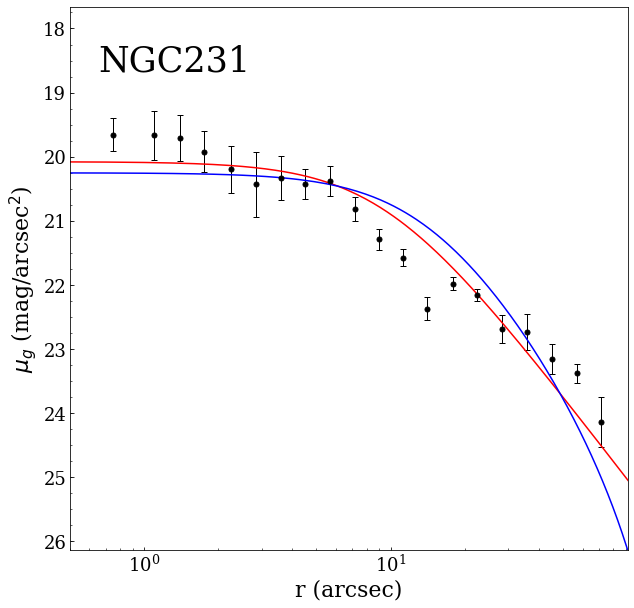}
    \includegraphics[width = 0.2\textwidth]{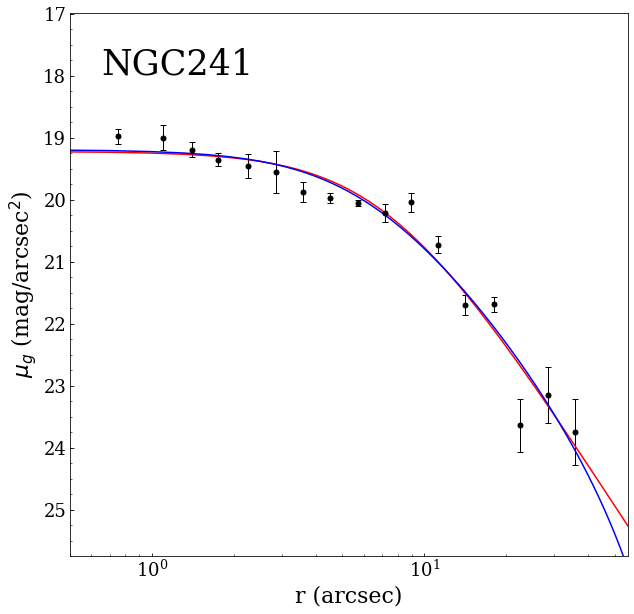}
    \includegraphics[width = 0.2\textwidth]{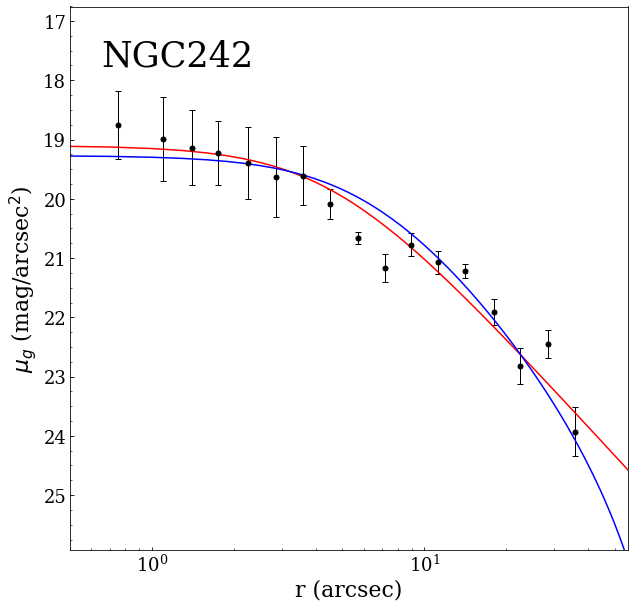}\\
    
    \hspace{-1.cm}
    
    \contcaption{}
\end{figure*}

\begin{figure*}
    \centering
    \hspace{-1.cm}
    \includegraphics[width = 0.2\textwidth]{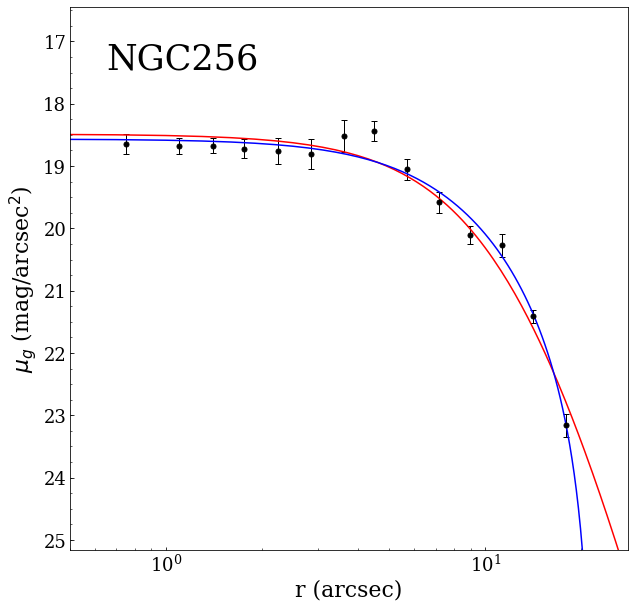}
    \includegraphics[width = 0.2\textwidth]{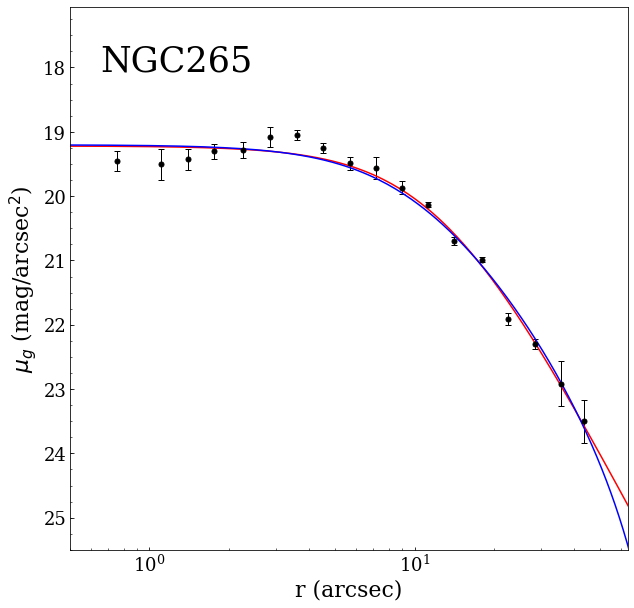}
    \includegraphics[width = 0.2\textwidth]{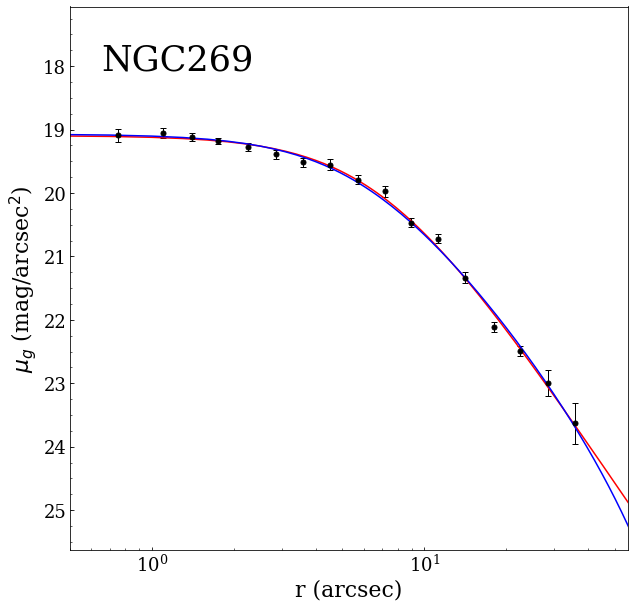}
    \includegraphics[width = 0.2\textwidth]{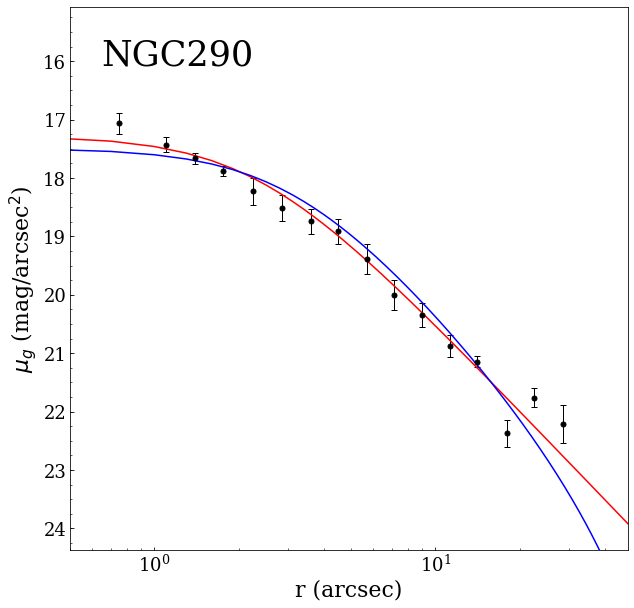}
    \includegraphics[width = 0.2\textwidth]{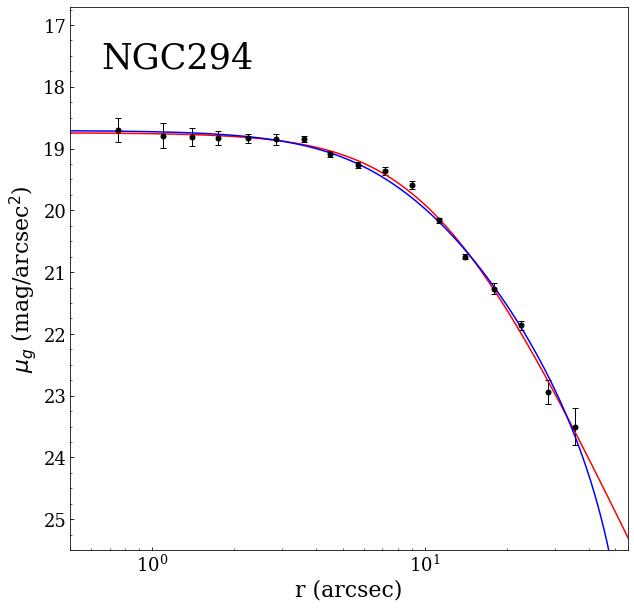}\\
    \hspace{-1.cm}
    \includegraphics[width = 0.2\textwidth]{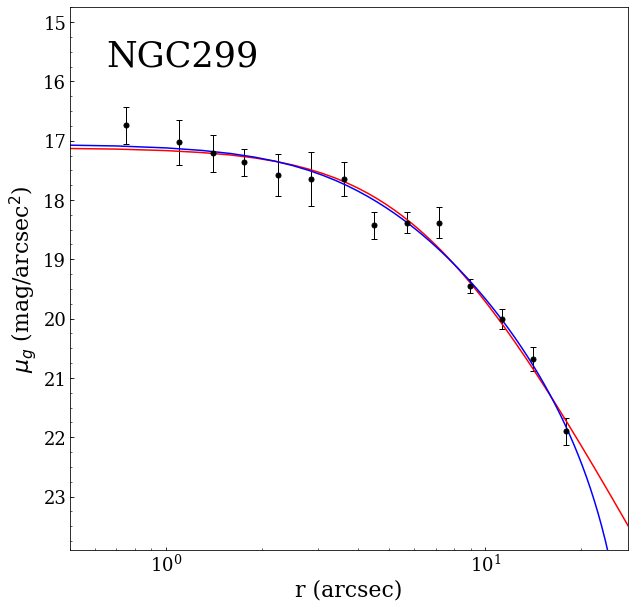}
    \includegraphics[width = 0.2\textwidth]{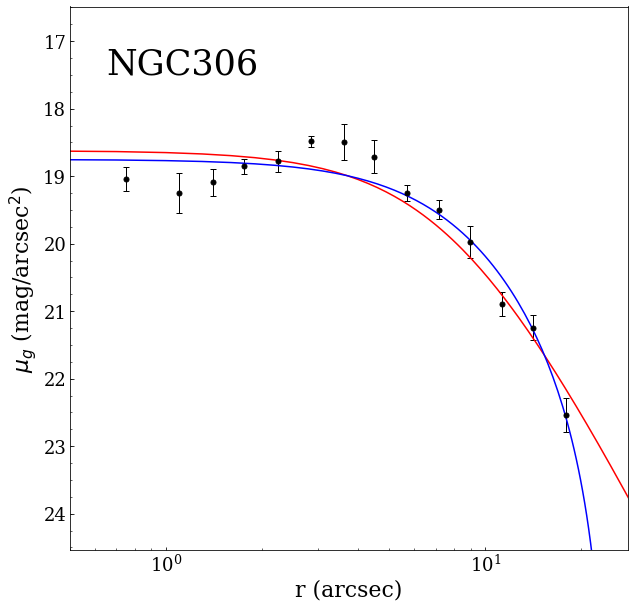}
    \includegraphics[width = 0.2\textwidth]{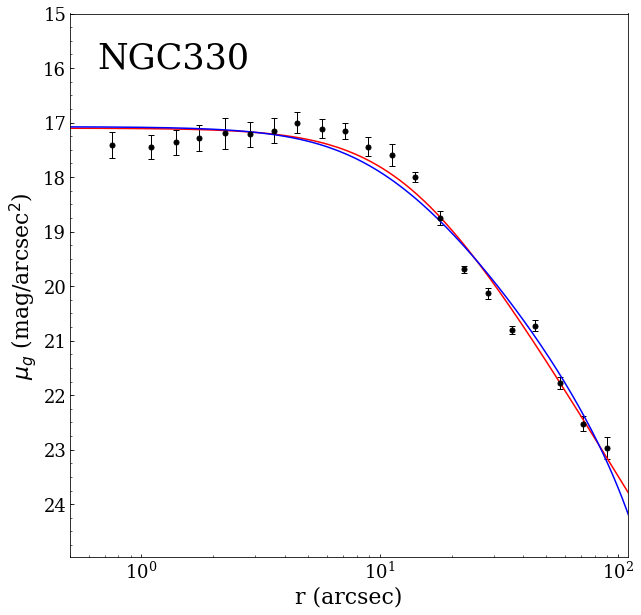}
    \includegraphics[width = 0.2\textwidth]{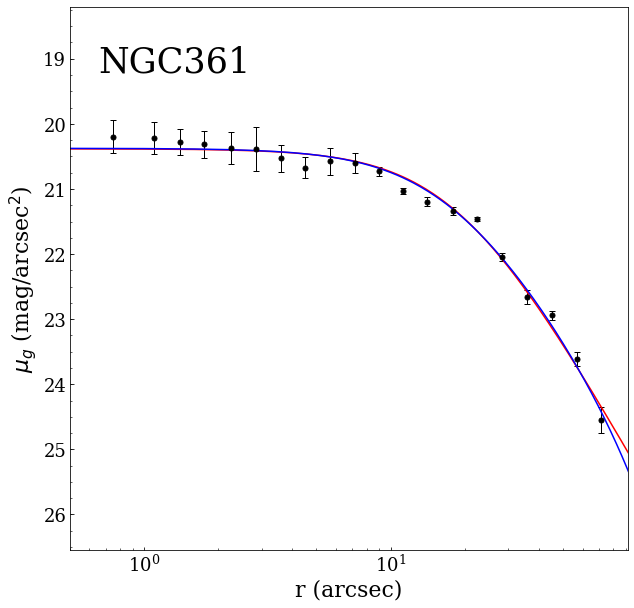}
    \includegraphics[width = 0.2\textwidth]{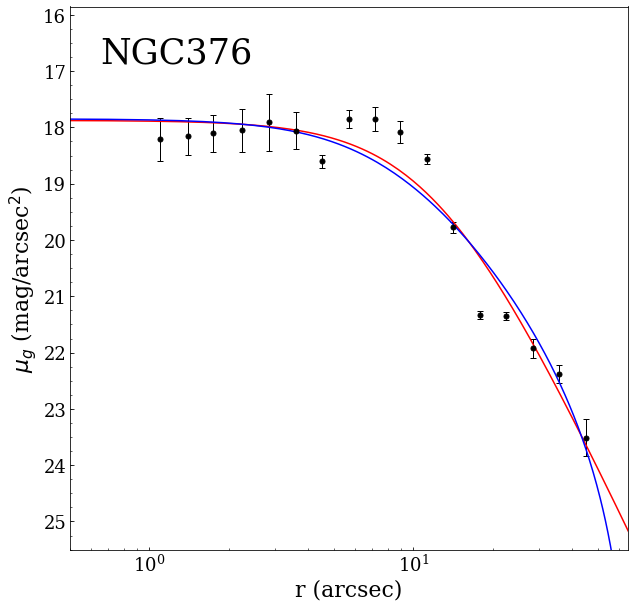}\\
    \hspace{-1.cm}
    \includegraphics[width = 0.2\textwidth]{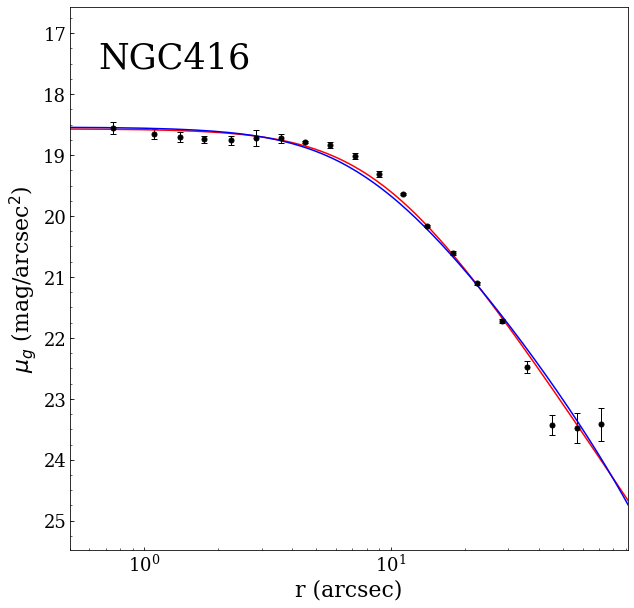}
    \includegraphics[width = 0.2\textwidth]{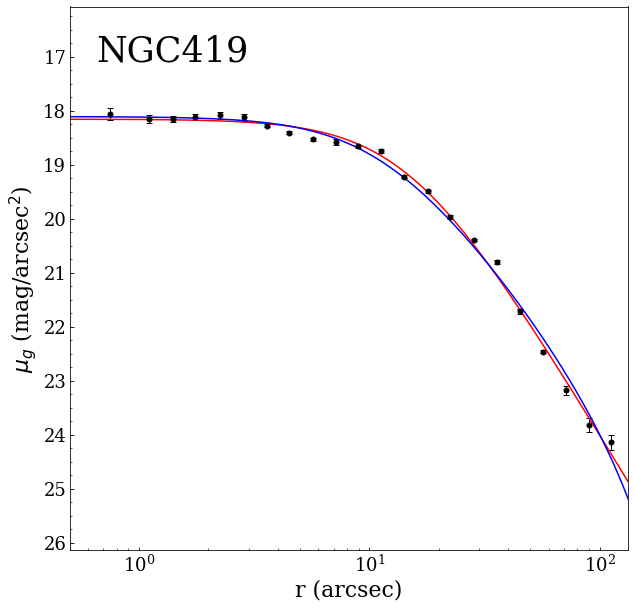}
    \includegraphics[width = 0.2\textwidth]{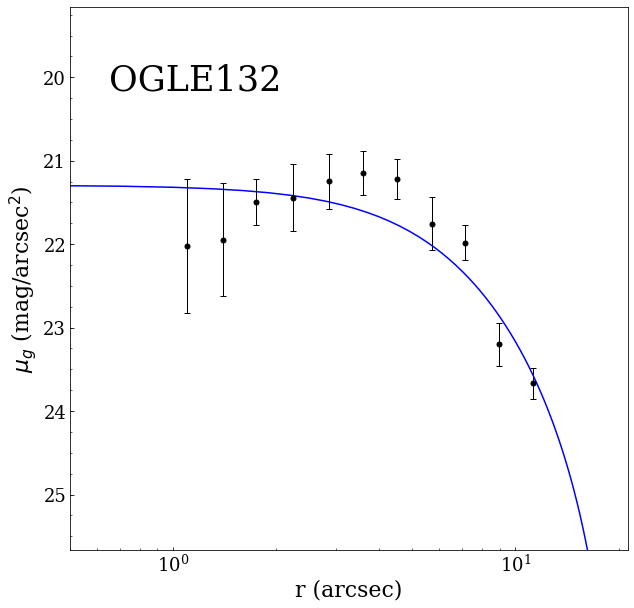}
    \includegraphics[width = 0.2\textwidth]{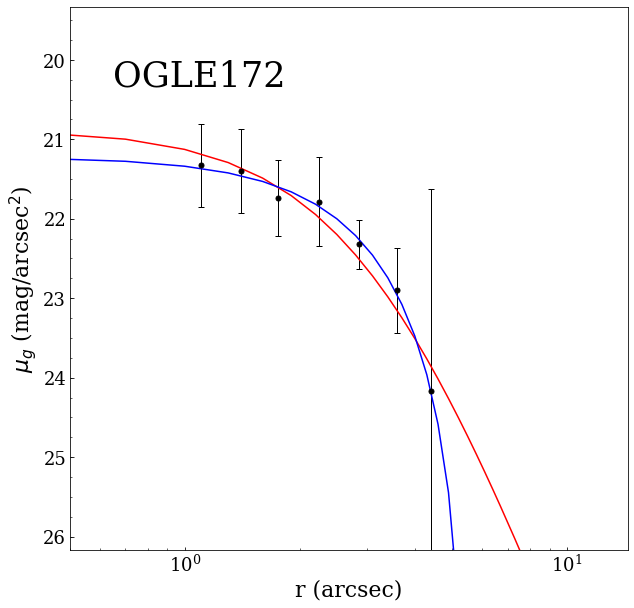}
    \includegraphics[width = 0.2\textwidth]{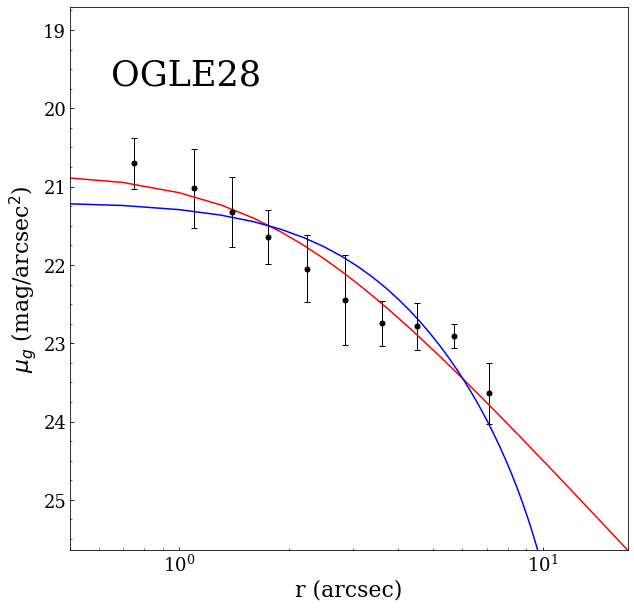}\\
    \hspace{-1.cm}
    \includegraphics[width = 0.2\textwidth]{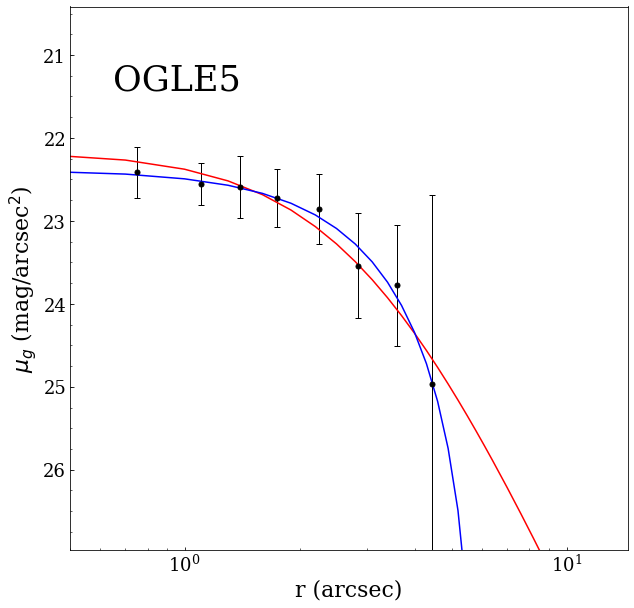}
    \includegraphics[width = 0.2\textwidth]{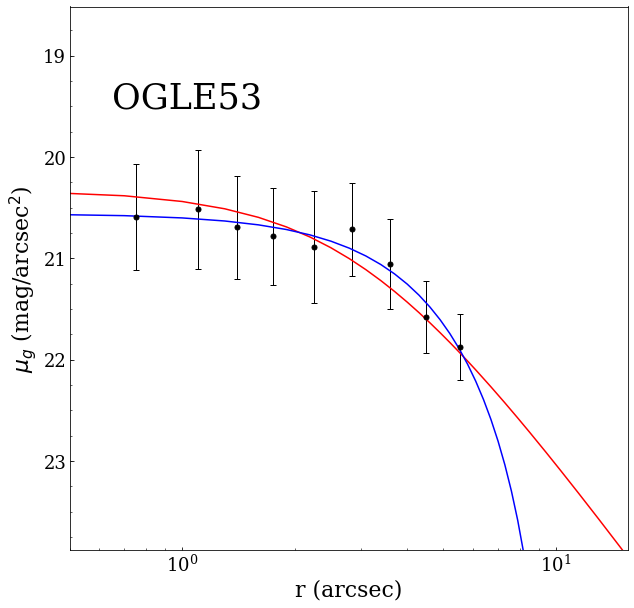}
    \includegraphics[width = 0.2\textwidth]{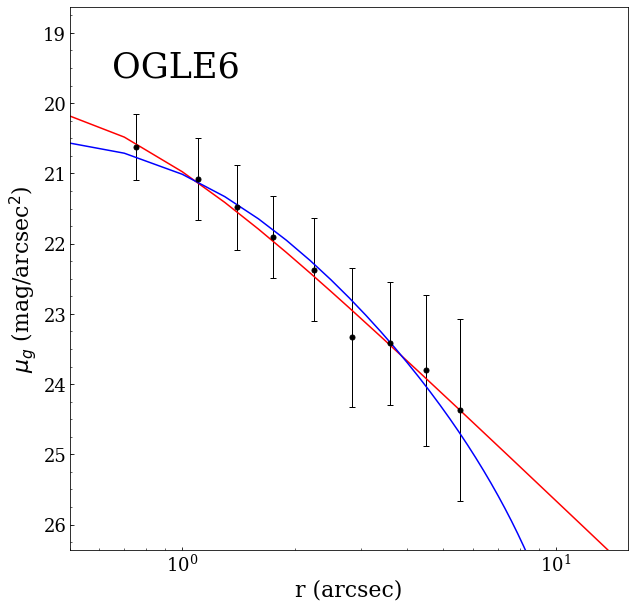}
    \includegraphics[width = 0.2\textwidth]{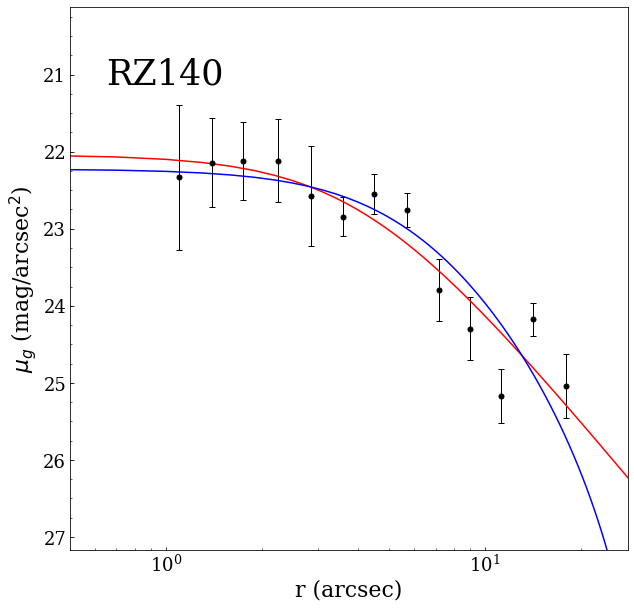}
    \includegraphics[width = 0.2\textwidth]{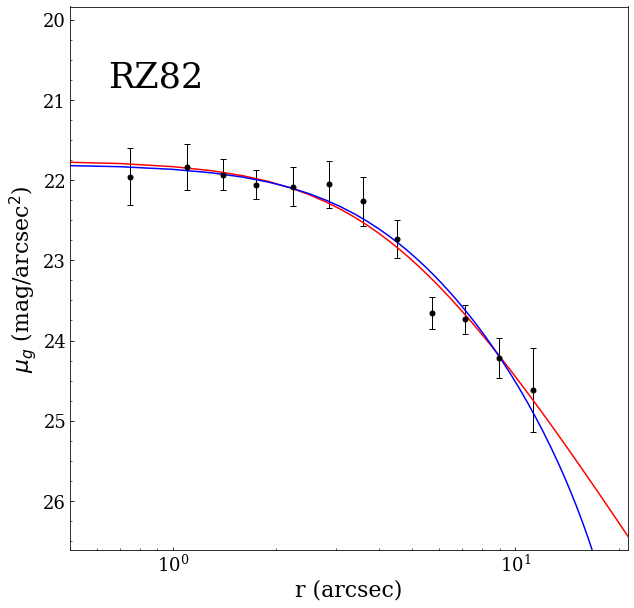}\\
    \contcaption{}
\end{figure*}

\bsp	
\label{lastpage}
\end{document}